\documentstyle[12pt,rotating]{article}
\input{psfig}
\newdimen\fullhsize\newdimen\play\newdimen\hmargins\newdimen\vmargins
\def\newmargins#1#2{\hmargins=#1\vmargins=#2
   \play=8.5truein\advance\play by-2\hmargins\global\fullhsize=\play
   \advance\play by-4\hmargins\divide\play by3\global\hsize=\play
   \play=11truein\advance\play by-2\vmargins\global\vsize=\play
   \play=-1truein\advance\play by\hmargins\global\hoffset=\play
   \play=-0.65truein\advance\play by\vmargins\global\voffset=\play}

\newmargins{0.3truein}{-0.25truein}
\rm
\newskip\humongous \humongous=0pt plus 1000pt minus 1000pt

\newif\ifdtup

\def\beq{\begin{equation}}
\def\eeq{\end{equation}}

\relax

\def\kpnn{{\mbox{$K_L \rightarrow \pi^0 \nu \overline{\nu}\ $}}}
\def\kpee{{\mbox{$K_L \rightarrow \pi^0e^+e^- $}}}
\def\kpmm{{\mbox{$K_L \rightarrow \pi^0 \mu^+\mu^- $}}}
\def\kpme{{\mbox{$K_L \rightarrow \pi^0\mu^\pm e^\mp $}}}
\def\kppee{{\mbox{$K_L \rightarrow \pi^+\pi^-e^+e^-$}}}
\def\kmm{{\mbox{$K_L \rightarrow \mu^+\mu^-$}}}
\def\kme{{\mbox{$K_L \rightarrow \mu^\pm e^\mp $}}}
\def\kee{{\mbox{$K_L \rightarrow e^+ e^- $}}}
\def\kpipi{{\mbox{$K_L \rightarrow 2\pi^0\ $}}}
\def\kpipipi{{\mbox{$K_L \rightarrow 3\pi^0\ $}}}

\def\pnn{{\mbox{$\pi^0 \nu \overline{\nu}\ $}}}

\def\pmm{{\mbox{$\pi^0 \mu^+\mu^- $}}}
\def\pme{{\mbox{$\pi^0\mu^\pm e^\mp $}}}
\def\ppee{{\mbox{$\pi^+\pi^-e^+e^-$}}}
\def\mm{{\mbox{$\mu^+\mu^-$}}}

\def\ee{{\mbox{$e^+ e^- $}}}

\newcommand{\kl}{{\rm K}_{L}}
\newcommand{\piz}{ \pi^{0} }
\newcommand{\piplus}{ \pi^{+} }
\newcommand{\eminus}{{\rm e}^{-}}
\newcommand{\kethree}{\kl \rightarrow \piplus \eminus \nu}
\newcommand{\klneut}{\kl \rightarrow \piz \piz}
\newcommand{\kpiznunu}{{\kl} \rightarrow \piz \nu \overline{\nu}}

\begin{document}       % End of preamble and beginning of text.

\title{
      {\bf An Expression of Interest to Detect and Measure
the Direct CP Violating Decay $K_L \rightarrow \pi^0 \nu \overline{\nu}$
and Other Rare Decays at Fermilab Using the Main Injector}} 

\date{September 22, 1997}
\maketitle

\begin{center}

\vspace*{0.3in}
\large
\bf The KAMI Collaboration
\vspace*{0.3in}

\small
E. Cheu, S.A. Taegar \\
\scriptsize
University of Arizona, Tucson, Arizona 85721 \\
\small
\vspace*{0.1in}

K. Arisaka, J. Park, W. Slater, A. Tripathi \\
\scriptsize
University of California at Los Angeles, Los Angeles, California 90095 \\
\small
\vspace*{0.1in}

C.O.~Escobar \\
\scriptsize
Universidade de Campinas, UNICAMP, CP 6165, Campinas 13083-970, SP, Brasil \\
\small
\vspace*{0.1in}

E. Blucher, R. Kessler, A. Roodman,  N. Solomey, C. Qiao,
Y. Wah, B. Winstein, R. Winston \\
\scriptsize
The Enrico Fermi Institute, The University of Chicago, Chicago Illinois
60637 \\
\small
\vspace*{0.1in}

A. R. Barker\\
\scriptsize
University of Colorado, Boulder, Colorado 80309 \\
\small
\vspace*{0.1in}

E. C. Swallow \\
\scriptsize
Department of Physics, Elmhurst College, Elmhurst, Illinois, 60126 and \\
The Enrico Fermi Institute, The University of Chicago, Chicago, Illinois
60637
\\
\small
\vspace*{0.1in}

G. Bock, R. Coleman, R. Ford, 
Y. B. Hsiung, D. A. Jensen, V. O'Dell, \\
E.~Ramberg, R.~E.~Ray, R.~Tschirhart, H.~B.~White \\
\scriptsize
Fermi National Accelerator Laboratory, Batavia Illinois 60510 \\
\small
\vspace*{0.1in}

K. Hanagaki, S. Mochida,  T. Yamanaka \\
\scriptsize
Department of Physics, Osaka University, Toyonaka, Osaka, 560 Japan \\
\small
\vspace*{0.1in}

M. D. Corcoran \\	
\scriptsize
Rice University, Houston, Texas 77005 \\
\small
\vspace*{0.1in}

%S. Averitte, J. Belz, E. Halkiadakis, A. Lath, S. Schnetzer,
%S. Somalwar,
%\\ R. Tesarek, \mbox{G. B. Thomson} \\
%\scriptsize
%Rutgers University, Piscataway, New Jersey 08855 \\
%\small
%\vspace*{0.1in}

M. Arenton, G. Corti, B. Cox, A. Ledovskoy, K.S. Nelson  \\
\scriptsize
University of Virginia, Charlottesville, VA 22901 \\
\small
\vspace*{0.1in}

A. Alavi-Harati, T. Alexopoulos, A.R. Erwin, M. A. Thompson \\
\scriptsize
University of Wisconsin, Madison, Wisconsin 53706 \\
\normalsize

\end{center}

\newpage
\parindent=0.25in
\vspace*{0.7in}

\begin{center}
{\bf ABSTRACT}
\end{center}
\vspace*{0.25in}

We, the KAMI (Kaons At the Main Injector) collaboration, express our interest
to continue the experimental program of rare kaon decay physics at Fermilab
using the Main Injector.
The 120 GeV Main Injector beam will provide neutral kaon beams which
are two orders of magnitude more intense than those currently available
at Fermilab. 
This dramatic increase in flux will allow us to study direct CP 
violation and other rare decay processes with unprecedented precision and
reach.

\vspace*{0.2in}
KAMI's primary physics goals shall be:

\begin{enumerate}

\item The first detection of the rare, direct CP violating decay \kpnn 
and measurement of its branching
ratio with an accuracy of 10\%, corresponding to a measurement of
the CP violation parameter $\eta$ with 5\% accuracy; and

\item Studies of various other rare decay processes such as 
\kpee, \pmm, \pme, \ppee, \mm, \ee, {\it etc.}
with sensitivities of approximately $10^{-13}$.

\end{enumerate}

To accomplish the above goals, it is necessary to upgrade the existing
KTeV detector in two major areas; 
hermetic photon vetos for \kpnn, and fiber
tracking for decay modes with charged particles.

This document summarizes the status of a feasibility study
and our R \& D plan for the near future. 

\newpage
\tableofcontents

\listoffigures

\listoftables
\newpage
%---------------------------------------------------------------------------

\section{Motivations}

\subsection{Theoretical background and motivation} 

    The origin of the matter/antimatter asymmetry manifest in our world
is of fundamental interest and remains outside the scope of the
now ``Standard Model" of particle interactions.  The theoretical structure
of the Standard Model can accommodate matter/antimatter asymmetries, but
the dynamical origin of these effects must reside at a level of understanding
beyond the Standard Model.  After 33 years of hard work since the original
observation of these asymmetries in the neutral kaon system we are now
at the threshold of performing  measurements of striking new asymmetry
effects expected in the Standard Model.  These effects are observed through
``CP violation" in the mixing and decay amplitudes of K and B meson decays.
The Standard model predicts large CP violation effects in the decay
amplitudes of rare B meson and very rare kaon decays.  More importantly,
the Standard Model predicts effects in the B and K systems with a common
formalism, so that matter/anti-matter asymmetries observed in these two
different systems must agree if the Standard Model is on the right track.

\vspace*{0.15in}
\parindent=0.25in

    To date, CP violation has only been observed through the window of
$K^0\leftrightarrow \overline{K^0}$ oscillations.
The effect is manifest as a difference in the
rate of $K^0 \rightarrow \overline{K^0}$ and
$\overline{K^0} \rightarrow K^0$ mixing.  Experiments at Fermilab and
CERN are now underway to study this difference in precise detail, with the
possibility of extracting a signal for CP violation in the decay amplitude
of $K \rightarrow 2\pi$ decays.   An observation of CP violation in
a decay amplitude (known as ``Direct CP Violation") would be the
first really new piece of information about CP violation since the original
discovery 33 years ago and the first significant insight into
its origin.  While this would be of immense significance,
the theoretical predictions for the magnitude of direct CP violation in
$K \rightarrow 2\pi$ decays are plagued with large uncertainties,
making it difficult to extract Standard Model parameters from the measurement.
In contrast, the theoretical predictions for rare
B and very rare K decays are much more reliable and provide a laboratory
to quantitatively measure the fundamental CP violating parameters of the
Standard Model.  The reliability and magnitude of the predicted asymmetries
in rare B and very rare K decays has stimulated an ambitious world-wide
effort to measure these effects.

\vspace*{0.15in}
\parindent=0.25in

    The very rare kaon decays of greatest interest are
$K_L \rightarrow \pi^0 \nu \overline{\nu}$ and
$K^+ \rightarrow \pi^+ \nu \overline{\nu}$.
The importance of these measurements have been discussed at
length in the literature for some 
\mbox{time \cite{buchalla1}\cite{buchalla2}\cite{buras2}\cite{buras3}.}
In the context of the Standard Model, measurement of these
two branching fractions can uniquely determine the two fundamental
CP violation parameters of the model.  These two parameters are referred
to as $\rho$ and $\eta$, where $\eta$ directly sets the scale of
CP violation within
the model.  Likewise, measurements in the system of B meson decays can
determine the $\rho$ and $\eta$ parameters uniquely.  Comparison of
$\rho$ and $\eta$
in the K and B systems provides a very powerful cross-check of our
understanding of CP violation within the Standard Model.
The $K_L \rightarrow \pi^0 \nu \overline{\nu}$
branching fraction is proportional to $\eta^2$ and provides a direct probe of
CP violation within the Standard Model.   The
$K^+ \rightarrow \pi^+ \nu \overline{\nu}$  branching fraction
is proportional to $(1.3-\rho)^2 + \eta^2$ and hence is sensitive to both
$\rho$ and $\eta$.  
The $K_L \rightarrow \pi^0 \nu \overline{\nu}$ and
$K^+ \rightarrow \pi^+ \nu \overline{\nu}$ processes are expected to occur
with branching fractions at the level of $3 \times 10^{-11}$ and 
$1 \times 10^{-10}$, respectively.
Measurement of these processes is extremely challenging due to the very
low branching fractions and the presence of unmeasurable neutrinos in
the final states.  

\vspace*{0.15in}
\parindent=0.25in

Experimental rare kaon decay programs that
can meet these technical challenges demand instrumentation that is at or
beyond state-of-the-art in the field.  These high performance kaon
beam and detector systems enable the precision study of less rare kaon decays
that are of interest in their own right, as well as providing key performance
milestones along the way.

\vspace*{0.15in}
\parindent=0.25in

The primary goal of the KAMI collaboration will be to detect the decay
\kpnn, measure its branching ratio, and extract a value for $\eta$ which
is accurate to approximately 5\%.  In addition, there are a number of
other rare kaon decays of interest to the collaboration.  Some of these
decays are sensitive to direct CP violation,
while others probe critical regions of the Standard Model.  
It is possible to address these other modes in KAMI without compromising
the \kpnn study.

In this Expression of Interest, we describe a detailed plan for how we
intend to address these compelling physics issues.  We first describe 
our current activities
with the KTeV (Kaons at TeVatron) experiment at Fermilab.  
After a summary of the status of KTeV, we will describe our
plans and goals for KAMI.  The status of detailed simulations, 
an R\&D plan, as well as preliminary budgets and schedules are presented.

\subsection{Experimental status}

All attempts to measure the decay \kpnn thus far have relied on 
observation of the Dalitz decay mode of the $\pi^0$ to
$e^+e^-\gamma$.  The charged vertex from the $e^+e^-$ provides kinematical
constraints which allow for simple reconstruction of the $\pi^0$ and
effective rejection of backgrounds to the sensitivity levels reached
thus far.  The best published limit to date for the decay is
$5.8 \times 10^{-5}$ (90\% CL) from Fermilab experiment E799-I~\cite{weaver}.

The 2$\gamma$ decay mode of the $\pi^0$ provides more than two 
orders of magnitude higher
sensitivity per unit time than the Dalitz mode,
but at the cost of increased background
due to fewer kinematical constraints.  Attempts to measure this decay
in the future will almost certainly migrate in the direction of the
2$\gamma$ mode to take advantage of this increased sensitivity.
The focus of future experiments will therefore be to understand how to
reduce backgrounds in the face of reduced constraints.

The most recent attempt to measure \kpnn has been made by the KTeV
experiment.  KTeV has used both the Dalitz and the 2$\gamma$ decay modes
of the $\pi^0$.  The 2$\gamma$ mode was used primarily to begin understanding
the backgrounds which will ultimately have to be confronted by KAMI. 
Preliminary results for the 2$\gamma$ decay mode are presented
in Section~\ref{special_run}.

Considering the fact that the current best limit is six orders-of-magnitude
higher than the predicted level,
we believe that a programmatic, step-by-step approach will be necessary
to eventually achieve signal detection.  The KTeV run in 1997 was a
significant step towards this goal.  We expect to make another
significant step in a 1999 run of KTeV, as described in Section~\ref{ktev99}.
This will provide us with an ideal opportunity to study this mode with 
much better 
sensitivity than has been possible in the past with a minimum investment.

\subsection{Current status of the KTeV experiment}

The KTeV collaboration was formed with the goal of probing the most
relevant questions relating to CP violation which are accessible via the
neutral kaon system, using the currently available TeVatron beam
and state-of-art detector technology~\cite{arisaka1}.
A new physics program to measure the value of $\epsilon^\prime / \epsilon$
with unprecedented precision 
was approved as E832 in 1992
as the primary goal of the KTeV project.
A program to investigate rare CP violating kaon decays,
originally approved as E799 in 1988, evolved into E799-II. 

After several years of construction, the experiment
was successfully commissioned in the summer of 1996.
Data collection began in the Fall of 1996 with 
a new detector, a new experimental hall and a new beamline. 
The run concluded
in September of 1997 and intensive analysis of data is currently underway.
By all accounts it was a very successful
run made possible, in part, by the substantial experience gained over time
from previous efforts.  

E832 collected data at a rate which was 10 times faster
than E731, its predecessor.  E731's measurement of 
$\epsilon^\prime / \epsilon$ gave a result
of ($7.4 \pm 5.2 \pm 2.9) \times 10^{-4}$~\cite{gibbons}, for an overall
uncertainty of about 7 x $10^{-4}$, dominated by
the statistical error of $5.2 \times 10^{-4}$.  KTeV expects to 
significantly improve on this result.

\begin{figure}
  \centerline{ \psfig{figure=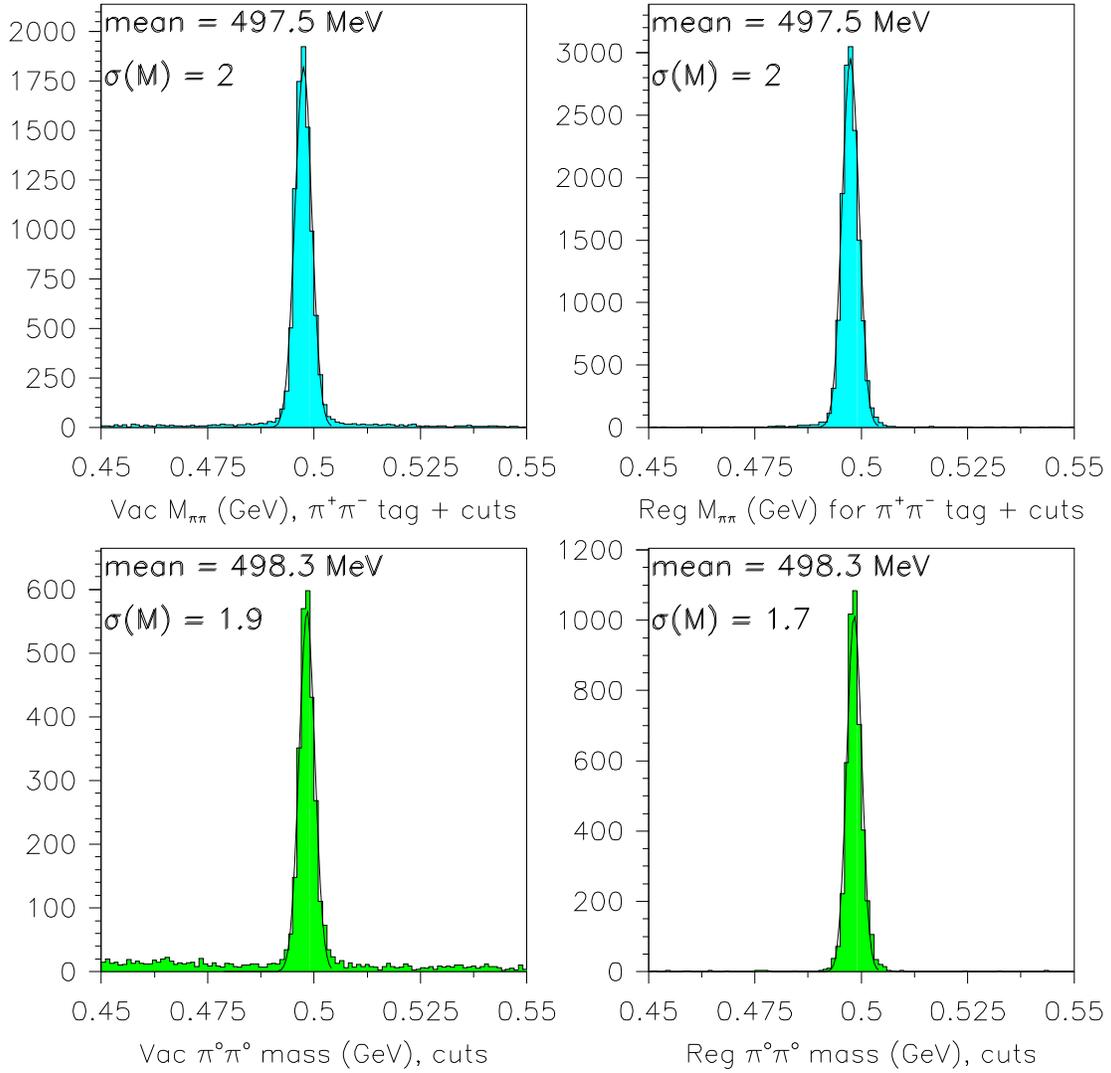,width=16cm} }
  \caption{$2\pi$ mass plots from E832 after loose online cuts.  
The plots in the right
column show the spectra from the beam containing the regenerator
and the plots in the left column show the spectra from the beam
without the regenerator.\label{pimass}}
\end{figure}

Online mass plots from E832, after loose online cuts, are shown in 
Figure \ref{pimass}.  
$\pi^+ \pi^-$ and $\pi^0 \pi^0$ decays from both $K_L$ and $K_S$
are shown.  Both charged and neutral modes have a similar mass resolution
of about 2 MeV.

A single day of E799-II data is equivalent
to half of the entire run of E799-I in 1992 for many decay modes.  
As an example of what can be expected from E799-II, we summarize
in Table \ref{tab:ses} the expected SES and
90\% confidence limit for three CP violating rare decay modes
in the data collected during 1997.  These numbers are extrapolated
from a detailed analysis of one-day of E799 data.

\begin{table}[here,top,bottom]
\begin{center}
\begin{tabular}{|lr|r|}
\hline
Decay Mode
        & & Results	\\
\hline\hline
$K_L\rightarrow\pi^0e^+e^-$ &SES
        &$  5.0 \times 10^{-11}$ \\
 	&90\% CL
        &$ <2.5 \times 10^{-10}$ \\
        &       &          \\
$K_L\rightarrow\pi^0\mu^+\mu^-$ &SES
        &$  7.0 \times 10^{-11}$ \\
 	&90\% CL
        &$ <1.6 \times 10^{-10}$ \\
        &       &                 \\
$K_L\rightarrow\pi^0\nu\overline{\nu}$ (Dalitz mode) &SES
        &$  7.5 \times 10^{-8}$  \\
 	&90\% CL
        &$ <1.7 \times 10^{-7}$  \\
\hline
\end{tabular}
\caption{Projected single event sensitivities and 90\% Confidence
Limits for CP violating decays
from E799 running in 1997.}
\label{tab:ses}
\end{center}
\end{table}

Both E832 and E799-II have made considerable advances over their immediate
predecessors and represent the continuous progression of a neutral kaon
program at Fermilab.
KTeV is the latest in a long series of successful
neutral kaon experiments at Fermilab.  The next major step in this continuing
program will be KAMI (Kaons At the Main Injector)
where we expect to make similarly impressive gains over time.

\subsection{New Results from KTeV}

New results have begun to emerge from KTeV on numerous fronts, even
before the completion of data taking.  These new results include
the first measurement of the rare decay 
\mbox{$K_{L}\rightarrow\pi^{+}\pi^{-}e^{+}e^{-}$,} the first observation of 
the $\Xi^0$ beta decay, the first direct search for the supersymmetric R$^0$
particle, and the first search for \kpnn using the 2$\gamma$ decay
mode of the $\pi^0$.  These results are briefly described below.  Many
more new results will be available soon.

\subsubsection{First measurement of $K_{L}\rightarrow\pi^{+}\pi^{-}e^{+}e^{-}$}
\label{sec-pipiee}

\begin{figure}
  \centerline{ \psfig{figure=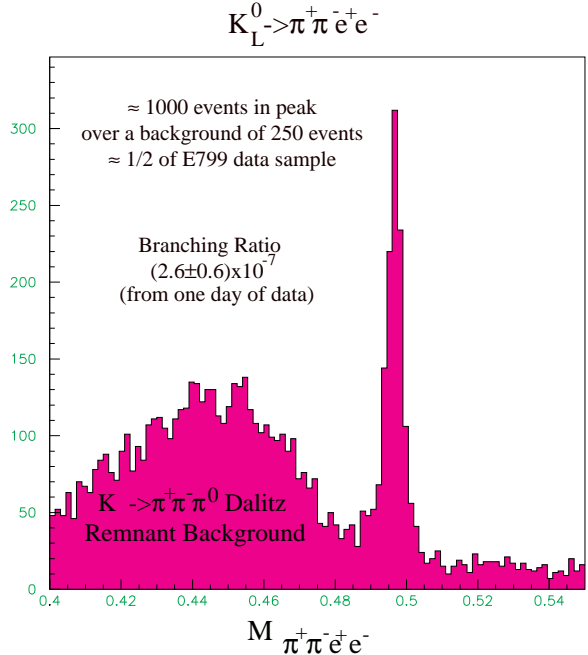,width=12cm} }
  \caption{Mass peak from the first measurement of
        $K_L \rightarrow \pi^+ \pi^- e^+ e^-$. \label{pipiee}}
\end{figure}

In the 1997 run of KTeV, the previously undetected decay
$K_{L}\rightarrow\pi^{+}\pi^{-}e^{+}e^{-}$ has been definitively
observed~\cite{odell}.  We show in Fig. \ref{pipiee}
the mass peak from approximately
one half of the data accumulated thus far.  Approximately 1000
events are observed in the peak over a background of 250 events for this data
sample.  A preliminary branching ratio of 
\mbox{(2.6$\pm$0.6)$\times$ 10$^{-7}$} has been measured based on one day of
data taking.

One reason for the strong interest in the decay
$K_{L}\rightarrow\pi^{+}\pi^{-}e^{+}e^{-}$ is the prospect for observing
CP violation~\cite{seghal}.  Interference of the
indirect CP violation Bremsstrahlung process with the
CP conserving M1 emission of a virtual photon
is expected to generate an asymmetry in the angle $\phi$ between the
normals to the decay planes of the $e^{+}e^{-}$ and the $\pi^{+}\pi^{-}$ in
the $K_{L}$ center of mass.  In addition, direct CP violation effects,
albeit small, can occur in this mode via the interference
between various amplitudes.  Detailed investigation of this asymmetry 
is currently in progress.

\subsubsection{First observation of the $\Xi^0$ beta decay}

\begin{figure}
  \centerline{ \psfig{figure=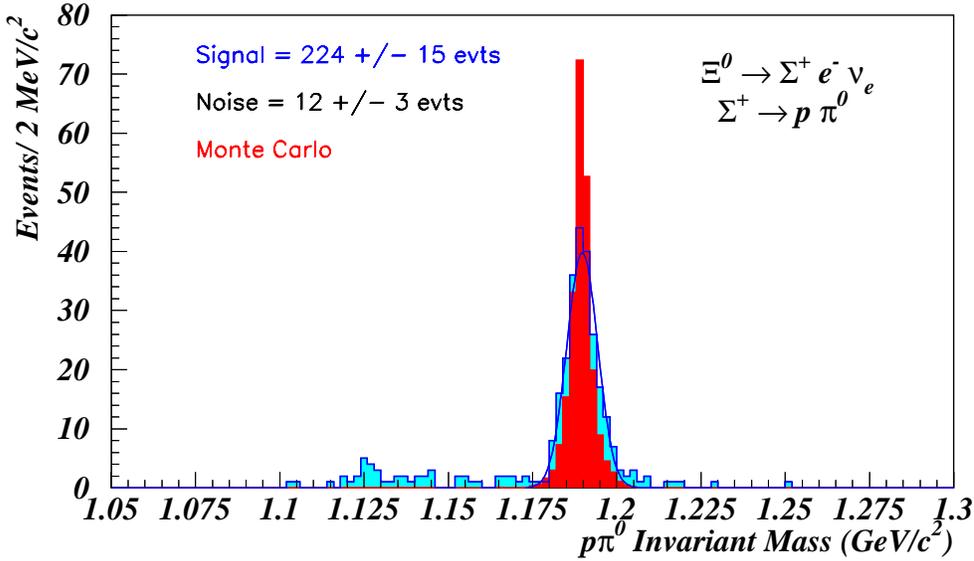,width=14cm} }
  \caption{Evidence for the first observation of cascade Beta
        decay $\Xi^0 \rightarrow \Sigma^+ e^- \overline{\nu}$, with
        \mbox{$\Sigma^+ \rightarrow p \pi^0$.} The reconstructed $\Sigma^+$
        mass is plotted along with a Monte Carlo overlay (dark region).
         \label{casbeta}}
\end{figure}

The $\Xi^0$ beta decay, $\Xi^0 \rightarrow \Sigma^+ e^- \overline{\nu}$ with
$\Sigma^+ \rightarrow p \pi^0$, has been observed for the first time by KTeV. 
The asymmetry of the electron is particularly interesting, as it
offers a fundamental test of the V-A structure of the weak interaction.
We have looked for double vertex events where there is a $\Sigma^+$
reconstructed from a proton and a $\pi^0$ ($\pi^0$ mass
constrained) downstream of the vertex formed by an electron track
and the `track' from the reconstructed
$\Sigma$. 
Figure \ref{casbeta} shows the reconstructed $\Sigma$ mass (with a Monte
Carlo overlay).  Work is progressing toward obtaining a branching ratio,
normalizing to $\Xi \rightarrow \Lambda \pi^0$.  Asymmetry
measurements will also eventually be made.

\subsubsection{Search for the supersymmetric R$^0$}

\begin{figure}
  \centerline{ \psfig{figure=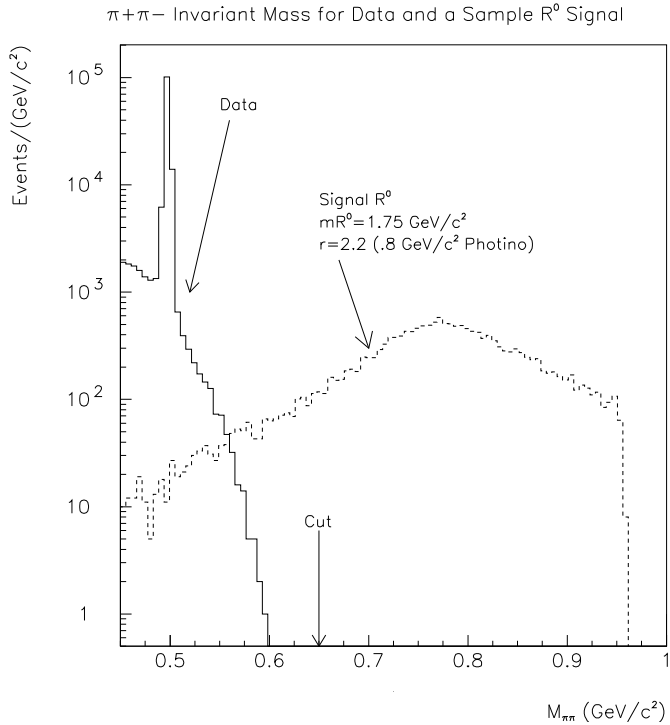,width=12cm} }
  \caption{$\pi^+\pi^-$ invariant mass distribution used in
        the R$^0$ search.  The solid line represents data and the
dashed line is from a Monte Carlo of the R0 signal. \label{R0}}
\end{figure}

A search for a light gluino, called the R$^0$, through its dominant decay mode
R$^0 \rightarrow \tilde{\gamma} \rho$ with $\rho \rightarrow \pi^+ \pi^-$,
has been performed on a one-day E832 data sample.  This search is motivated
by recent predictions in the literature~\cite{farrar1}\cite{farrar2}.
The photino in this SUSY scenario is a cold dark matter candidate.
This is the first time a direct search for such a decay has been performed.
Figure \ref{R0} shows
the $\pi^+\pi^-$ invariant mass distribution for the data (solid) and an
R$^0$ Monte Carlo (dashed).  The R$^0$ search region is above
\mbox{650 MeV/c$^2$.}
With one day's data, we are sensitive to an R$^0$ mass between
\mbox{1.5 - 4.5 GeV/c$^2$} and an R$^0$ lifetime between \mbox{1 - 5000 ns,}
with an R$^0/K_L$
production ratio below $10^{-4}$ to 2.5x$10^{-7}$ and an upper limit on
the R$^0$ production cross section times branching ratio of the order of
\mbox{$10^{-35}$ cm$^2$/(GeV$^2$/c$^3$)} at $x_F$=0.1.  Since this search is
quite clean, more data will be analyzed for this mode in the near future.

\subsubsection{Preliminary result for \kpnn from KTeV 97 special run}
\label{special_run}

The best published limit to date for the decay \kpnn is 
$5.8 \times 10^{-5}$ (90\% CL) from Fermilab experiment E799-I~\cite{weaver}.
One of KTeV's many goals was to extend this limit by several orders
of magnitude.  

Although the best limit for \kpnn from KTeV
in the 1997 run will come from the
full analysis of the $\pi^0$ Dalitz mode, 
we are also investigating the 2$\gamma$ decay mode.
The 2$\gamma$ mode provides us with more than two orders of magnitude higher
sensitivity per unit time, but at the cost of increased background
due to fewer kinematical constraints.
This study is an important input to the design of the KAMI detector.

\vspace*{0.15in}
\parindent=0.25in

To understand the type and level of backgrounds we will ultimately
be confronted with in KAMI, a special
half-day of data was taken in December 1996.  During this special run,
one beam was further collimated down to 4 cm x 4 cm (at the CsI)
in order to obtain better $P_t$ resolution on the decay.  The second beam
was completely closed off.  From a preliminary analysis, we have obtained
an upper limit on the branching ratio of \mbox{1.8 x $10^{-6}$ at a 
90\%~CL~\cite{nakaya}}.  This represents a factor of 30 improvement over the
best existing limit, obtained by E799-I using the Dalitz decay mode
of the $\pi^0$.

\begin{figure}
  \centerline{ \psfig{figure=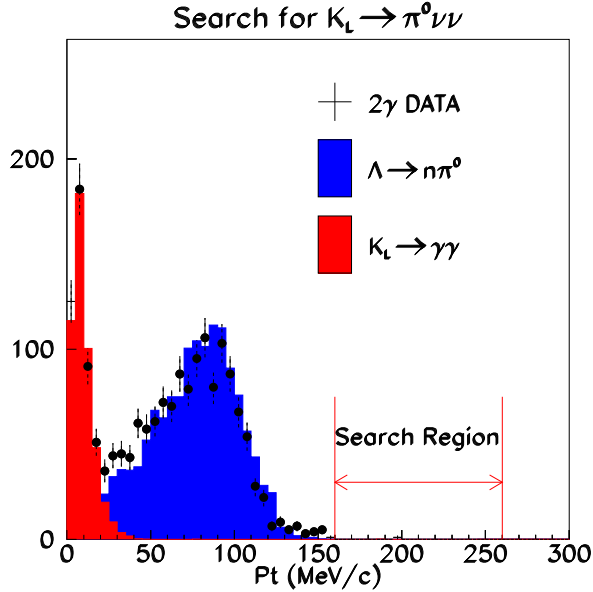,width=10cm} }
  \caption{$P_t$ distribution of $K_L \rightarrow \pi^0 \nu
        \overline{\nu}$ candidate events using the 2$\gamma$ decay
        mode of the $\pi^0$ during a special 1 day run in December
        of 1996. \label{pi0nn}}
\end{figure}

Figure \ref{pi0nn} shows the $P_t$ distribution of candidate events after
the final cuts. As is shown here, the observed $P_t$ distribution
can be well reproduced by $K_L\rightarrow 2\gamma$ and
$\Lambda\rightarrow n \pi^0$.  For $P_t$ values above 160 MeV/c$^2$,
one event still remains.  This event is consistent with a neutron
or $K_L$ 
interaction in the detector.  This understanding is based on detailed
studies of beam interactions in the vacuum window using KTeV data. 
From these studies we know that these interactions are the source of
high-$P_t$ $\pi^0$s which strike the CsI calorimeter.
This will not be a source of
background for KAMI as the neutral beam will see only vacuum until
it encounters the vacuum window immediately upstream of the CsI calorimeter.

\subsubsection{\kpnn at KTeV 99} 
\label{ktev99}

KTeV expects to run for a second time in 1999~\cite{loi99}.
The kaon flux from the combined
FY97 and FY99 runs will allow us to complete our measurement of
$\epsilon^\prime / \epsilon$ and to reach our proposed sensitivities
for a wide array of rare decay modes.  It will also play a significant
role in helping us to plan for KAMI.

In order to study the decay mode $K_L\rightarrow\pi^0\nu\overline{\nu}$
using $\pi^0 \rightarrow 2\gamma$,
we have proposed a dedicated run with a single small beam, similar to the
short study done by KTeV at the end of 1996~\cite{loi99}.
This short run has resulted in the best limit to date,
as reported in the previous section.

Table \ref{tab:factors_pi0nn} summarizes the expected single event
sensitivity(SES) with four weeks of running time,
projected from half a day of data taken in December 1996.
We expect to achieve a sensitivity of $3 \times 10^{-9}$ with
the same beam size used for the 1996 run  \mbox{(4 cm x 4 cm at
the CsI).}

Since the KTeV Letter of Intent for the FY99 run  was submitted
in June of 1997, we have performed
extensive Monte Carlo simulations of the expected background level.
Table~\ref{tab:background_pi0nn} summarizes the results.
The interaction of beam neutrons with the detector is likely to be the
most serious background, as was the case in the 1996 data.
This background can be reduced by detecting the beam neutrons.
An hadronic section of the Back-Anti, six nuclear interaction lengths deep,
will be installed just downstream of the existing EM section of the Back
Anti in order to address this issue.
With a 99\% detection efficiency for the beam neutron,
we expect that this background will appear around a sensitivity level of
$1\times10^{-8}$, as shown in Table~\ref{tab:background_pi0nn}.

\begin{table}[here,top,bottom]
\begin{center}
\begin{tabular}{|l|r|r|c|}
\hline
                & KTeV 97        & KTeV 99        & Improvement   \\
\hline\hline
Proton Intensity & $3\times10^{12}$ & $1\times10^{13}$ & 3.33          \\
Repetition Cycle & 60 sec.      & 80 sec.       & 0.75          \\
Beam Size        & 4.0 x 4.0 cm$^2$ & 4.0 x 4.0 cm$^2$   & 1.00 \\
Running Time     & 11 hours     & 4 weeks       & 61.1          \\
\hline
Improvement(KTeV 99/KTeV 97)
                &             &               & 150           \\
\hline\hline
SES               & 4.6 x $10^{-7}$ &3.0 x $10^{-9}$ &              \\
\hfill (no $\gamma$ conversion)    &    &(4.4 x $10^{-9})$ & \\
\hfill (with at least one $\gamma$ conversion)  &    &(1.0 x $10^{-8})$ & \\
\hline
\end{tabular}
\caption{The expected sensitivity from KTeV 97 and KTeV 99 for
                $K_L\rightarrow\pi^0\nu\overline{\nu}$
                with $\pi^0 \rightarrow \gamma \gamma$.}
\label{tab:factors_pi0nn}
\end{center}
\end{table}

In order to further distinguish signal events from this type of background,
we are also considering the possibility of using an active photon
converter located between the vacuum window and the first drift chamber. 
This will provide verification that the decay originated within
the fiducial volume of the detector.
The converter will also improve the kinematical constraints on the decay,
allowing for reconstruction of the $\pi^0$ invariant mass,
which should help in rejecting other types of backgrounds.

A converter, consisting of a \mbox{1 mm} thick lead sheet sandwiched
between two sheets of scintillator,
will convert at least one photon from a $\pi^0$ decay 30\%
of the time.
Thus, the expected sensitivity when at least one gamma converts will be
about $1 \times 10^{-8}$, as given in Table~\ref{tab:factors_pi0nn}.

\begin{table}[here,top,bottom]
\begin{center}
\begin{tabular}{|l|r|}
\hline
Decay Mode & Background Level      \\
\hline\hline
$K_L \rightarrow \pi^0 \pi^0$           & $\sim 5\times 10^{-10}$ \\
$K_L \rightarrow \pi^0 \pi^0 \pi^0$     & $< 5\times 10^{-9}$ \\
$K_L \rightarrow \gamma \gamma$         & $ < 9\times 10^{-10}$ \\
$\Lambda \rightarrow n \pi^0$           & $< 7\times 10^{-10}$ \\
$\Xi^0  \rightarrow \Lambda \pi^0, \Lambda \rightarrow n \pi^0$
                                        & $\sim 2\times 10^{-10}$ \\
nA $\rightarrow \pi^0$A                 & $\sim 1\times 10^{-8}$ \\
\hline
\end{tabular}
\caption{Expected background levels for the \pnn\ search using $\pi^0
\rightarrow \gamma \gamma$ at KTeV 99.}
\label{tab:background_pi0nn}
\end{center}
\end{table}

\subsection{Future prospects} 

Because of the tremendous physics importance associated with the decay
\kpnn, dedicated searches have been proposed at BNL~\cite{chiang} 
and KEK~\cite{inagaki} along with
our proposal to focus on this mode in KAMI.

The BNL group proposes to execute the experiment using a micro-bunched
proton beam with a 45 degree targeting angle to produce a kaon beam
with a mean momentum of 700~MeV/c.  The low kaon momentum allows
for a measurement of the momentum of the decaying $K_L$ using time-of-flight.
Sufficient kinematical constraints are available in order to reconstruct
all four-vectors, including those of the missing neutrinos.  The timescale
for the BNL experiment is similar to the timescale for KAMI and the 
sensitivity expected by the BNL group is also similar to that expected
by KAMI.

The KEK experiment is very similar to KAMI conceptually in that
extremely good photon veto efficiency is required.  
It is expected that a sensitivity of $10^{-10}$ will be achieved by 1999
during the first phase of the experiment, using the existing 12~GeV 
proton machine at KEK.  The ultimate goal of this group 
is to perform this measurement with a new 50~GeV
high-intensity machine, JHP, which is expected to be operational in 2003.

\section{The KAMI Experiment}

The next major step in the continuing neutral kaon program at Fermilab
will be KAMI (Kaons At the Main Injector).  As in the case of each
previous effort, KAMI will benefit greatly from our experience with its
predecessor; in this case KTeV.  KTeV's success 
gives us every reason to be
optimistic about the prospects for KAMI.

\subsection{Detector concepts}
 
A clear observation of the decay \kpnn will be the highest priority for KAMI.
The KAMI detector must be optimized with this principle in mind.
At the same time, we would like to explore other important rare decay
modes such as $K_L\rightarrow \pi^0 e^+e^-$ and $\pi^+\pi^- e^+e^-$.
In order to achieve these physics goals with minimum investment,
the KAMI detector has been designed according to the following principles: 

\begin{enumerate}
 
\item Utilize the existing KTeV infrastructure to the greatest extent possible. 
   This includes the experimental hall, the CsI calorimeter,
   and significant parts of the readout electronics and 
   the DAQ system.

\item For \kpnn, install newly constructed hermetic photon veto
   detectors.

\item For charged decay modes, and to maintain our ability to calibrate
the CsI calorimeter to the required accuracy, insert scintillating fiber 
tracking planes inside of the vacuum tank.  The vacuum tank will pass through
the gap of the analysis magnet. 

\end{enumerate}

\begin{figure}[htb]
\centerline{
        \begin{turn}{90}
        \psfig{figure=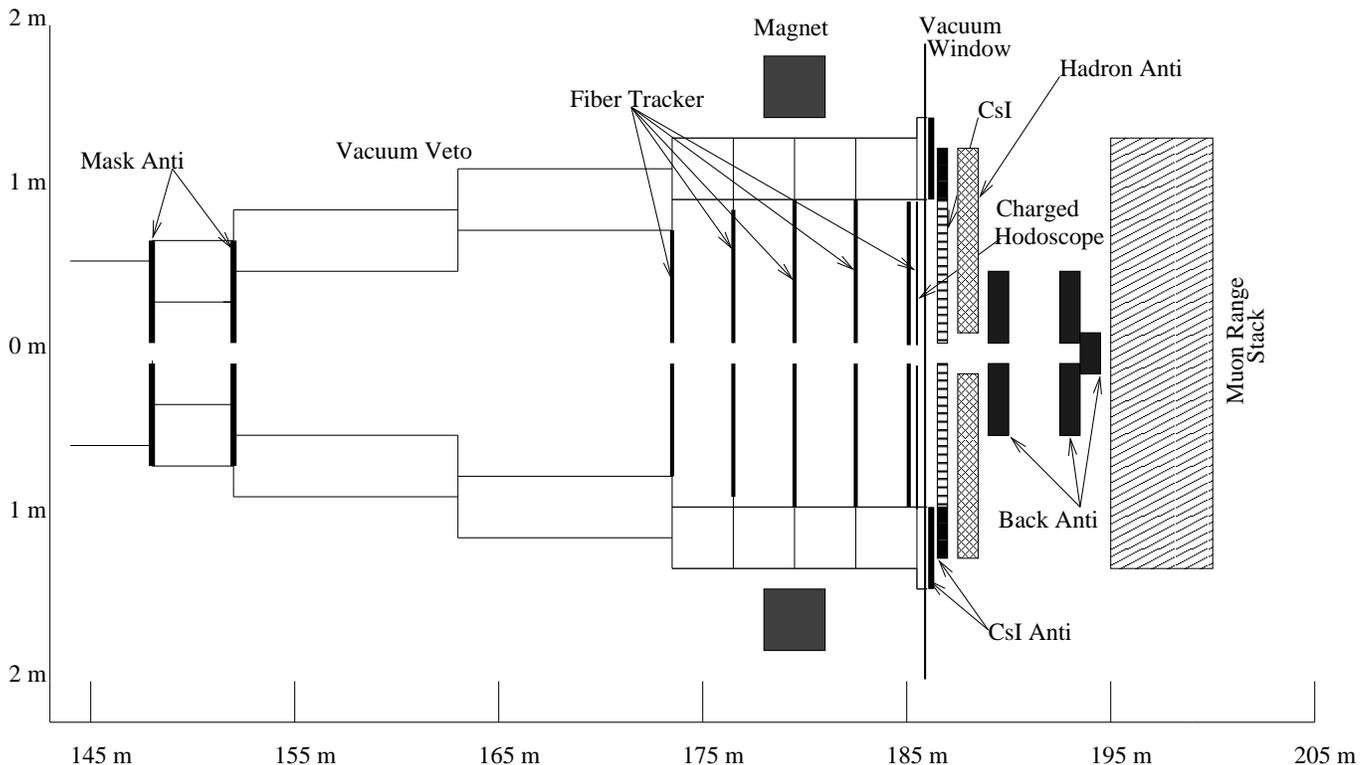,width=11cm}
        \end{turn}
}
  \caption{Schematic of the KAMI detector. \label{kami_det}}
\end{figure}

\parindent=0.in
Figure \ref{kami_det} shows the layout of a possible detector 
design based on the above concepts.

\parindent=0.25in
With a 24 mrad targeting angle, 
the average momentum of decaying kaons will be between 10-15 GeV, 
depending on the distance of the decay volume from the production target.
This is about a factor
of five lower than the kaon spectrum at KTeV and is due primarily 
to the lower beam energy of the Main Injector. 
 
Because of the lower energy kaons in KAMI, the overall detector geometry must 
be compressed longitudinally in order to maintain good acceptance. 
The fiducial decay region for \kpnn will be about 21~m long.  The
following 12~m 
of the decay volume contains a charged particle spectrometer. 
The most striking  feature is that the inner wall of all of the vacuum
pipes are lined with an hermetic system of photon veto detectors.

\subsection{Physics sensitivity}
\label{sens}

The signal sensitivity for \kpnn is determined by the kaon flux, geometrical
acceptance and effective running time.
In order to achieve a high sensitivity, the large flux associated with 
a large beam size is desirable.
However, in order to obtain good $P_t$ resolution for 
the reconstruction of $\pi^0 \rightarrow 2\gamma$, 
the transverse beam size must be restricted.
We have tuned the beam size in order to collect at least 
30 \pnn signal events per year (Standard Model BR assumed)
while maintaining the required $P_t$ resolution. 

\vspace*{0.15in}
\parindent=0.25in

Table \ref{tab:kpnn_sens} lists the \kpnn sensitivities
expected by KAMI as well as for proposed experiments at BNL and KEK.
For comparison, the sensitivity at KTeV is also listed
for the $\pi^0 \rightarrow 2\gamma$ decay mode.
Two possible scenarios are listed for KAMI; KAMI-Far and KAMI-Near.
KAMI-Far uses the existing KTeV production target and
neutral beam line with minor modifications,
where as KAMI-Near requires a new target station located 120~m
downstream of the existing target station.  This allows for a higher kaon
flux per unit physical beam size at the calorimeter.
The obvious advantage of the KAMI-Far scenario is the cost savings, as well
as the reduction of lambda decay induced backgrounds
such as $\Lambda \rightarrow \pi^0 n$.

\begin{table}[here,top,bottom]
\begin{center}
\begin{tabular}{|l|r|r|r|r|r|r|}
\hline
	 &KTeV 97 &KTeV 99  &KAMI-Far &KAMI-Near &BNL &KEK \\
\hline\hline
Proton Energy   &800 GeV   &800 GeV     &120 GeV   &120 GeV &24 GeV  &13 GeV  \\
Intensity/pulse & $3\times10^{12}$ & $1\times10^{13}$   & $3\times10^{13}$
     & $3\times10^{13}$   & $5\times10^{13}$   & $2\times10^{12}$  \\
Repetition cycle&60 s    &80 s      &2.9 s   &2.9 s &3.6 s &4.0 s \\
Flat top        &20 s    &40 s      &1.0 s   &1.0 s &1.6 s &2.0 s \\
Targeting angle &4.8 mrad  &4.8 mrad    &24 mrad   &24 mrad &45 deg  &6 deg   \\
Beam x width    &0.22 mrad  &0.22 mrad   &0.6 mrad &1 mrad  &4 mrad  &4 mrad  \\
Beam y width    &0.22 mrad  &0.22 mrad   &0.6 mrad &1 mrad  &125 mrad& 4 mrad  \\
Beam solid angle&0.048 $\mu$str &0.048 $\mu$str  &0.36 $\mu$str&1 $\mu$str  
		&500 $\mu$str& 16 $\mu$str \\
\hline
Kaons/pulse   		& $6.7\times10^{6}$  & $2.3\times10^{7}$  
& $2.8\times10^{7}$  	& $1.1\times10^{8}$ & $2.5\times10^{8}$  
& $1.5\times10^{6}$  \\
Kaon flux at BA         &0.3 MHz &0.6 MHz &28 MHz  &110 MHz  &150 MHz
&0.75 MHz \\
Neutron flux/pulse      & $1.3\times10^{7}$  	& $4.4\times10^{7}$  
& $2.0\times10^{8}$  	& $5.5\times10^{8}$  	& $7.5\times10^{9}$  
& $2.3\times10^{6}$  \\
Neutron flux at BA      &0.8 MHz &1.2 MHz &200 MHz &550 MHz &5 GHz   
&1.2 MHz \\
\hline
Ave. kaon mom.   	&70 GeV/$c$  &70 GeV/$c$  &13 GeV/$c$   &10GeV/$c$   
			&0.7 GeV/$c$ &2 GeV/$c$ \\
Z decay region          &38 m    &38 m    &34 m    &34 m    &3.5 m   &2.7
m   \\
Decay probability       &2.1\%  &2.1\%  &10\%   &10\%   &16\%   &4.3\%  \\
Kaon decay /pulse       & $1.4\times10^{5}$  	& $4.8\times10^{5}$  
& $2.8\times10^{6}$  	& $8.2\times10^{6}$  	& $4.0\times10^{7}$  
& $6.5\times10^{4}$  \\
Kaon decay /sec         & 7.1 kHz & 12 kHz  & 2.8 MHz & 8.2 MHz & 25 MHz  
			& 32 kHz  \\
\hline
Running time            &0.46 day&28 days &365 days&365 days&365 days
&84 days \\
DAQ live time           &0.65   &0.65   &0.65   &0.65   &0.65   &0.65   \\
Live time               &0.7    &0.7    &0.7    &0.7    &0.7    &0.7    \\
Kaon decays/day  	& $9.4\times10^{7}$  	& $2.4\times10^{8}$  
& $3.8\times10^{10}$ 	& $1.5\times10^{11}$ 	& $2.1\times10^{11}$ 
& $6.3\times10^{7}$  \\
Total kaon decay        & $4.3\times10^{7}$  	& $6.6\times10^{9}$  
& $1.4\times10^{13}$ 	& $5.6\times10^{13}$ 	& $7.7\times10^{13}$ 
& $2.3\times10^{10}$ \\
Acceptance              &5\%    &5\%    &7.1\%  &7.4\%  &1.6\%  
&8\% \\
\hline\hline
Single Event Sens.      & $4.6\times10^{-7}$  & $3.0\times10^{-9}$  &
$1.0\times10^{-12}$     & $2.4\times10^{-13}$ & $8.2\times10^{-13}$ &
$5.4\times10^{-10}$ \\
No. of \pnn & $7\times10^{-5}$ &0.01   &30     &124     
&37     &0.06  \\
\hline

\end{tabular}
\caption{Parameters and sensitivities for several proposed \kpnn searches.}
\label{tab:kpnn_sens}
\end{center}
\end{table}

\begin{table}[here,top,bottom]
\begin{center}
\begin{tabular}{|l|r|r|r|r|}
\hline

 			&KTeV 97 &KTeV 99	&KAMI-Far &KAMI-Near	\\
%			&	&	&-Far	&-Near	\\
\hline\hline	
Proton Energy		&800 GeV	&800 GeV &120 GeV	&120 GeV	\\
Intensity/pulse		&$3\times 10^{12}$	&$1\times 10^{13}$	
&$3\times 10^{13}$	&$3\times 10^{13}$	\\
Repetition cycle	&60 s	&80 s	&2.9 s 	&2.9 s	\\
Flat top		&20 s	&40 s	&1.0 s	&1.0 s	\\
Targeting angle		&4.8 mrad&4.8 mrad&8 mrad	&8 mrad	\\
Beam x width		&0.5 mrad&0.6 mrad&0.6 mrad & 2.5 mrad \\	
Beam y width		&0.5 mrad&0.6 mrad&0.6 mrad&2.5 mrad \\
Beam solid angle	&0.5 $\mu$str&0.72 $\mu$str&0.36 $\mu$str&6.3 $\mu$str \\
\hline	
Kaon production/pulse	&$2.3\times 10^8$	&$8.4\times 10^8$	
&$7.0\times 10^7$	&$1.5\times 10^9$	\\
Average kaon momentum	&70 GeV/$c$	&70 GeV/$c$	&21 GeV/$c$
&15GeV/$c$  \\
Z decay region		&38 m	&38 m	&23 m	&23 m	\\
Decay probability	&2.1\% 	&2.1\%	&10\%	&10\%	\\
Kaon decay /pulse	&$9.1\times 10^6$	&$3.3\times 10^7$	
&$7.1\times 10^6$	&$1.2\times 10^8$	\\
Kaon decay /sec		&0.45 MHz&0.82 MHz	&7.1 MHz	&120 MHz	\\
\hline
Running time		&98 days	&112 days&365 days&365 days \\	
DAQ live time           &0.65  	&0.65 	&0.65	&0.65    \\
Live time		&0.7	&0.7 	&0.7	&0.7	\\
No. of kaon decays/day	&$6.0\times 10^9$	&$1.6\times 10^{10}$	
&$9.7\times 10^{10}$ 	&$1.7\times 10^{12}$ \\	
Total Kaon Decays	&$5.8\times 10^{11}$	&$1.8\times 10^{12}$	
			&$3.5\times 10^{13}$	&$6.1\times 10^{14}$ \\
\hline\hline
$K_L \rightarrow \pi^0 e^+e^-$ (Br=$5\times10^{-12}$ exp'd) & & & & \\
Acceptance		&5.2\%	&5.2\%	&2.1\%	&2.1\%	\\
Single Event Sens.	&$3.3\times 10^{-11}$ &$1.1\times 10^{-11}$
&$1.4\times 10^{-12}$ &$7.8\times 10^{-14}$ \\
No. of events 		&0.1	&0.5	&4	&64	\\
\hline

$K_L \rightarrow \pi^0 \mu^+\mu^-$ (Br=$1\times10^{-12}$ exp'd) & & & & \\
Acceptance              &5.5\%  &5.5\%  &2.4\%  &2.4\%  \\
Single Event Sens.      &$3.1\times 10^{-11}$ &$1.0\times 10^{-11}$
&$1.1\times 10^{-12}$ &$6.8\times 10^{-14}$ \\
No. of events           &0.04    &0.1    &0.8      &15     \\
\hline

$K_L \rightarrow \pi^+\pi^- e^+e^-$ (Br=$2.6\times10^{-7}$)  & & & & 	\\
Acceptance		&1.7\%	&1.7\%	&0.7\%	&0.7\%	\\
Single Event Sens.	&$1.0\times 10^{-10}$ & $3.4\times 10^{-11}$ 
& $4.1\times 10^{-12}$ & $1.1 \times 10^{-13}$ \\
No. of events 		&2500	&7700	&64 k	&1.1 M	\\
\hline

\kpme    & & & &                 \\
Acceptance              &5.4\%  &5.4\%  &2.3\%   &2.3\%   \\
Single Event Sens.      &$3.1\times 10^{-11}$    &$1.0\times 10^{-11}$
&$1.2\times 10^{-12}$   &$7.1\times 10^{-14}$ \\
\hline

$K_L \rightarrow \mu^+\mu^-$ (Br=$7\times10^{-9}$)  & & & &             \\
Acceptance              &27\%   &27\%   &10\%   &10\%   \\
Single Event Sens.      &$6.3\times 10^{-11}$   &$2.1\times 10^{-12}$
                        &$2.1\times 10^{-13}$   &$1.2\times 10^{-14}$ \\
No. of events           &1.1 k  &3.4 k  &25 k   &427 k  \\
\hline

$K_L \rightarrow e^+e^-$ (Br=$3\times10^{-12}$) & & & &                 \\
Acceptance              &21\%   &21\%   &10.8\%   &10.8\%   \\
Single Event Sens.      &$8.2\times 10^{-12}$   &$2.6\times 10^{-12}$
&$2.6\times 10^{-13}$	&$1.5\times 10^{-14}$ \\
No. of events           &0.4   &1   &11  &197  \\
\hline

\end{tabular}
\caption{Parameters and sensitivities for charged mode rare decays
for KTeV and KAMI.}
\label{tab:kpee_sens}
\end{center}
\end{table}

As indicated in Table \ref{tab:kpnn_sens},
KAMI-Far is sensitive enough to detect 30 events per year.
After three years of operation, we expect to have on the order of 100 
signal events.
With KAMI-Near, more than 100 signal events could be collected within
one year. 
Since this branching ratio is proportional to $\eta^2$,
the statistical error on 100 events corresponds to an accuracy of 
1/2$\sqrt{100}$ = 5\% on $\eta$, assuming no background. 
A detailed study of background levels is given in 
Section~\ref{background_study}. 
 
\vspace*{0.15in}
\parindent=0.25in

Table \ref{tab:kpee_sens} shows the sensitivity for some charged decay modes.
Here, we assume a more aggressive 8 mrad targeting angle (instead of 24 mrad)
in order to obtain the highest sensitivity possible.
Generally speaking, KAMI-Far is more than a order of magnitude
improvement over KTeV, and KAMI-Near could add another order
of magnitude improvement in sensitivity.
As shown in the table, 
the kaon flux at KAMI-Near will reach levels exceeding 
100M decays per second.  
This will result in a sensitivity of better than
$10^{-13}$ per year for most of the charged decay modes.

\section{Neutral Kaon Beam at the Main Injector}

\subsection{Primary proton beam}

A study for modifying a limited portion of the 800 GeV Switchyard to run
120 GeV protons from the Main Injector was recently completed~\cite{brown}.
This effort describes the work needed to deliver 120 GeV protons
to the Meson Area and KAMI.  
The optics for KAMI require the addition of
3(4) quadrupole doublets in existing beamline enclosures for the KAMI-Far
(Near) targets in NM2 (NM3).  
This report emphasizes the need to complete the
installation of beamline elements for 120 GeV running in the upstream portion 
of the Switchyard before the start of the TeVatron 1999 fixed target run.  
This would allow confirmation of the phase space occupied by the proton beam 
from the Main Injector well before the final installation
of the downstream portions of the new 120 GeV beamlines.
In addition, it appears feasible to transport low-intensity 120 GeV 
protons to KTeV 
during the 1999 run without changes to the 800~GeV KTeV primary 
beamline.
Losses on two of the dipoles in the beamline restrict high
intensity running~\cite{kob}.

  To best utilize the full intensity available from the Main Injector,
KAMI is being designed to operate in a debunched beam.  Debunching of the
RF structure of the Main Injector allows a uniform spill structure,
improving the ability of the KAMI detector to handle high rates.
Discussions with the Main Injector experts indicate debunching is feasible
but care must be taken with beam loading effects~\cite{kamiconcept}.

\subsection{Target station and neutral beamline}

The KAMI-Far option uses the existing KTeV NM2 target station and
neutral beamline~\cite{coleman}.  The last three dipoles just upstream of 
the KTeV target
would need to be modified to provide a steeper targeting angle. An additional
1.5~ft. of earth shielding would be required over NM2.

The KAMI-Near option requires a new target station located in NM3.
The KTeV beam and experimental hall design included sufficient transverse
space for shielding a KAMI target station in NM3 and the building has a
``shelf" on which crane rails can be installed.
The design of a near target station for 120 GeV was first considered in
the KAMI CDR~\cite{kamiconcept}. It consists of a production target, 
a ``hyperon" sweeper
magnet, a magnetized proton dump, a beam defining collimator, and a 
final sweeper dipole as shown in Figure~\ref{fig:detector_bbb2}.
Many of the existing KTeV target station elements will 
be reused.  Much work remains to optimize these concepts and to incorporate
our experience from KTeV.

\vspace*{0.15in}
\parindent=0.25in

Table~\ref{tab:flux}  summarizes the kaon and neutron flux
as well as the average momentum of decaying kaons for various targeting angles
ranging from 8 to 24 mrad.
The flux is normalized in units of $10^{-6}$ per incident proton,
and per $\mu$str solid angle.
In this flux estimation, we have assumed a 3 inch thick lead absorber
to reduce the photon flux, as is the case in KTeV.
A 20 inch thick Be absorber was used in KTeV
during E832 data taking to improve the
neutron/kaon ratio, but is not considered in this table.
The Be absorber would attenuate the neutron and kaon flux by a factor of 0.19
and 0.35, respectively.

\begin{table} [here,top,bottom]
\begin{center}
\begin{tabular}{|l|c|c|c|c|}
\hline

{\bf Targetting angle}                  &8mrad  &12mrad &16mrad &24mrad \\
\hline\hline

                                        & & & & \\
{\bf Kaon flux}\hfill ($\times 10^{-6}$/proton/$\mu$str)   & & & & \\
\hspace*{0.5in} at $z =  0$ m             &9.05   &7.50   &6.12   &4.37 \\
\hspace*{0.5in} at $z = 40$ m (KAMI-Near)&8.21   &6.62   &5.39   &3.78 \\
\hspace*{0.5in} at $z =160$ m (KAMI-Far) &6.47   &5.06   &3.98   &2.60 \\

                                        & & & & \\
{\bf Neutron flux}\hfill ($\times 10^{-6}$/proton/$\mu$str)
                                        &184.   &79.8   &43.1   &18.6 \\
                                        & & & & \\
{\bf Neutron/Kaon ratio} \hfill (at $z = 0$ m)
                                        &20.3   &10.6   &7.0    &4.3 \\
                                        & & & & \\
\hline\hline
                                        & & & & \\
{\bf Kaon average momentum}\hfill (GeV/$c$)     & & & & \\
\hspace*{0.5in} Generated kaon          &27.1   &23.1   &19.8   &15.4 \\
\hspace*{0.5in} Decayed at KAMI-Near    &15.4   &13.6   &12.1   &10.1 \\
\hspace*{0.5in} Decayed at KAMI-Far     &20.5   &18.0   &16.1   &13.4 \\
                                        & & & & \\
{\bf Neutron average momentum}\hfill (GeV/$c$)
                                        &48.9   &36.4   &27.8   &18.3 \\
                                        & & & & \\
\hline
\end{tabular}
\caption{Kaon and neutron flux and average momentum
at different targeting angles.}
\label{tab:flux}
\end{center}
\end{table}

\begin{figure}
\centerline{\psfig{figure=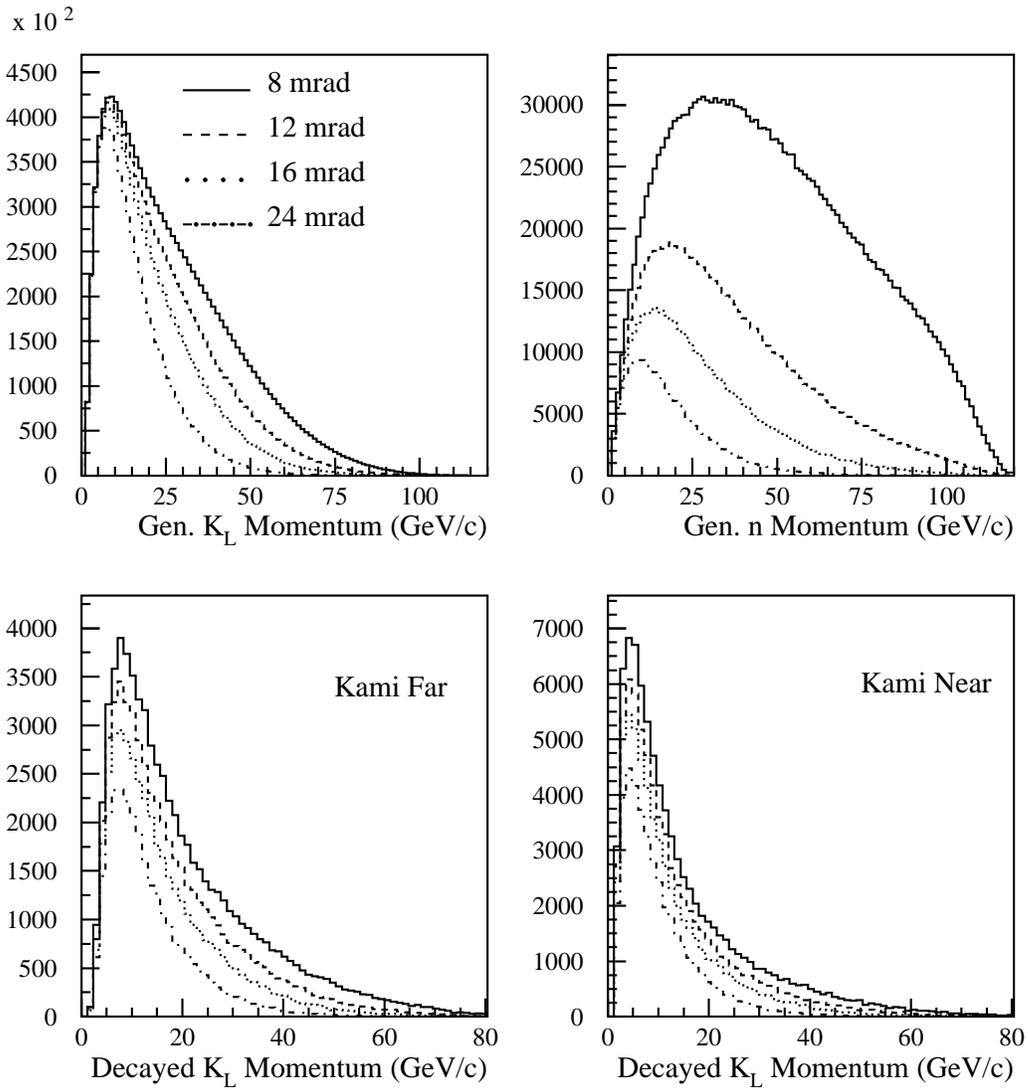,width=16cm}}
\caption{The Kaon and neutron momentum spectra at the Main Injector
for targeting angles of 8, 12, 16 and 24 mrad.}
\label{momenta}
\end{figure}

The kaon flux and momentum spectrum
are calculated using the Malensek parametrization~\cite{malensek}.
The momentum spectra for both neutrons and kaons at various targeting
angles are shown in Fig.~\ref{momenta}.

The kaon momenta are presented in three different  
ways in Table~\ref{tab:flux} and Fig.~\ref{momenta}:
the momentum of kaons generated at the target; the momentum
of kaons which decay in the KAMI-Far configuration; and the momentum
of kaons which decay in the KAMI-Near configuration.
The spectrum for KAMI-Near is softer because lower energy kaons
will decay closer to the production target.

The neutron spectrum in Fig.~\ref{momenta} has been generated using the 
parametrization of Edwards~{\it et al.}~\cite{edwards}.  
The forward neutron 
production has been modified with a $P_t$ dependence from 
Engler~{\it et al.}~\cite{engler}.
The flux predicted by this generator 
was a factor of 5.7 too low compared to the value measured in KTeV.
We have therefore decided to multiply the simulated flux by this factor
in order to be conservative.

\vspace*{0.15in}
\parindent=0.25in

\begin{figure}
\centerline{\psfig{figure=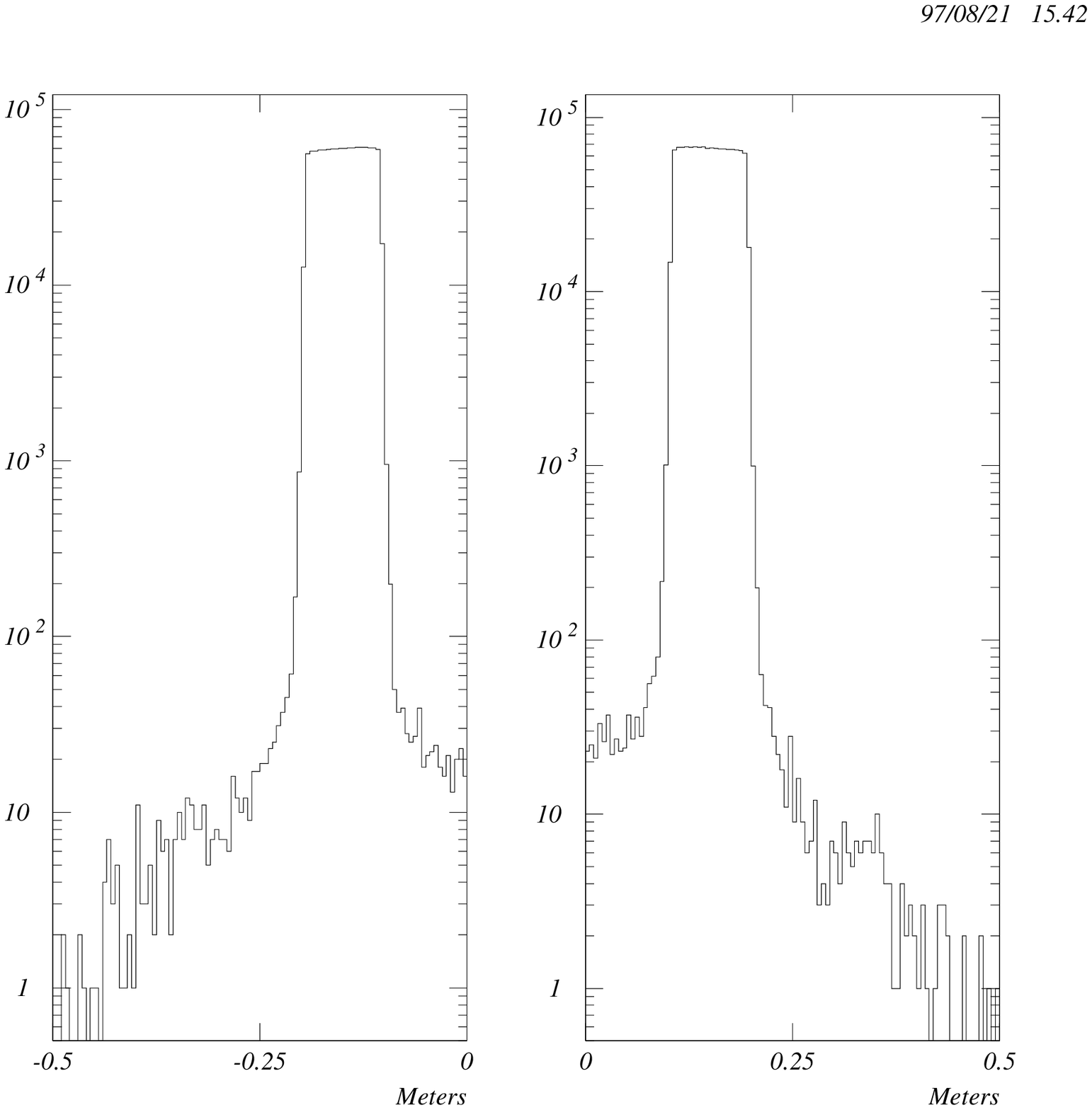,width=14cm,height=10cm}}
\caption{The beam profile of the KTeV $K^0$ beam, reconstructed from 
$K_{e3}$ decays from E832.}
\label{fig:beam_profile}
\end{figure}

Currently, we are studying the beam profile which results from the
existing KTeV collimator system.
Fig.~\ref{fig:beam_profile} shows the Kaon beam profile reconstructed
from $K_{e3}$ events in E832. (One should note that KTeV has two
parallel beams.)
The beam halo is a possible source of accidental background.  We are
currently studying data from KTeV to better understand the origin of
the beam halo
and possible ways in which it might be further reduced.

\section{The KAMI Detector}

\subsection{Overall geometry}

The detector geometry is primarily governed by the stringent requirement
of photon rejection efficiency, which must be fully hermetic
along the entire kaon decay volume.

The most upstream section is a 5 m long veto region which is
surrounded by the two Mask Anti detectors and vacuum photon veto
detectors to reject all upstream decays.
This section is followed by a 21 m long fiducial region,
again completely covered by vacuum photon veto detectors
located inside of the vacuum tank. 
 
The next 12 m contains a charged spectrometer,
consisting of five fiber tracking modules, a wide-gap analyzing magnet,
and four sections of photon veto detectors. 
A charged hodoscope is located
downstream of the last tracking module,
followed by a vacuum window.
 
The KTeV pure CsI calorimeter will be re-used and will sit just downstream
of the vacuum window.  The gap between the vacuum window and the CsI 
is filled by two sets of CsI Anti counters to cover any cracks between
the vacuum veto system and the CsI.
Behind the CsI, the neutral beam is dumped onto a beam-hole calorimeter known
as the Back Anti, designed to veto photons going down the CsI beam hole.
Finally, there are multiple layers of iron shielding and muon counters
for muon identification.

\begin{figure}[htb]
\centerline{
       \begin{turn}{90}
        \psfig{figure=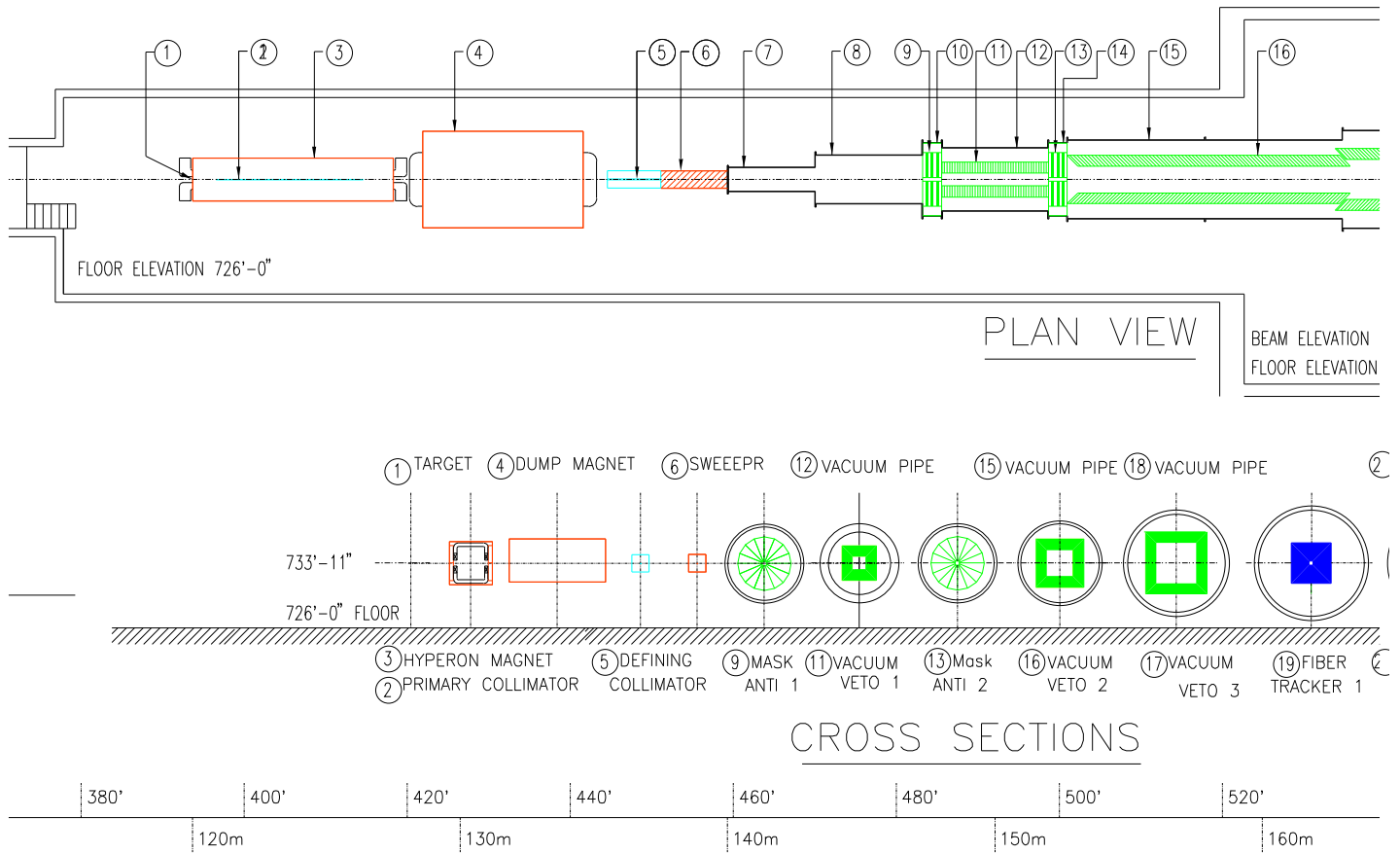,width=14cm}
       \end{turn}
}
\caption{Detailed plan view and cross sections of the
upstream section of the KAMI detector. The neutral
beam line for the KAMI-Near option is also shown.}
\label{fig:detector_bbb2}
\end{figure}

\begin{figure}[htb]
\centerline{
        \begin{turn}{90}
        \psfig{figure=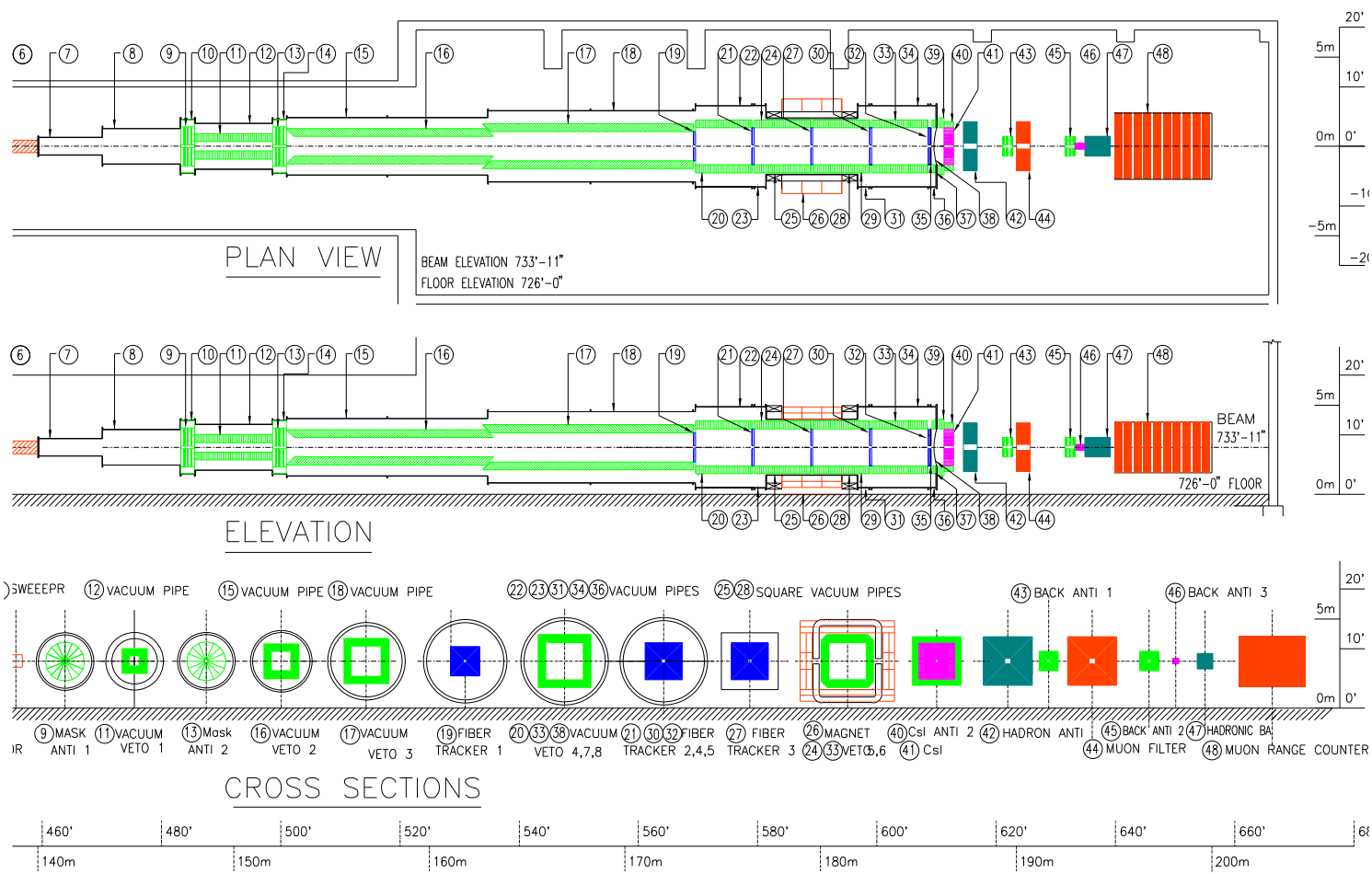,width=14cm}
        \end{turn}
}
\caption{Detailed plan view and cross sections of
the KAMI detector.}
\label{fig:detector_bbb3}
\end{figure}

Figure~\ref{fig:detector_bbb2} and Figure~\ref{fig:detector_bbb3}
are detailed schematics of the upstream and downstream 
sections of the detector. The neutral
beam line for the KAMI-Near option is also shown in 
Figure~\ref{fig:detector_bbb2}.
Table \ref{tab:geometry} shows the locations of all the detector elements.

\begin{table}[here,top,bottom]
\begin{center}
\begin{tabular}{|l|r|r|r|r|r|r|}
\hline

                        &z(up)   &z(down) &x(in)   &x(out)  &y(in)   &y(out) \\
			& (m)      & (m)  & (m)    & (m)    & (m)  & (m)  \\
\hline\hline
[KAMI-Far]     & & & & & &\\
Production target       &0.00	 &0.30  &-       &0.0015  &-
&0.0015 \\
Primary collimator      &9.60   &11.60  &-       &0.0029  &-
&0.0029 \\
Defining collimator     &85.00  &88.00  &-       &0.0255  &-  
&0.0255 \\
\hline\hline
[KAMI-Near]     & & & & & &\\
Production target       &120.00	 &120.30  &-       &0.0015  &-
&0.0015 \\
Primary collimator      &120.40  &128.00  &-       &0.0043  &-
&0.0043 \\
Defining collimator     &135.50  &137.50  &-       &0.0078  &-
&0.0078 \\
\hline\hline
Mask Anti 1             &147.30  &148.00  &0.07    &0.70    &0.07    &0.70 \\
Vacuum Veto 1           &148.00  &153.00  &0.25    &0.65    &0.25    &0.65 \\
Mask Anti 2             &152.00  &152.70  &0.07    &0.70    &0.07    &0.70 \\
Vacuum Veto 2           &152.70  &163.00  &0.50    &0.90    &0.50    &0.90 \\
Vacuum Veto 3           &163.00  &173.50  &0.75    &1.15    &0.75    &1.15 \\
\hline
Fiber Tracker 1 (x/y)   &173.50  &173.60  &0.07    &0.70    &0.07    &0.70 \\
Vacuum Veto 4           &173.60  &176.50  &0.95    &1.35    &0.85    &1.35 \\
Fiber Tracker 2 (x/y)   &176.50  &176.60  &0.07    &0.85    &0.07    &0.85 \\
Vacuum Veto 5           &176.60  &179.50  &0.95    &1.35    &0.95    &1.35 \\
Magnet (gap)            &178.00  &181.00  &-       &1.45    &-       &1.45 \\
Fiber Tracker 3 (x)     &179.50  &179.60  &0.07    &0.95    &0.07    &0.95 \\
Vacuum Veto 6           &179.60  &182.50  &0.95    &1.35    &0.95    &1.35 \\
Fiber Tracker 4 (x/y)   &182.50  &182.60  &0.07    &0.95    &0.07    &0.95 \\
Vacuum Veto 7           &182.60  &185.50  &0.95    &1.35    &0.95    &1.35 \\
Fiber tracker 5 (x/y)   &185.50  &185.60  &0.07    &0.95    &0.07    &0.95 \\
Charged Hodoscope       &185.60  &185.65  &0.07    &0.95    &0.07    &0.95 \\
Vacuum Veto 8           &185.70  &185.90  &0.95    &1.45    &0.95    &1.45 \\
Vacuum window           &185.90  &185.95  &-       &1.95    &-       &1.95 \\
\hline
CsI Anti 1              &186.05  &186.30  &0.95    &1.45    &0.95    &1.45 \\
CsI Anti 2              &186.30  &186.70  &0.95    &1.25    &0.95    &1.25 \\
CsI                     &186.30  &186.80  &0.075   &0.95    &0.075   &0.95 \\
Hadron Anti		&187.30  &188.00  &0.15	   &1.25    &0.15   &1.25 \\
Back Anti 1             &189.30  &189.80  &0.075   &0.50    &0.075  &0.50 \\
Muon Filter             &190.00  &190.70  &0.15    &1.25    &0.15  &1.25\\
Back Anti 2             &192.50  &193.00  &0.075   &0.50    &0.075 &0.50\\
Back Anti 3 (Pb/Quartz) &193.00  &193.50  &-       &0.15    &- &0.15 \\
Hadronic BA             &193.50  &194.80  &-       &0.50    &- &0.50 \\
Muon range counter      &195.00  &200.00  &-       &1.70    &-     &1.30 \\
\hline
\end{tabular}
\caption{Locations and inner/outer dimensions of all the KAMI detector
elements.}
\label{tab:geometry}
\end{center}
\end{table}

\subsection{CsI calorimeter}
 
The KTeV CsI calorimeter is the most advanced, high-precision
electromagnetic calorimeter currently in use.
The calorimeter consists of 3100 pure CsI crystals and is described
in detail in Section~\ref{csi}.  The crystals
are read out using photomultiplier tubes and the signals are
digitized at the PMT base in 19 ns time slices, in synch with the
RF structure of the beam.  The digitizer is
a multi-ranging device with 16 bits of dynamic range in the form
of an 8-bit mantissa and a 3-bit exponent.  The noise per channel
is about 0.8 MeV.  The energy resolution of the calorimeter 
is better than 1\% over the energy
region of 5 - 100 GeV.  Resolution at this level will be necessary in 
KAMI in order to reject backgrounds to decays such as
\mbox{$K_L \rightarrow \pi^0 \nu \overline{\nu}$} and
\mbox{$K_L \rightarrow \pi^0 e^+ e^-$}.  

Because the KAMI beam will likely
be debunched with no real RF structure, the digitization scheme
for the readout electronics will have to be modified.  Additionally,
the array will have to be re-stacked from the two beam hole configuration
currently used by KTeV to the single beam hole configuration 
(\mbox{15 cm $\times$ 15 cm})
required by KAMI.  No additional modifications should be necessary.

\subsection{Photon veto detectors}
\label{phot_veto}
   
One of the most challenging detector issues facing KAMI is the
efficient detection of all photons produced by background events
along the 34 m long vacuum decay region.  Complete hermeticity
and efficient photon detection down to energies as low as a few MeV
is required.  

The inefficiency at low energy is dominated
by sampling effects, where a fraction of the shower electrons
are absorbed in the lead, while at high energy it is  
more dominated by photonuclear absorption.  
In the latter
case, it is possible for a photon to experience a photonuclear
absorption interaction before it begins to shower.  If all of the
secondary products in the interaction are neutrons, the interaction
may escape detection.  Photonuclear absorption has been studied extensively
in the past in various energy regions~\cite{inagaki}.

A photon veto detector for KAMI will likely be based on the existing KTeV
veto design.  However, in order to improve detection efficiency
for low energy photons, both finer sampling and more scintillation
light are required.  The cost of such a detector is of primary concern
and a good deal of effort has gone into designing a low-cost device.

As discussed in Section 5.2, 
a GEANT simulation of a possible photon veto design shows that with
1~mm lead sheets and 5 mm thick scintillator tiles,
better than 80\% detection efficiency for photons with energies
between 2-20~MeV can be achieved.
For high energy photons, 
photonuclear absorption effects need to be taken into
account. More detailed study is still necessary.
 
The photon veto system for KAMI consists of three major elements; 
the Mask Anti, the vacuum veto and the Back Anti.  Each detector
system is described in the following sections.

\subsubsection{Mask Anti detector} 

Background events from upstream decays must be rejected by
the active mask photon detector, called the Mask Anti (MA).
A detailed simulation of backgrounds originating from $K_L \rightarrow 3\pi^0$ 
decays indicates the need for two stages of detector, as shown in 
Figure~\ref{kami_det} and Figure~\ref{fig:detector_bbb3}.
Each Mask Anti detector is a 20 radiation length deep sampling calorimeter,
consisting of alternating layers of 1~mm thick lead sheets and 5~mm thick
scintillator sheets.

\subsubsection{Vacuum veto detector}

The vacuum veto is a fine sampling calorimeter consisting of 
1~mm thick lead sheets and 5~mm thick plastic scintillator.
All of the materials are located on the inside wall of the vacuum tank
to avoid any dead material at the boundary of the tank and the
veto detector.

\begin{figure}
\centerline{\psfig{figure=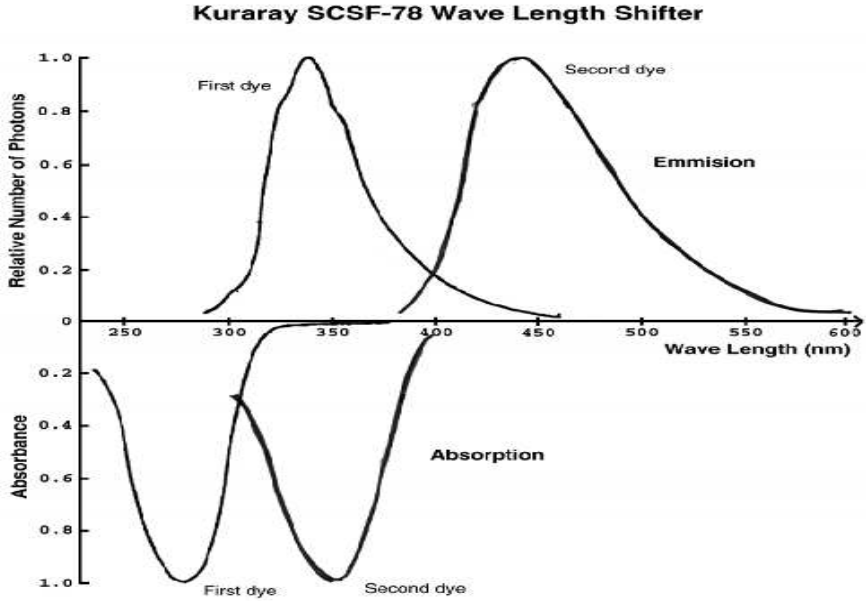,width=12cm,height=9cm}}
\caption{The absorption and emission spectra of the two dyes proposed
for the KAMI photon veto detectors.
The first dye will be used in the scintillator and the second dye will be
used as a wavelength shifter in the fibers.}
\label{fig:wls_spectrum}
\end{figure}

Wave length shifting fibers are inserted in the scintillator
to efficiently collect photons and to transport them
to photomultipliers which are to be mounted outside of the vacuum tank
for easy access.
We are investigating the possibility of using the two wave length shifting
dyes which are used in Kuraray SCSF78 scintillator.
The absorption and emission spectrum of two dyes are shown in
Figure~\ref{fig:wls_spectrum}.
We plan to use the first dye in the scintillator plates, and
the second dye in the scintillator fibers.

At 20~cm intervals in the z direction, about 8000 WLS fibers will be
bundled together and viewed by 2-3 inch PMTs at both ends of the fibers.
In and near the spectrometer magnet, fine mesh photomultipliers
must be used.
Signals from both ends of the fibers will be meantimed for better
timing resolution.

The lead sheets in the vacuum veto modules will be tilted by 45 degrees 
relative to the beam direction (z-axis) in the fiducial decay region
to provide the best sampling
ratio for 45 degree incident photons and to avoid any cracks for angles
of incidence up to 90 degrees.
Table \ref{tab:degree} shows the sampling ratio and total depth as a function
of the opening angle of photons.

\begin{table}[here,top,bottom]
\begin{center}
\begin{tabular}{|r|r|r|r|}
\hline

Photon opening  &Sampling       & Total 		&Total 	\\
angle (degree)   &frequency (mm)  &depth ($X_0$)	&depth ($\lambda_0$) \\
\hline\hline
0               &1.41           &(Infinite)	&(Infinite)	\\
5		&1.31		&138		&4.3	\\
10		&1.22		&69		&2.3	\\
15              &1.15           &46		&1.5	\\
20		&1.10		&35		&1.2	\\
30              &1.04           &24		&0.8	\\
45              &1.00           &16.9		&0.56	\\
60              &1.04           &13.8		&0.46	\\
75              &1.15           &12.4		&0.41	\\
90              &1.41           &12.0		&0.40	\\
\hline
\end{tabular}
\caption{Sampling ratio and total depth (in radiation lengths and nuclear
interaction lengths) of vacuum veto counters for
various angle of incidence photons.}
\label{tab:degree}
\end{center}
\end{table}

For high energy photons (above 1 GeV), the inefficiency of the vacuum vetos
must be smaller than $3\times10^{-6}$ to reduce the number of $2\pi^0$ background events
to a manageable level. As mentioned previously, a small fraction of photons
will undergo a photonuclear interaction and produce only neutrons.
As shown in Table~\ref{tab:degree}, the proposed vacuum veto detector
has multiple nuclear interaction lengths for shallow-angle, high-energy
photons. This should allow detection of secondary neutrons and 
minimize the inefficiency
which results from photonuclear interactions.

In Table \ref{tab:veto}, the vacuum veto counters are compared to the photon
veto detectors for KTeV and BNL-E787. 
The KAMI design is based on the same WLS fiber readout scheme
as KTeV, but has the same fine sampling ratio as BNL-E787.
  
Currently, inexpensive scintillator is under investigation
by the MINOS Collaboration. By extruding polystyrene,
the cost can be reduced by an order of magnitude compared to
the conventional commercial product.
D0 has already made such extruded scintillator for their pre-shower detector.
Table \ref{tab:scint} summarizes the scintillator types and performance
for D0, MINOS and KAMI.

\begin{table}[here,top,bottom]
\begin{center}
\begin{tabular}{|l|c|c|c|}
\hline
			&KTeV		&BNL-E787	&KAMI	\\
\hline\hline
Lead sheet thickness	&2.8 mm		&1 mm		&1 mm	\\
Scintillator thickness	&2.5 mm		&5 mm		&5 mm	\\
Total Depth		&16 $X_0$	&14.3 $X_0$	& $>$ 20 $X_0$ \\
Number of Layers	&24-32		&75		& $>$ 100	\\
Light Guide		&WLS fiber	&Clear light	&WLS fiber	\\
			&(1 mm $\phi$	&pipe		&(1 mm $\phi$	\\
			& single clad	&		& double clad	\\
			& 3 cm spacing)	&		&1-2 cm spacing) \\
\hline	
No. of p.e. /MIP/layer	&2.4 pe		&10 pe		&10 pe		\\
No. of p.e. /MIP	&60 pe		&750 pe		& $>$ 1000 pe	\\
No. of p.e. /MeV	&0.3 pe/MeV	&5 pe/MeV	&5 pe/MeV	\\
\hline
\end{tabular}
\caption{Comparison of the KAMI vacuum veto counters with photon veto counters
from KTeV and BNL.}
\label{tab:veto}
\end{center}
\end{table}

\begin{table}[here,top,bottom]
\begin{center}
\begin{tabular}{|l|c|c|c|}
\hline

			&D0		&MINOS		&KAMI	\\
\hline\hline
Cross section		&Triangle	&Rectangular	&Rectangular	\\
			&5 mm x 9 mm	&1 cm x 2 cm	&2 cm x 5 mm	\\
Length			&2.8 m		&8 m		&2 m		\\
\hline
WLS fiber		&0.84 mm $\phi$	&1 mm $\phi$	&1 mm $\phi$	\\
Type of WLS		&3HF		&BCF91		&SCSF78		\\
Emission lambda		&550 nm		&520 nm		&450 nm		\\
Photo detector		&VLPC		&PMT		&PMT		\\
			&(QE=60\%)	&(QE=12\%)	&(QE=25\%)	\\
\hline\hline
No. of p.e. /MIP/layer	&15 p.e.	&8 p.e. (at 2m)	&10 p.e. exp'd	\\
\hline
\end{tabular}
\caption{Comparison of scintillator used by D0, MINOS and KAMI.} 
\label{tab:scint}
\end{center}
\end{table}

\begin{figure}[htb]
\centerline{
        \begin{turn}{-90}
        \psfig{figure=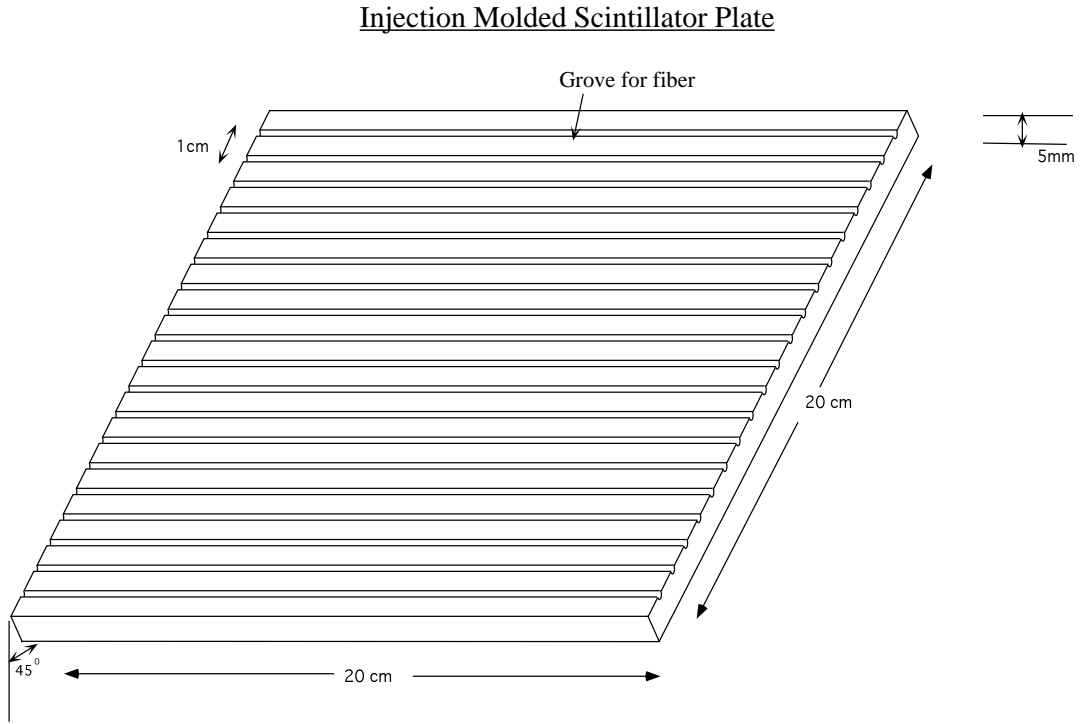,width=12cm}
        \end{turn}
}
\caption{Schematic of a typical injection molded scintillator plate.}
\label{fig:scinti_rejection}
\end{figure}

Another possibility is to use injection molded scintillator.
The PHOENIX collaboration at BNL has developed scintillating tile
for their Shashulik Calorimeter using this technology.
KEK's \pnn group is also developing such scintillator for
a similar veto system. A schematic of a typical scintillator is 
shown in Fig~\ref{fig:scinti_rejection}.

The total cost of the vacuum veto detector is estimated to be \$4.4M.
The cost is broken down in Table \ref{tab:scint_cost} in Section~\ref{cost}.

\subsubsection{Back Anti detector}
\label{ba}

The Back Anti resides in the neutral beam and detects photons which 
pass through the beam hole
of the CsI calorimeter. Due to the high counting rate and high radiation
dose at this position, a very fast, radiation~hard calorimeter is 
necessary. One well-established, radiation~hard material is quartz.
We are considering a tungsten/quartz fiber, or tungsten/quartz plate
sampling calorimeter as our base design.
This area will be exposed to \mbox{200 - 500 MHz} of neutron
interactions. In order to distinguish photon interactions from
neutron interactions, it is extremely important to make the Back Anti as 
transparent to neutrons as possible.
Using a Cherenkov radiator such as quartz will help in this regard. 
The current design has four
layers of longitudinal segmentation and allows for different thresholds
to be applied to each region. 
An hadronic section of the Back Anti could be installed just behind 
the electromagnetic section if it were determined that this would be useful.
  
In order to detect photons which pass through the CsI beam hole but which
miss the Back Anti, additional veto counters are installed
immediately upstream of the Back Anti.  These counters will be constructed
similar to the Mask Anti.

\subsubsection{CsI Anti detector} 

The gap between the vacuum window and the CsI calorimeter must be
filled by additional photon veto detectors in order to plug any
possible cracks which photons might pass through.  These are
known as the CsI Anti detectors.
There are two stages of CsI Anti; a small one inside of the blockhouse which
houses the CsI, and a large one just upstream of the blockhouse. 
Due to the limited space available inside of the
blockhouse, only a 20~cm ring outside of the CsI can be
covered. This is not large enough to detect all of the photons which escape
through the end of the vacuum tank.
Therefore, another veto detector with a larger outer dimension must be 
installed between the vacuum window and the blockhouse.
The CsI Anti for KAMI is very similar to the one currently in use by KTeV 
except for its finer sampling thickness and larger volume.
The KAMI CsI Anti consists of 1~mm thick lead sheets and 2.5~mm thick
plastic scintillators built into modules which are 25~cm (13 $X_0$) deep
due to the limited space available in the CsI blockhouse.

\subsection{Charged particle spectrometer}
 
The KAMI charged particle spectrometer consists of the KTeV spectrometer 
magnet and five tracking stations consisting of scintillating fiber planes.
The spectrometer is described in detail below.

\subsubsection{Scintillating fiber tracker}

The KAMI tracking detectors are made from 500~$\mu$m diameter
scintillating fibers.  
There are a total of 5 modules, spaced at 3~m intervals.
Four of the five modules are identically constructed of four sets of 
fiber planes in an x/x$^\prime$ y/y$^\prime$ configuration.
The fifth module is located at the middle of the magnet for redundant 
measurement of momentum and to reject background events with a kink. 
This is especially important in rejecting $K_{e3}$ and $K_{\mu3}$ backgrounds
to two lepton decay modes such as \mm and \ee. 
This module has only x/x$^\prime$ planes in order to minimize
the thickness of material.

Visible Light Photon Counters (VLPC) are currently under consideration
for the fiber readout.
The high quantum efficiency of VLPC detectors ($>$60\%) make them
particularly attractive, although they must operate at liquid Helium
temperature.

The fibers may be read out at just one end or both ends.  In the
former case, the far end of the fiber would be mirrored to improve
the light collection.  Single ended readout reduces the cost and
has been demonstrated to produce sufficient numbers of photo electrons.
However, there is significant time skewing associated with single
ended readout which will impact our timing resolution.  Reading the
fibers out at both ends allows mean-timing of the two signals for
good timing resolution and makes it possible to include the fibers
in a trigger with a narrow time window.  The cost of the readout, of course,
doubles.

The end of fiber to be read out
will be spliced to a clear fiber,
which is fed through to the outside of the vacuum
tank and then brought to the cryogenic VLPC system.
More than 5 p.e. per MIP is expected which, even in the worst situation,
results in a 99.3\% detection efficiency for each view 
(x/x$^\prime$ or y/y$^\prime$). 
The detection efficiency of a prototype device in the presence of a 
40~MHz background has been measured to be better than 98\% by
D0~\cite{ruchti}. 
A 15~cm~x~15~cm hole will occupy the central region of each tracking plane
in order to minimize neutron interactions from the beam.  The fibers in
this region will be cut in the middle and read out at both ends. 
  
A total of 98.7 k channels are required for the tracking system
in order to read the fibers out at both ends.
A breakdown of the channel count appears in Table \ref{tab:fiber_count}.
 
\begin{table}[here,top,bottom]
\begin{center}
\begin{tabular}{|l|r|r|r|r|}
\hline
 
Tracker No.      &Plane type     &Size (x/y)   & \#ch/plane  & \#plane \\
\hline\hline
Fiber Tracker 1    &xx$^\prime$yy$^\prime$ &140 x 140 cm     &4666    &4 \\
Fiber Tracker 2    &xx$^\prime$yy$^\prime$ &160 x 160 cm     &5332    &4 \\
Fiber Tracker 3    &xx$^\prime$            &160 x 160 cm     &5332    &2 \\
Fiber Tracker 4    &xx$^\prime$yy$^\prime$ &180 x 180 cm     &6000    &4 \\
Fiber Tracker 5    &xx$^\prime$yy$^\prime$ &180 x 180 cm     &6000    &4 \\
\hline
Total \# channels   &              &                 &98658      &18	\\

\hline
\end{tabular}
\caption{Breakdown of the number of channels required for the KAMI 
fiber tracking system.  The fibers are read out at both ends.}
\label{tab:fiber_count}
\end{center}
\end{table}

The total cost of the fiber tracking system is about \$6M
(for readout at both ends of the fiber),
dominated by the cost of the 
VLPC detectors and by the associated electronics and cryogenics. 
A cost breakdown appears in Table \ref{tab:fiber_cost} in Section~\ref{cost}. 

\subsubsection{Spectrometer magnet}
 
\begin{figure}
\centerline{ \psfig{figure=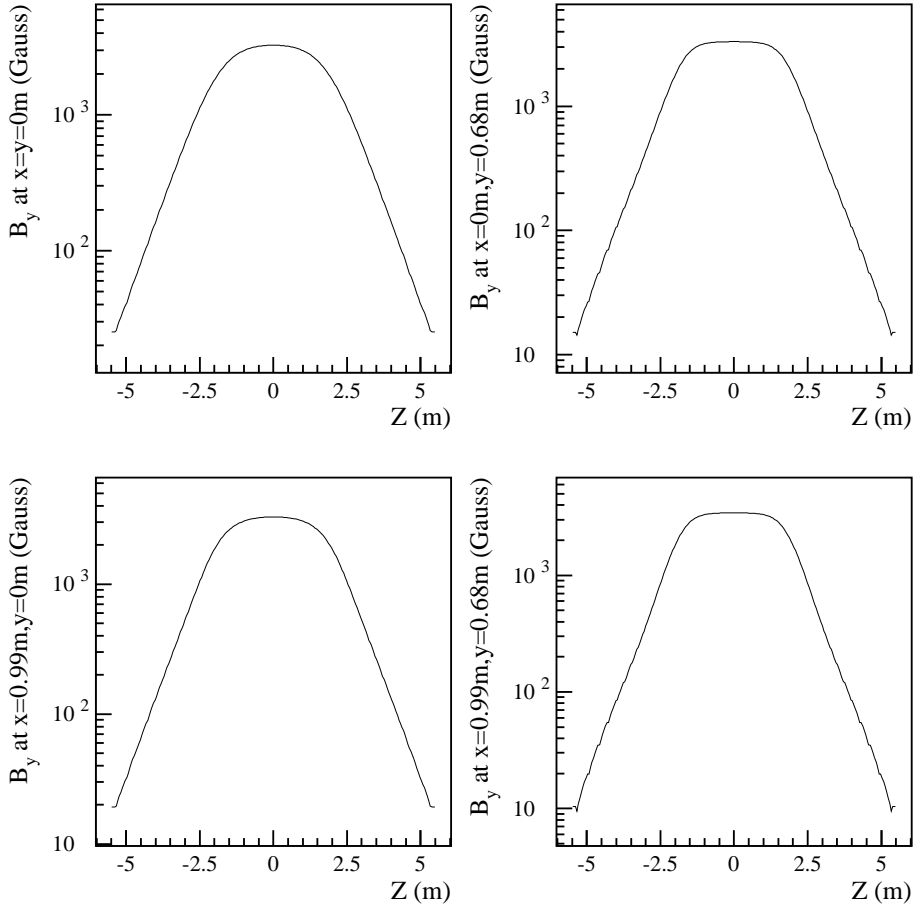,width=14cm} }
\caption{Vertical component of the expected magnetic field profile along
the z direction for the analysis magnet to be used in KAMI, shown for 
four different x and y positions.}
\label{magnet_field}
\end{figure}

The existing KTeV spectrometer magnet will be used for KAMI.  In order to
allow for the 2.8~m square vacuum pipe to pass through, the magnet will
have to be re-gapped from the current 2.1~m gap to 2.9~m.
The coil positions will have to be adjusted slightly in order to
reproduce the excellent field uniformity achieved in KTeV.
Figure~\ref{magnet_field} shows the expected y-component of the magnetic field
along the z-axis for four different x and y positions.

\subsection{Other detectors}

\subsubsection{Vacuum pipe and vacuum window}

Most of the vacuum pipe sections are of conventional cylindrical construction,
similar to the steel pipes currently used in KTeV.  The vacuum pipe
must be of sufficient strength to support the photon veto detectors.
The only exception is the vacuum pipe section which passes through the 
inside of the spectrometer magnet.  This particular pipe section will
have a square cross section and will be constructed of non-magnetic stainless
steel.  A detailed mechanical design is underway.

Since all of the momentum measurements of charged particles
are performed in vacuum, 
the vacuum window does not need to be super thin.
Any conversion of photons in the window can be detected
by the CsI calorimeter, in principle.
We do not see any technical difficulties here.
Several materials are under consideration which will satisfy both our 
technical and safety requirements.

We have already achieved a vacuum of $10^{-6}$ torr at KTeV.
We plan to use the same pumping system for KAMI, perhaps with some
modest upgrades.
With minor modifications, we anticipate $3\times10^{-7}$ torr for KAMI
which has a much smaller evacuated volume than KTeV.  The vacuum
pumping ports will all be located upstream of the Mask Anti detectors.
  
\subsubsection{Charged hodoscope}

The Charged Hodoscope is necessary as part of the charged particle trigger
and to veto charged particles in neutral triggers.
The inefficiency of the hodoscope should be controlled at the level of $10^{-5}$
for the latter purpose.  For this reason, it is located inside of the vacuum
tank just upstream of the vacuum window.

The charged hodoscope consists of two sets of plastic scintillator planes;
an x-view and a y-view.
Each counter has a dimension of 1~cm~(thick) x 2.5~cm~(wide) 
x 1.9~m~(long).
Both ends are viewed by fast, small photomultipliers from 
outside of the vacuum.
This configuration will provide accurate timing information 
($<$100~psec) 
for charged particles.

\subsubsection{Muon range counter}  

For better muon/pion identification than is achieved by KTeV at low energy, 
a muon range counter may be required.
Such systems have been used in other two-lepton mode experiments such as
BNL791/871, where the muon ranges were measured with 10\% accuracy.

\subsection{Electronics}

\subsubsection{Trigger}

The trigger for KAMI will be a three-tiered system, where
the first two levels will be realized in hardware and the third
level in software. The first level charged trigger
will use the charged hodoscope, located just upstream of the
vacuum window. This hodoscope consists of two crossed planes,
so that event topologies consisting of oppositely charged tracks
are easily distinguished.  We will also explore the
possibility of implementing a hit counting scheme at level-1
using the scintillating fiber tracker.
The neutral level-1 trigger will use
the CsI calorimeter to form a coherent energy sum from the entire 
calorimeter. The whole level-1 trigger system will be synchronized
to a global clock. However, since the KAMI beam will be debunched,
level-1 triggers can arrive anywhere within the period of this clock.
Therefore, we need to
perform further studies into how much smearing of the total energy
threshold occurs due to the debunched beam structure.

The neutral level-2 trigger will consist of a hardware cluster
counting scheme similar to the one used by KTeV. Since the anode
of each of the CsI phototubes is directly connected to the input of the
digitizer, the phototube dynode signals will be used to generate the bits 
required by the cluster counter. The cluster counter uses a parallel adding
scheme to quickly find the number of clusters in the calorimeter.
For two cluster events, the dominant trigger rate results from
$2 \pi^0$ events in which two of the photons are either missed
or fused. The event rate from $K_{e3}$ decays is greatly reduced
by a charged hodoscope veto. 

The cluster counter is particularly
sensitive to the effects of a debunched beam. A photon hitting
the CsI deposits its energy in many crystals. This, coupled
with the fluctuations in time of arrival of a photon relative to a 
fixed gate, can lead to large variations in the resulting pulse
used to generate a cluster counter bit. Since missing a single bit
can change the number of clusters found, we will have to study
the effects of the beam structure on the cluster counter trigger.

For the third level trigger, we will use filter software to
fully reconstruct each event before writing it to tape. 
The filter code will make very loose cuts on events to quickly
determine whether events are consistent with the required
topology. We
expect that the reconstruction software will require between 3-5 ms per event
and will reduce the number of events written to tape by a factor of 5-10. 

\subsubsection{Readout electronics}

To avoid deadtime incurred during readout, we are exploring
the possibility of using a fully buffered readout system.  This will
require upgrading most of the current KTeV readout elements, including
the ADCs, latches, TDCs and trigger readout. For the KTeV
experiment buffered readout was not implemented and reduced the
livetime by approximately 20\%. Currently, we are exploring
the commercially available TDCs and ADCs which support buffered
readout. The cluster counter readout is already buffered, but
we must explore whether the readout depth is sufficient for 
the conditions at KAMI.

The CsI readout consists of a digital 
photomultiplier tube or DPMT. The DPMT contains two ASICs,
the QIE (charge integrating encoder) and the DBC (data buffering and
clocking) chip which provides
the clock signals and readout of the QIE. In its current 
configuration, the DBC begins transferring data from a 
level-1 FIFO to a level-2 FIFO after the receipt of a
valid level-1 trigger. At 53~MHz, this transfer requires
up to 5~$\mu$s for 32 time slices. A simple modification to
the DBC would remove the deadtime associated with this transfer.

\subsubsection{Data acquisition}

For the KAMI detector we plan to use the same data acquisition
system as was used in the KTeV detector.
The architecture of this system consists of a buffer memory
matrix. In this scheme, data is received by the data acquisition
system in multiple parallel streams. Each of these
streams writes data into dual-ported memories.
For the KTeV experiment, a total of six streams were used. 

In the memory matrix, the
rows orthogonal to each of the data streams are connected to
a processor plane which contains multiple CPUs. These 
CPUs perform level-3 filtering of the events, and
typically rejected 80-90\% of the events read out
from the detector in KTeV.
The data from each event is transferred to the processor plane via
a 64-bit VME DMA.
The system is flexible enough so that one
event can be sent to any of the processor planes, allowing
one to both split the data by trigger type and allocate one
of the processor planes as a monitoring plane. For example,
in the KTeV experiment three of the processor planes received
the beam trigger events in a round-robin fashion. The fourth
plane received approximately 10\% of the events and did a
detailed analysis of the data to monitor the data quality.

Each of the six data streams has a bandwidth of approximately
40~MB/s for a total bandwidth of 240~MB/s. The instantaneous bandwidth for
each plane is approximately 40~MB/s, the VME64 specification.
Note that because of the matrix architecture of the data
acquisition system, the whole system can readily be expanded,
depending upon the average event size and rate.

\section{Expected Detector Performance}
 
This chapter describes the expected performance of major detector
elements in detail based on our current studies.

\subsection{CsI calorimeter}
\label{csi}

The KTeV calorimeter consists of 3100 pure CsI crystals, 27 radiation
lengths long, prepared for optimal resolution and linearity. It is
instrumented with low-gain, highly linear PMTs, each with its own
digitizer. 

\begin{figure}
\centerline{\psfig{figure=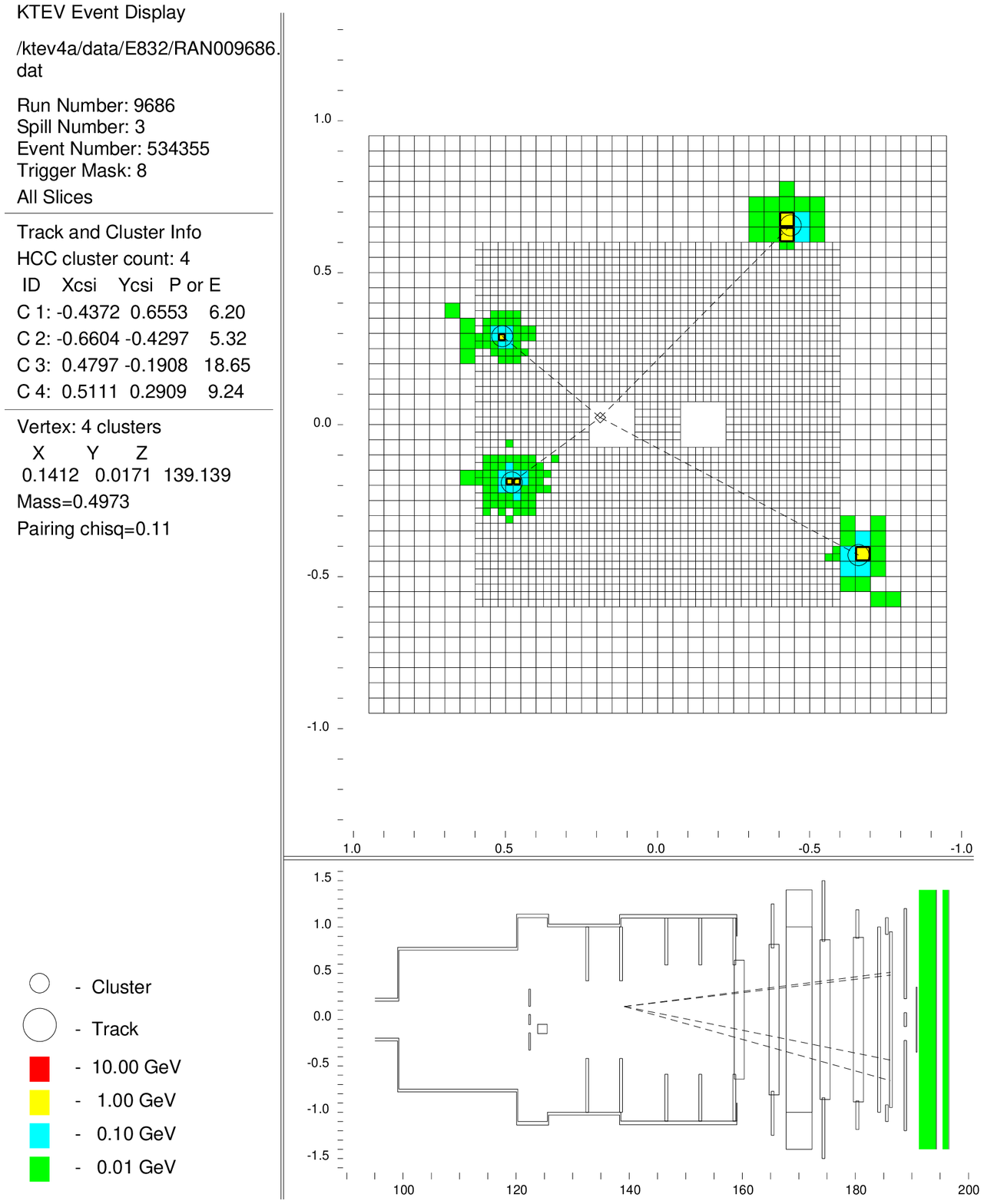,width=15cm}}
\caption{KTeV event display showing a typical 2$\pi^0$ event.}
\label{evtdisp}
\end{figure}

The layout of the CsI calorimeter for KTeV can be seen in 
Figure~\ref{evtdisp}, an event display of a $\klneut$ decay. 
The length of the crystals was chosen to be 27 radiation lengths (50 cm)
in order to achieve excellent energy
resolution and linearity.  There are two sizes of crystals, $2.5
\times 2.5 \times 50.0$ cm$^3$ crystals in the central region of
the calorimeter, and $5.0 \times 5.0 \times 50.0$ cm$^3$ crystals
in the outer region. The entire calorimeter is $0.95 \times 0.95$ m$^2$. 
The two transverse dimensions were chosen to optimize the
$\klneut$ mass resolution, while minimizing the number of channels.
To optimize the resolution and linearity of the calorimeter we
individually tuned the Aluminized-Mylar wrapping of each crystal so
that the scintillation response along the shower is uniform, to the
level of 5\%.  Finally, on average, the actual light yield for our
crystals is 20 photo-electrons/MeV, which corresponds to a
contribution to the energy resolution of $ 0.007/\sqrt{{\rm E (GeV)}} $.

\vspace*{0.125in}
\parindent=0.25in

The signal from each crystal is digitized by a ``Digital PMT" base, or
DPMT.  This DPMT is mounted directly behind each crystal.  The DPMT is
an auto-ranging device, with eight ranges.  The input current, I, is
split into eight binary ranges, ie. into I/2, I/4, I/8, I/16, I/32,
I/64, I/128, and I/256.  Then each of the eight split currents are
integrated, with a clock speed of up to 53 MHz. After integration,
it is determined
which of the eight binary ranges is {\it in range}, and the integrated
current from this range is digitized with an 8-bit FADC.  Thus the
DPMT produces an 8-bit mantissa and a 3-bit exponent for 16 bits of
dynamic range.

Advantages of the DPMT include extremely low noise, multiple samples
per crystal and a wide dynamic range.  The noise level is approximately
4~fC, or less than 1 MeV.  The multiple samples allow for additional
rejection of out-of-time accidental activity.  In addition, the time
resolution of the DPMT has been found to be 150~psec, using a
calibration laser flasher system.  

\vspace*{0.125in}
\parindent=0.25in

For the KAMI experiment, the important calorimeter considerations
include energy resolution, resistance to radiation damage, efficiency
to separate nearby photons, and gaps between crystals. Let us discuss
these issues in turn.

\begin{figure}
\centerline{\psfig{figure=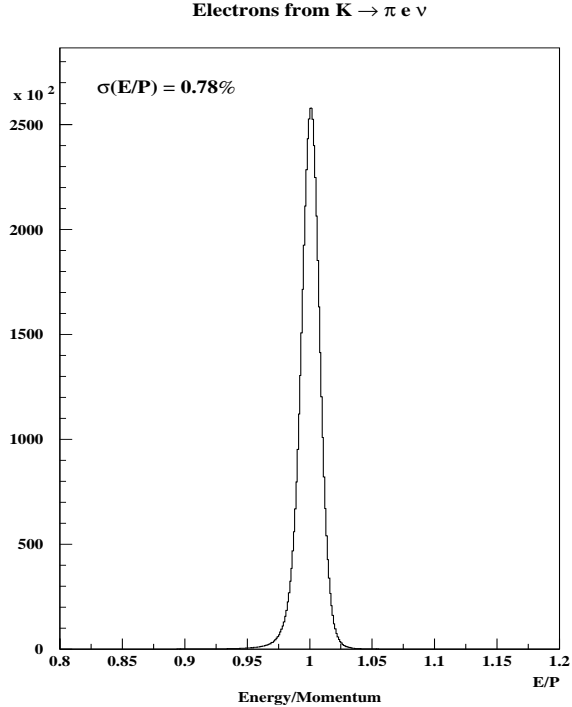,width=8cm,height=10cm}}
\caption{The E/P ratio for electrons from $K_{e3}$ decays in the KTeV 
CsI calorimeter.}
\label{eop}
\end{figure}

\begin{figure}
\centerline{\psfig{figure=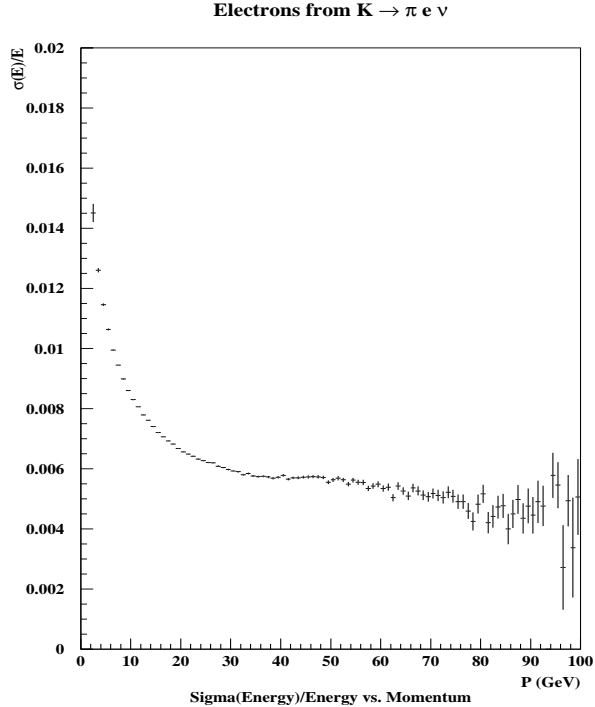,width=8cm,height=10cm}}
\caption{Energy resolution vs. momentum for electrons from 
$K_{e3}$ decays in the KTeV
CsI calorimeter.  The estimated contribution from the momentum resolution
has been removed.}
\label{resol}
\end{figure}

The energy and position resolution of the calorimeter can be
determined from momentum analyzed electrons from $\kethree$ decays.
The ratio of energy measured in the calorimeter to momentum measured
in the spectrometer is shown in Figure~\ref{eop}.  For electrons
incident over the entire calorimeter, and having momentum from 4-100~GeV,
the resolution is $\sigma(E/P) = 0.78\%$. The resolution as a
function of momentum, with the estimated contribution from the
momentum resolution removed, is shown in Figure~\ref{resol}.  The
resolution is 1.3\% at the KAMI mean photon energy of roughly 3
GeV. Some improvement in the resolution at low energies will be
possible, since in KTeV we have masked off typically 50\% of the
scintillation light to improve the linearity of our PMTs.  For KAMI,
where sub 1\% linearity is not critical, we will remove these masks
to increase the scintillation light output.

\vspace*{0.125in}
\parindent=0.25in

In one year of running KAMI-Near, we expect that the radiation dose in the
center of the calorimeter will be approximately 60~kRads.  At this dose we 
expect to
see some degradation of the calorimeter's resolution.
%
%  KA estimates x20 dose compared to E799.  E799 has 0.5~kRad/month of
%  actually running.  So Dose = 0.5~kRad/month * 20 * 6 months = 60~kRad
%
The radiation resistance
of the CsI crystals has been tested in two ways.  First, controlled
tests of a small sample of crystals were performed in which doses of
10~kRads were applied, and the change in scintillation response was
monitored.  The results varied significantly from crystal to crystal,
with changes in the expected energy resolution ranging from 10\% to
50\% at 10~kRads.  Second, by the middle of the 1996-1997 KTeV run,
the center of the CsI had been exposed to a dose of 1~kRad.  Although
the scintillation light response along the shower has been observed to
change in roughly 100 of the 3100 crystals, no significant degradation 
of the resolution has been observed in these crystals or in the array
as a whole.  By the end of the 1997 KTeV run the dose will have
doubled, and a further evaluation of its effect will be made.

\vspace*{0.125in}
\parindent=0.25in

\begin{figure}
\centerline{\psfig{figure=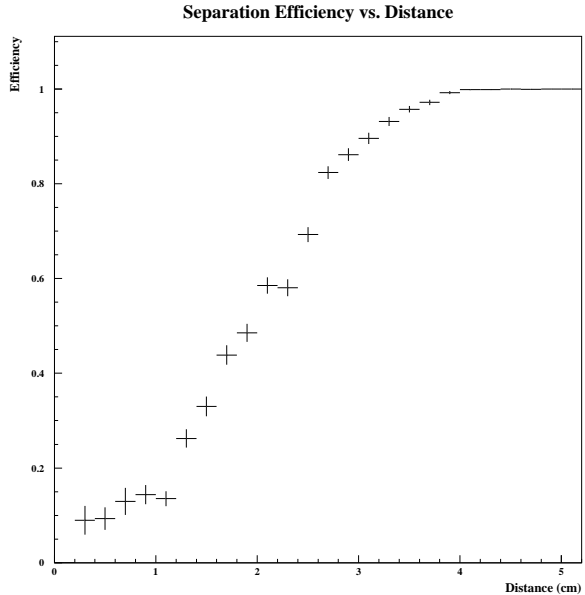,width=10cm}}
\caption{Monte Carlo of the inefficiency of separating fused 
photons vs. the distance between the two photons.}
\label{shape}
\end{figure}

One of the ways in which $\klneut$ events appear as background to
$\kpiznunu$ is for two of the photons to overlap or fuse in the calorimeter.
This background is suppressed by requiring that the photon's transverse
distribution of energy in the calorimeter be consistent in shape with typical
electro-magnetic clusters.  A shape $\chi^{2}$ is formed using the
measured position of the cluster as a lookup for the mean and rms of
the energy in each crystal in the cluster. Typically a 7x7 array is
used in the 2.5~cm crystals, and a 3x3 array is used in the 5.0~cm
crystals. The efficiency of the shape $\chi^{2}$ requirement 
from a Monte Carlo simulation of the \mbox{$\klneut$} background is shown
in Figure~\ref{shape} as a function of the distance between photons
for the case of photons in the 2.5~cm $\times$ 2.5~cm crystals.  
Nearby photons can be
distinguished 50\% of the time when they are separated by roughly the
crystal's transverse dimension.

\vspace*{0.125in}
\parindent=0.25in

Finally, the CsI calorimeter was constructed to minimize the amount of
inactive material between crystals and the size of gaps between
crystals.  The crystals are wrapped with one layer of $12.5~\mu{\rm
m}$ Aluminized-Mylar (and two overlapping layers on one face).  In
addition, the wrapping was secured with a thin layer of backless
transfer tape.  The inactive material thus comprises only 0.22\% of
the array in cross-sectional area. Also, gaps between crystals were
minimized by pushing horizontally on each individual row of the array. 
However, on lines
extending radially from the beam holes, there are trajectories on which
photons will only go through the mylar wrapping.  The probability that
a photon will lie on such a trajectory is roughly $7\times 10^{-6}$. A
more complete simulation is needed to estimate the probability that a
photon will convert in the mylar, and lead to a significant energy
deposition. In KTeV data itself, the inactive material and gaps between
crystals has been seen in an enhancement in the number of events with
low electron E/P which occur at the boundaries of
crystals. In the 2.5~cm crystals, the E/P ratio is between 0.80
and 0.95 due to crystal boundaries for $9 \times 10^{-5}$ of all
electrons.

\subsection{Photon veto detectors}

\begin{figure}
\centerline{\psfig{figure=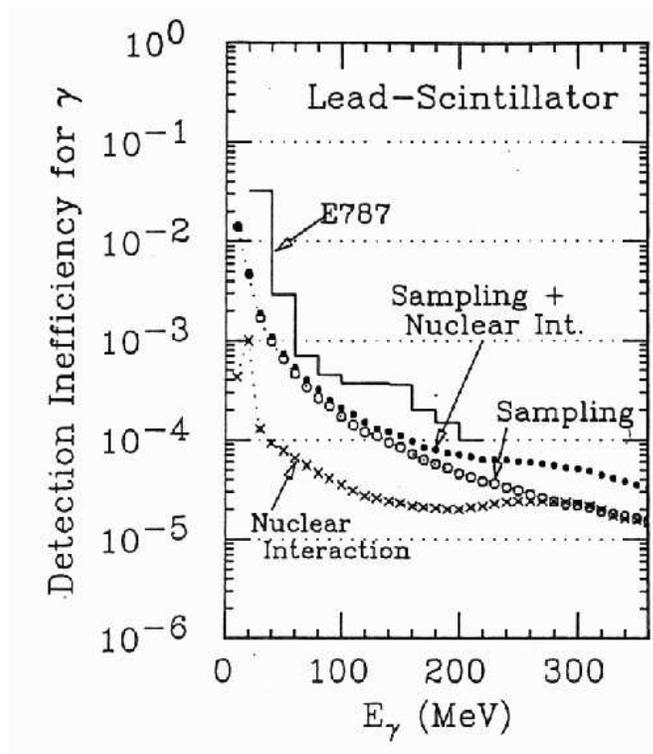,height=10cm}}
\caption{Photon detection inefficiency as a function of energy in
lead-scintillator detectors from Ref.~\cite{inagaki}.}
\label{inagaki_phot}
\end{figure}

The detection efficiency of the photon veto detectors ultimately
determines the background level for the \kpnn search.
A direct measurement of the inefficiency for a similar detector was 
performed by BNL E787 in a $K^+ \rightarrow \pi^+ \nu \overline{\nu}$
search, using $K^+ \rightarrow \pi^+ \pi^0$.
The results of this study are shown in Fig.~\ref{inagaki_phot}.
A group at KEK has done detailed
simulations of photon veto inefficiencies for their proposal to measure
\kpnn~\cite{inagaki}. 
Their result is also plotted in the same figure.

Photon detection efficiency is, of course, a strong function of energy
with the lower energies being more problematic.
Inefficiencies at low energy ($<$ 30 MeV) are dominated by sampling effects 
where a 
significant fraction of the energy is absorbed in the inactive material.
Inefficiencies at higher energies are dominated by the small probability
of photonuclear interactions.

At KAMI, the energy of a typical photon is higher than 1 GeV
resulting in inefficiency expectations of less than $10^{-5}$.
If two missing photons 
from the $K_L \rightarrow 2\pi^0$ background process
both have energies above 1 GeV, the resulting background would be less than
$10^{-10}$.  Unfortunately, this is not always the case as the energies of the
two missing photons can be distributed quite asymmetrically.  This is shown
in Figure~\ref{phv_2pi0} in Section~\ref{det_geo}.

\vspace*{0.125in}
\parindent=0.25in

A detailed study, described in Section~\ref{background_study}, indicates that
the most serious background events result when the lower-energy
missing photon carries less than 20~MeV. 
To estimate the inefficiency for such photons,
a GEANT simulation was performed, assuming the detector geometry
described in Section 4.3.2.
An initial study was done with a 17 $X_0$ deep sampling calorimeter 
consisting of either 0.5~mm,  
1~mm or 2~mm thick lead sheets with 5~mm thick scintillator plates.
  
\begin{figure}
\centerline{
        \begin{turn}{-90}
        \psfig{figure=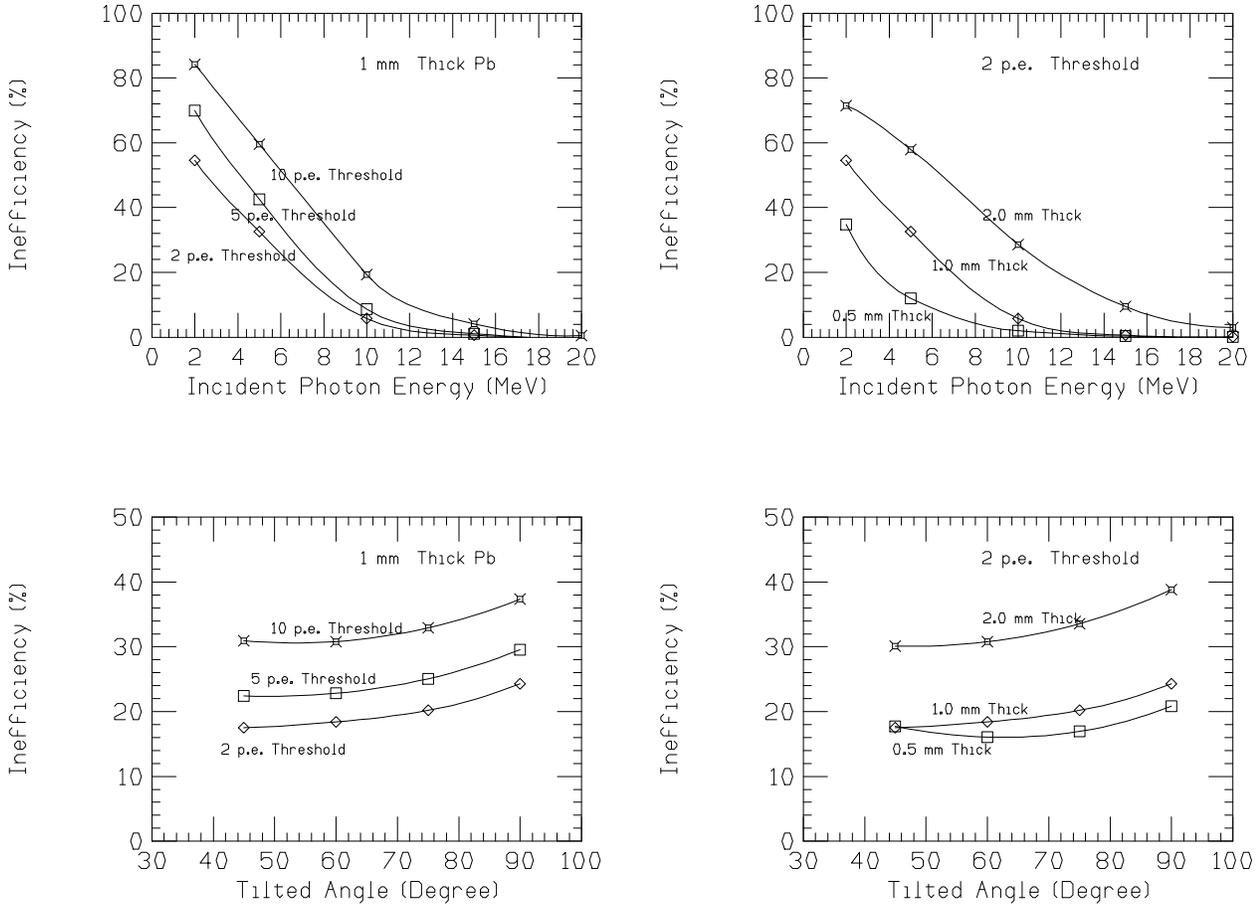,width=12cm}
        \end{turn}
}
\caption{Photon detection inefficiency 
in the vacuum veto counters.
The top two plots 
show the inefficiency as a function of photon energy,
for various photo electron thresholds and lead sheet thicknesses.
The bottom two plots show the average inefficiency for photons
below 20~MeV as a function of the lead sheet tilt angle.} 
\label{ineff_tilt}
\end{figure}

The top two plots in Fig.~\ref{ineff_tilt}
show the inefficiency as a function of photon energy,
for various photo electron thresholds and lead sheet thickness.
Below 10 MeV, detection of photons becomes increasingly difficult.

In reality, low energy photons tend to have larger opening angles
which increases the sampling inefficiency,
unless the lead sheets are tilted to compensate. 
To address this, a more detailed simulation was 
performed based on the real four vectors of the two photons from 
background events which miss the CsI.
The inefficiency was measured as a function of the tilt angle
of the lead sheets relative to the beam direction (z-axis).
As is shown in the bottom two plots in
Fig.~\ref{ineff_tilt}, the inefficiency is a weak function
of the tilt angle. A 45 degree tilt angle gives the best
result, as expected.
With a 2-3 p.e. threshold, a 20\% inefficiency appears to be
feasible, even with 1~mm lead sheets.

\vspace*{0.125in}
\parindent=0.25in

In order to better understand the inefficiency at higher energies, we plan 
to perform a detailed simulation
of photonuclear effects to determine the ultimate limitations 
of the photon veto detectors.
Due to the large nuclear interaction length of the photon veto detector 
for shallow angle photons, we expect to achieve an inefficiency of better than
$3\times10^{-6}$ for photons $>$1~GeV.

\subsection{Scintillating fiber tracker}

Thanks to the tremendous effort by the D0 collaboration and others at 
Fermilab, 
scintillating fiber tracking is becoming a mature technology.
The D0 fiber tracker consists of 2~m long, 830~$\mu$m diameter scintillating
fibers spliced to 5~m long, 1~mm diameter clear fibers~\cite{ruchti}.  
The light emerging
from the clear fibers is viewed by VLPC detectors.
The D0 prototype detector produces 9~p.e. per MIP with the expectation of 
12~p.e.
per MIP with their final detector. 
Position resolution of 92~$\mu$m has been achieved using double
layers of fibers. 

We plan to use the same technology for the KAMI changed particle spectrometer.
As mentioned in Section 4.4.1, 
we are considering reading out 
both ends of the fibers to obtain better time resolution
and triggering capabilities.  
Our plan is to use smaller diameter fibers than the D0 fiber tracker.
A double layer of 500~$\mu$m fibers will result in 60~$\mu$m position 
resolution, better than that currently achieved by KTeV using drift chambers.
The clear fiber light guide used by KAMI would be shorter than the one 
used by D0,
resulting is less attenuation.  Thus, even with thinner fibers, we expect 
to obtain more than 5~p.e. per MIP.
A detailed study of the momentum resolution is currently underway.

\begin{figure}
\centerline{
        \begin{turn}{-90}
        \psfig{figure=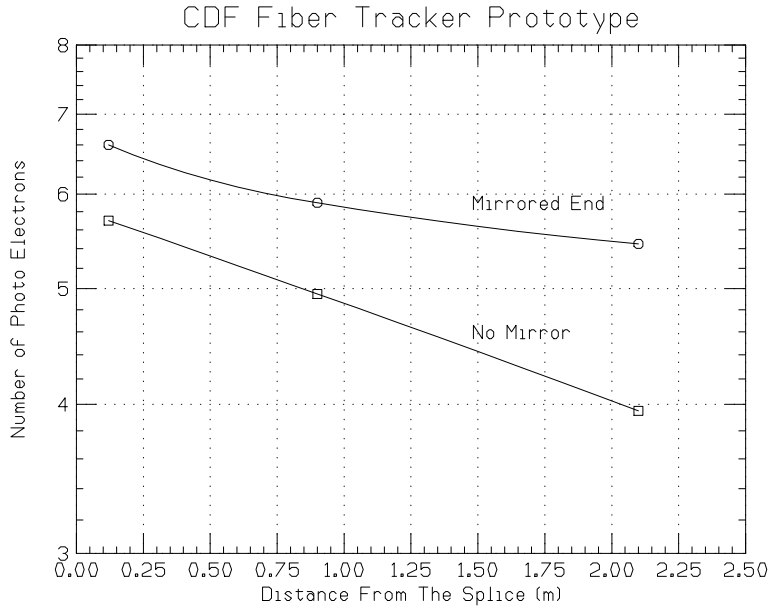,width=8cm}
        \end{turn}
}
\caption{The light yield obtained by CDF from a prototype fiber tracker with
500 $\mu$m diameter fibers~\cite{atac}.}
\label{cdf_fiber}
\end{figure}

CDF has also developed a prototype fiber tracker~\cite{atac}.  
Using 500 $\mu$m diameter fibers, they have obtained more
than 4 p.e. per MIP, as shown in 
Fig.~\ref{cdf_fiber}.

\section{\kpnn Background Study}
\label{background_study}

The kaon flux at KAMI will reach levels which should 
result in the observation of approximately 
100 \kpnn decays, as shown in Section~\ref{sens}.
In order to optimize KAMI's
detector geometry and to understand the detector performance required 
to suppress backgrounds at such a high sensitivity level,
we have performed detailed Monte Carlo simulations. 
This section describes the results of these studies.

\vspace*{0.15in}
\parindent=0.25in

It is first necessary to restrict the transverse size of the beam 
for the \kpnn decay 
search for two reasons:

\begin{enumerate}

\item To obtain good $P_t$ resolution for the reconstructed $\pi^0$; and

\item To minimize the geometrical acceptance for photons
   which pass through the beam hole in the CsI calorimeter.
   The photons in the beam hole must be vetoed by the Back Anti
   detector which is exposed to a high flux of neutrons and kaons.

\end{enumerate}

We have optimized the beam sizes for the two different target positions
shown in Table~\ref{tab:kpnn_sens}.  A beam size of 0.36 $\mu$str is
optimum for KAMI-Far, and 1 $\mu$str is optimum for KAMI-Near.
With these beam solid angles, we can expect 30 and 124
signal events per year for KAMI-Far and KAMI-Near, respectively, as
listed in Table~\ref{tab:kpnn_sens}.
These event rates are adequate in order to extract a meaningful 
value for $\eta$.

\begin{figure}
\centerline{\psfig{figure=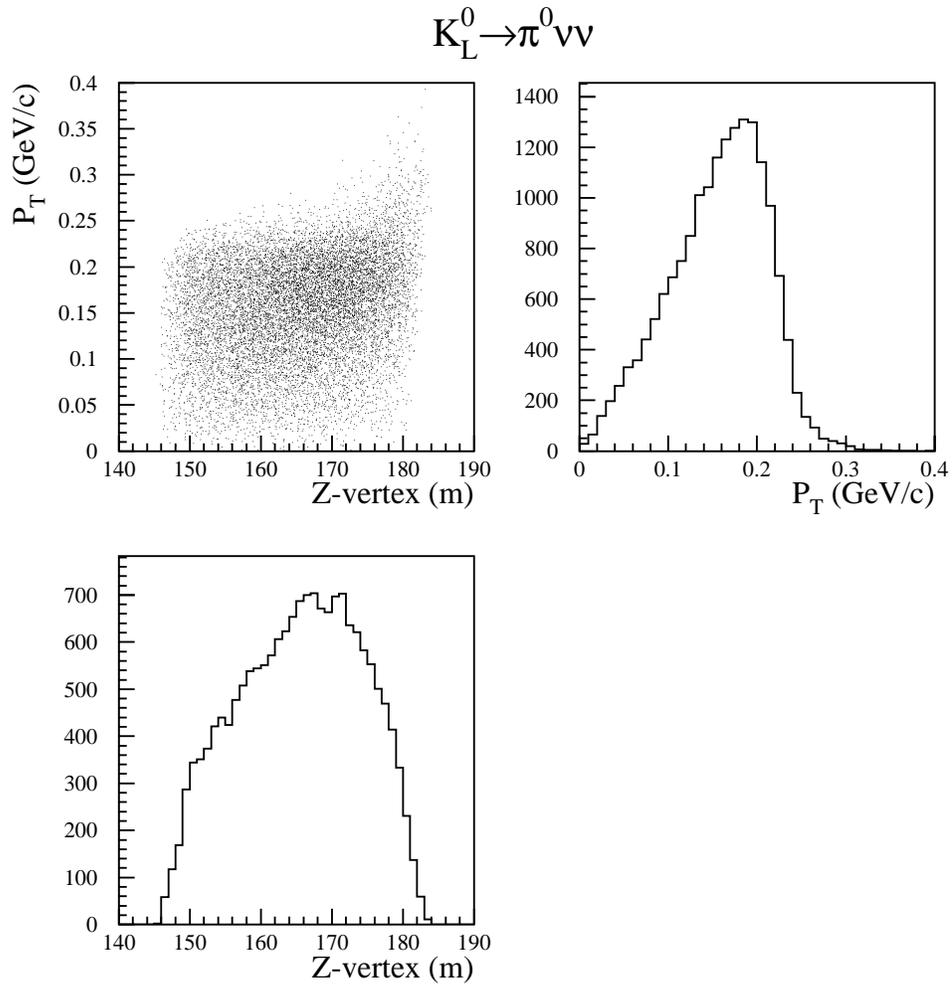,width=14cm}}
\caption{The z vertex and $P_t$ distributions for $\pi^0 \nu \bar{\nu}$ 
events from a Monte Carlo simulation.}
\label{zpt_pi0nn}
\end{figure}

Once the beam size is determined, 
the kinematical distribution of signal events may be studied. 
The $P_t$ distribution and  
vertex $z$ distribution are shown in Fig~\ref{zpt_pi0nn}.
 
\vspace*{0.125in}
\parindent=0.25in

The background level is a strong function of
photon detection inefficiency.
Understanding this relationship is the main focus of this chapter.
There are several possible sources of background events.
From kaon decays, $K_L \rightarrow 3\pi^0$, $2\pi^0$ and $\gamma \gamma$ 
are the major concerns.
From Hyperon decays, $\Lambda \rightarrow \pi^0$n, and 
$\Xi \rightarrow \Lambda \pi^0$
(with sequential $\Lambda \rightarrow \pi^0$n decays) are the major
contributors.  In addition, neutron interactions with residual gas
or with detector elements may produce $\pi^0$s with high $P_t$.

So far, most of our studies have been focused on the study of 
$K_L \rightarrow 2\pi^0$
backgrounds, since this will be the ultimate physical background
source which can produce a single $\pi^0$ with high $P_t$ 
in the CsI calorimeter.

\subsection{$K_L \rightarrow 2\pi^0$ background estimate}
 
One of the difficulties of studying backgrounds in rare decay experiments is
the excessive amount of CPU time which is required to generate a sufficient 
number of background events.  This problem can be solved
by artificially degrading the photon veto efficiency, as described below.

In order for $K_L \rightarrow 2\pi^0$ to become a background event, two 
photons must be detected by the CsI and the other two photons must
go undetected.  There are many different ways for photons to escape
detection.  The photons can go undetected either 
in the vacuum veto counters,
the Back Anti or the CsI.  Additionally, two photons can fuse in the CsI and
be detected as a single cluster.  

If we artificially reduce the efficiency of the various detectors, 
the probability of missing two photons increases, thus, we gain statistics.
As long as we keep a detailed record of why each photon is missed,
we can impose better detector efficiences analytically as additional
offline cuts without losing statistics.

\subsubsection{Kaon beam generation}

The neutral kaon beam for these studies is generated using the standard 
KTeV/KAMI
Monte Carlo simulation software, assuming the following conditions:

\begin{enumerate}
    
\item	For KAMI-Far, the target is located at z=0~m, while
   for KAMI-Near, the target is located at z=120~m.

\item	The targeting angle is set to 24~mrad to reduce the neutron flux.

\item	For KAMI-Far, the beam solid angle is
	0.6~mrad~$\times$~0.6~mrad~= 0.36~$\mu$str.
   	For KAMI-Near, the beam solid angle is 
	1~mrad~$\times$~1 mrad = 1$\mu$str.

\item	Kaon decays are generated in the 34 m long fiducial region from
z=152 - 186 m
   (from the position of the Mask Anti to the CsI calorimeter).

\item The kaon momentum is generated in the range of 5-120 GeV
using the Malensek parameterization~\cite{malensek}.

\end{enumerate}

\subsubsection{Detector geometry}
\label{det_geo}

\begin{figure}[htb]
\centerline{
        \begin{turn}{-90}
        \psfig{figure=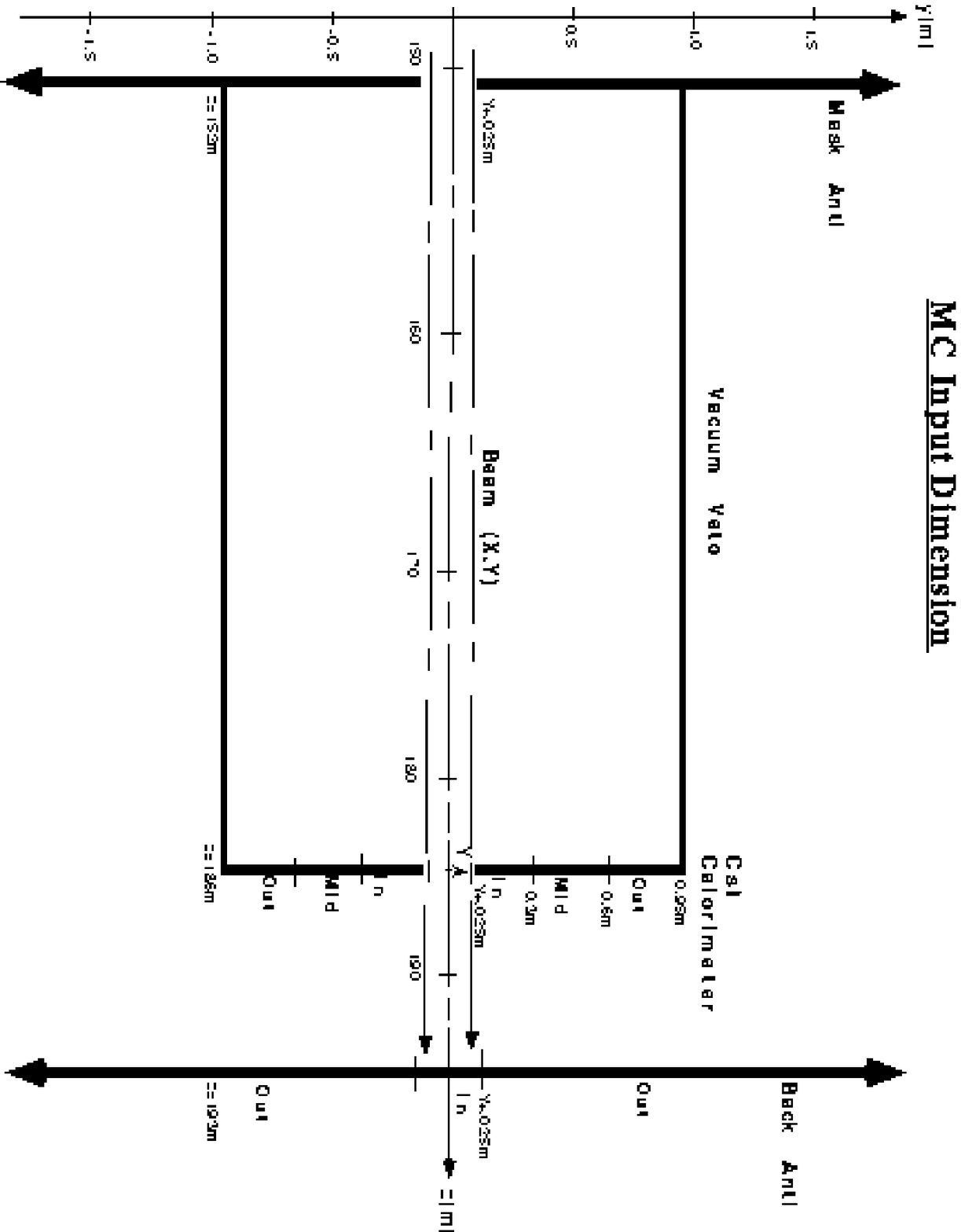,width=10cm}
        \end{turn}
}
\caption{The detector geometry used for the Monte Carlo simulation
background study.}
\label{mcinput}
\end{figure}

For simplicity, we have modeled the detector geometry as shown in
Figure.~\ref{mcinput}.  The relevant details of the detector model are:

\begin{enumerate}

\item At z = 152 m, there is one Mask Anti(MA) which is flat and
   infinitely wide in the x and y directions. 
   Later, this section will be replaced by an additional MA
   with a box geometry to make it finite in size.

\item The Vacuum Photon Veto (PV) is 
   1.9~m~(x)~$\times$~1.9~m~(y)~$\times$~34~m~(z) and has a simple 
   box shape. It is located between
   the MA and the CsI calorimeter.

\item There is no charged spectrometer included in the simulation. 
   The Charged Veto is not considered in this study either,
    since we are simulating only photons.

\item The CsI is divided into three regions;  CsI-In, CsI-Mid and CsI-Out.
   CsI-In is the central region of 
$\pm$30~cm~$\times$~$\pm$30~cm where the efficiency
   is low for low-energy photons due to accidental activity. 
   CsI-Mid is the middle region of the calorimeter, bounded by 
   $\pm$60~cm~$\times$~$\pm$60~cm.
   Both CsI-In and CsI-Mid consist only of small (2.5~cm $\times$ 2.5~cm) 
   crystals.  CsI-Out is the outer-most part of the calorimeter, and consists 
   only of large (5~cm $\times$ 5~cm) crystals.
   Both CsI-Mid and CsI-Out are expected to have high efficiency, even for 
   low energy photons.
   CsI-Out is expected to have poorer rejection power for fused photons
   than CsI-In and CsI-Mid because of the larger transverse size of the 
   crystals.    

\item The Back Anti (BA) is located at z=193~m and 
   divided into two regions; BA-In (inside of the
   the neutral beam) and BA-Out (outside of the beam).
   The x/y dimentions of
   BA-In is 2.5~cm larger than the beam size (at z=186~m) in both x and y. 
   This region is exposed to a high neutron flux and is 
   expected to have poor photon detection efficiency.
   BA-Out is infinitely large and is expected to have a high efficiency
   for detecting photons.

\end{enumerate}

The exact dimensions of the various detector elements used in the simulation
are listed in Table~\ref{tab:mcdim}.  
The beam size is defined as $\pm$X and $\pm$Y at Z=186~m.
 
\begin{table}[here,top,bottom]
\begin{center}
\begin{tabular}{|l|r|r|r|r|r|r|}
\hline

         &Material        
&Z (m)  &X$_{in}$ (m) &X$_{out}$ (m) &Y$_{in}$ (m) &Y$_{out}$ (m) \\
\hline\hline	
MA      &Pb/Scint.       &152     &X+.025  &Inf.    &Y+.025  &Inf.	\\
PV      &Pb/Scint.       &152-186 &0.95    &-       &0.95    &-       	\\
	&		 &	  &        &	    &	     &		\\
CsI-In  &CsI(2.5 cm$^2$)   &186     &X+.05   &0.30    &Y+.05   &0.30	\\
CsI-Mid &CsI(2.5 cm$^2$)   &186     &0.30    &0.60    &0.30    &0.60	\\
CsI-Out &CsI(5.0 cm$^2$)   &186     &0.60    &0.95    &0.60    &0.95	\\
	&		 &	  &	   &	    &	     &		\\
BA-In   &Pb/Quartz       &193     &0       &X+0.025 &0       &Y+0.025 	\\
BA-Out  &Pb/Scint.       &193     &X+0.025 &Inf.    &Y+0.025 &Inf.	\\
\hline
\end{tabular}
\caption{Detector dimensions used in a Monte Carlo simulation of the $2\pi^0$
background.
The beam size is defined as $\pm$X and $\pm$Y at Z=186~m.  All dimensions
are in meters.}
\label{tab:mcdim}
\end{center}
\end{table}

It is important to categorize the reasons why photons go undetected
in as much detail as possible.
In most cases, the photons will be lost due to the inefficiency of the 
detector.
However, in case of the CsI, photons can be lost due to fusion as well.
The way in which photons may be missed can be broken down into the following
seven categories:
\vspace*{0.125in}
\parindent=0.in

00) Inefficiency in BA-In			\hfill  [BAI]

10) Inefficiency in BA-Out   			\hfill	[BAO]

20) Inefficiency in PV      			\hfill	[PV]

30) Fusion in CsI-In/Mid (small crystals) 	\hfill	[CIf]

40) Fusion in CsI-Out  (Large crystals)   	\hfill	[COf]

50) Inefficiency in CsI-In (central $\pm$30 cm region)  \hfill [CI]	

60) Inefficiency in CsI-Mid/Out (outer region).	\hfill	[CO]
\vspace*{0.125in}

In the case of loss due to inefficiency, each category can be further
divided 
into the following ten energy regions:
\vspace*{0.125in}
\parindent=0.in

	0) 0.00 - 0.02 GeV	

        1) 0.02 - 0.04 GeV 

        2) 0.04 - 0.06 GeV

        3) 0.06 - 0.10 GeV

        4) 0.10 - 0.20 GeV

        5) 0.20 - 0.40 GeV

        6) 0.40 - 1.00 GeV

        7) 1.00 - 3.00 GeV

        8) 3.00 - 10.0 GeV

        9) 10.0 - infinite.

\vspace*{0.125in}

In the case of loss by fusion ([CIf] [COf], or ID = 30, 40), each category
is further divided into the following ten distance categories:
 
\vspace*{0.125in}
\parindent=0.in 

        0) 0   - 2.5 cm

        1) 2.5 - 5.0 cm

        2) 5.0 - 7.5 cm

        3) 7.5 -10.0 cm

        4) 10.0 -12.5 cm

        5) 12.5 -15.0 cm

        6) 15.0 -17.5 cm

        7) 17.5 -20.0 cm

        8) 20.0 -22.5 cm

        9) 22.5 -infinite.
\vspace*{0.125in}
\parindent=0.25in

By combining all of the categories listed above, there are a total of 70
possible reasons for losing photons.
This means that for two photons there are 70 x 70 = 4900 possible reasons.
Photons are not equally distributed throughout these 4900 bins.  Studies
indicate that there are several key parameters which contribute most to
the background. These are:

\begin{enumerate}  

\item Vacuum veto inefficiency 
   for very low energy photons ($<$ 20 MeV);
\item Vacuum veto inefficiency 
   for high energy photons (1-3 GeV);
\item Inefficiency in the CsI for high energy photons (3-10 GeV);
\item Inefficiency in the BA for very high energy photons 
   ( $>$ 10 GeV);
\item Inefficiency due to fusions in the small CsI crystals.
 \end{enumerate}
\parindent=0.25in
 
\begin{figure}
\centerline{\psfig{figure=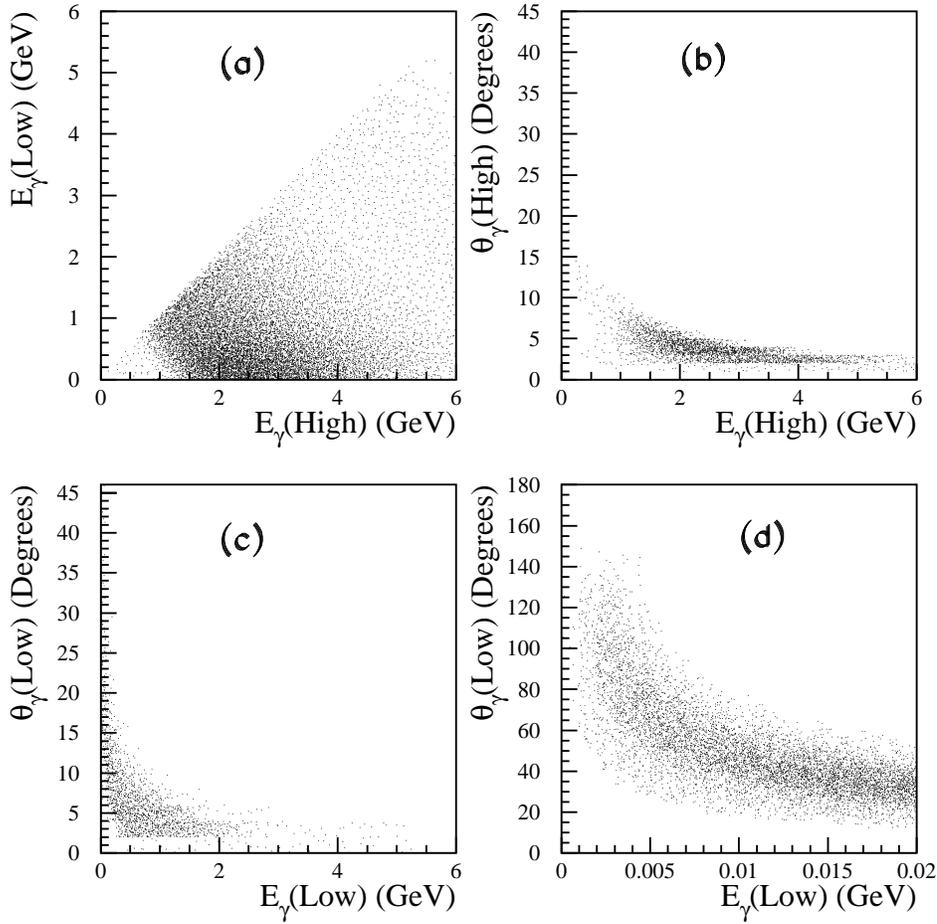,width=14cm}}
\caption{The energy and angle between two missed photons 
in \mbox{$K_L \rightarrow 2\pi^0$}decays.
(a) is a scatter plot of the two photon energies;
(b) shows the correlation of opening angle vs. energy
for the higher energy photon;
(c) shows the correlation of opening angle vs. energy
for the lower energy photon;
(d) is the same as plot (c),
except the photon energy
is restricted to below 20 MeV.}
\label{phv_2pi0}
\end{figure}

When two photons are missed, one photon tends to
have a low energy and a large opening angle, while the other photon has a high
energy and a small opening angle.
Figure~\ref{phv_2pi0} shows the correlation between two missed photons.
Figure~\ref{phv_2pi0}-a is a scatter plot of the two photon energies.
Figure~\ref{phv_2pi0}-b shows the correlation of opening angle vs. energy
for the higher energy photons.
Figure~\ref{phv_2pi0}-c shows the correlation of opening angle vs. energy
for the lower energy photons.
Figure~\ref{phv_2pi0}-d is the same as plot Figure~\ref{phv_2pi0}-c, 
except the photon energy
is restricted to below 20 MeV.

In order to derive the correlations between detector parameters,
we start with completely inefficient detectors, except for the CsI.
We assume that the CsI has a 1\% inefficiency for photons above 1~GeV.
For energies less than 1~GeV in the CsI, and for photons in all other 
detectors, (i.e. MA, PV or BA), 100\% inefficiency is assumed.
Table \ref{tab:ineff} shows the initial inefficiency for each detector
for each energy bin.

\begin{table}[here,top,bottom]
\begin{center}
\begin{tabular}{|c|c|r|r|}
\hline
ID      &Energy     &MA/PV/BA   &CsI		\\
       &Low-High (GeV) &Inefficiency  & Inefficiency		\\
\hline\hline
 0      &0.0   -  0.02    &1       &1	\\
 1      &0.02  -  0.04    &1       &1 	\\
 2      &0.04  -  0.06    &1       &1	\\
 3      &0.06  -  0.10    &1       &1	\\
 4      &0.10  -  0.20    &1       &1	\\
 5      &0.20  -  0.40    &1       &1	\\
 6      &0.40  -  1.00    &1       &1	\\
 7      &1.00  -  3.00    &1       &$1\times 10^{-2}$ \\
 8      &3.00  -  10.0    &1       &$1 \times 10^{-2}$ \\
 9      &10.0  -  Inf.    &1       &$1 \times 10^{-2}$ \\
\hline
\end{tabular}
\caption{Starting point for detector inefficiencies, binned by energy,
for studying the \mbox{$K_L \rightarrow 2\pi^0$} background.  The 
inefficiency for all detectors has been increased in order to obtain sufficient
statistics, as described in the text.}
\label{tab:ineff}
\end{center}
\end{table}

For fusions, we assume 100\% detection efficiency if the distance between
the two photons
is larger than the twice of crystal size.
Otherwise, we set the efficiency to 0.  The fusion inefficiencies are listed
in Table~\ref{tab:fuse}.

\begin{table}[here,top,bottom]
\begin{center}
\begin{tabular}{|r|r|r|r|}
\hline
ID      &Distance   &CsI-In/Mid  &CsI-Out	\\
        &Low-High (cm)   &Inefficiency &Inefficiency    	\\
\hline\hline
 0      &0.0  -  2.5   &1       &1       	\\
 1      &2.5  -  5.0   &1       &1		\\
 2      &5.0  - 7.5    &0       &1  		\\
 3      &7.5  - 10.    &0       &1		\\
 4-9    &10.  - Inf.   &0       &0		\\
\hline
\end{tabular}
\caption{The fusion separation inefficiencies in the CsI used for
studying the $K_L \rightarrow 2\pi^0$ background, binned according
to the distance which separates the two photons.  The inefficiency
has been increased in order to obtain sufficient statistics, as described
in the text.} 
\label{tab:fuse}
\end{center}
\end{table}
 
\parindent=0.in
These artificially high inefficiencies ensure  
high statistics for each type of background.
\parindent=0.25in

\subsubsection{Offline cuts}

After events are generated by the Monte Carlo, the following offline 
cuts are applied
prior to tightening the photon veto efficiency cuts:

\begin{enumerate}

\item Only two clusters in the CsI, each with E $>$ 1 GeV;
\item 5 GeV $<$ Total Energy in CsI $<$ 20 GeV;
\item $P_t$ $>$ 150 MeV/c;
\item No hits in the MA/PV/BA; and
\item A reconstructed z vertex between 152 m and 171 m.

\end{enumerate}

These cuts were carefully chosen from our early studies to reject various 
types of backgrounds ($\Lambda \rightarrow \pi^0$n, 
$K_L \rightarrow 3\pi^0$, etc.).
Under these conditions, the acceptance for the signal is 7.1\%
for KAMI-Far, and 7.4\% for KAMI-Near.

\subsubsection{Background estimation}

In a standard Monte Carlo job, one can generate an order of 
$10^{8}$ $K_L \rightarrow 2\pi^0$ events.
If one category of missed photon pairs, say PV(0-0.02 GeV) and 
BA($>$10GeV)
receives N events, the expected number of background events of this
type at signal sensitivity of $3\times10^{-11}$ 
(i.e. the Standard Model signal level)
can be estimated by the formula below, where an acceptance of
7\% is assumed for the signal.  First, the Single Event Sensitivity (SES) is
calculated: 
\vspace*{0.125in}
\parindent=0.5in

SES = Br($K_L\rightarrow 2\pi^0$) / (No. of generated events) / 
Acceptance(\pnn) $\times$ N

        = 9$\times10^{-4}$ / $10^8$ / 0.07 $\times$ N 

        = 1.3$\times10^{-11}\times$ N.	

\parindent=0.in
\vspace*{0.125in}

The number of Background events, $(NB)$, which result when 
a signal sensitivity of $3\times10^{-11}$ is achieved is given by 
\vspace*{0.125in}
\parindent=0.5in

        $NB$ = SES / 3$\times10^{-11}$ = 0.43(N $\pm \sqrt{N}$).	

\vspace*{0.125in}
\parindent=0.25in
In principle, we can calculate 4900 such numbers for all of the different
categories simultaneously.
This will form a matrix $NB(i,j)$ of background events,
where $i$ and $j$ both run over the 70 different categories of lost
photons.
 
It is highly desirable to have more than 100 events in each category
in order to minimize the statistical error.
This is very important for the case where the inefficiency 
can not be reduced by more than an order-of-magnitude from the initial
value listed in Table~\ref{tab:ineff}.

By looking at this matrix of 4900 entries, one can 
determine how to impose additional efficiency requirements
for each detector (i.e. smaller inefficiency).  One can also sum up each
column to determine the overall background contribution
from each of the 70 categories; 

\begin{displaymath}
NB_{sum}(i) = \sum_{j=1}^{70} NB(i,j).
\end{displaymath}

\parindent=0.in
If one imposes the additional reduction of inefficiency, Ineff$(i)$, 
for the $i$'th type of background, then
$NB(i,j)$ becomes $NB(i,j) \times$ Ineff$(i)$ and
$NB_{sum}(i)$ becomes $NB_{sum}(i) \times$ Ineff$(i)$.
\vspace*{0.125in}

The final goal is to determine the 70
parameters, Ineff$(i)$, which satisfy the condition 
 
\begin{equation} 
NB_{total} = \sum_{i=1}^{70} \sum_{j=i}^{70} NB(i,j) \times {\rm Ineff}(i) 
\times {\rm Ineff}(j) <1.
\end{equation}

One way to determine the optimum value of the parameter 
Ineff$(i)$ is to reduce it
until $NB_{sum}(i)$ becomes significantly smaller than one,
say less than 0.1.
After executing this procedure for all 70 categories, if one ends up with ten
categories each with 0.1 events, the total number of background events
would be on the order of 1.

\vspace*{0.125in}
\parindent=0.25in

By appling this method iteratively, we have derived 
the values in Table~\ref{tab:inef_far} as the default
veto efficiencies for various detectors.
This table is consistent with 
the measurements by BNL E787 and Inagaki's study~\cite{inagaki} for the
energy region between 20~MeV and 400~MeV.
To be conservative, the actual inefficiencies in this table were set 
higher than their numbers.
 
Below 20~MeV, a 20\% inefficiency is assigned to the vacuum veto system,
based on the result of the GEANT simulation described in 
Section 5.2.
Above 1 GeV, inefficiency values were optimized to achieve our
background reduction goals.

\begin{table}[here,top,bottom]
\begin{center}
\begin{tabular}{|l|r|r|r|r|r|}
\hline
ID      &Energy    &MA/PV   &CsI-Mid/Out &CsI-In/BA-Out  &BA-In	\\
        &Low-High (GeV) & Inefficiency & Inefficiency &Inefficiency  
& Inefficiency		\\
\hline\hline
 0      &0.0   - 0.02    &$2\times 10^{-1}$    &1       &1       &1\\
 1      &0.02  - 0.04    &$3\times 10^{-2}$    &$1\times 10^{-1}$    &1
&1		\\
 2      &0.04  - 0.06    &$3\times 10^{-3}$    &$1\times 10^{-2}$    &1
&1		\\
 3      &0.06  - 0.10    &$7 \times 10^{-4}$    &$1\times 10^{-3}$    &$1
\times 10^{-2}$    &1		\\
 4      &0.10  - 0.20    &$4\times 10^{-4}$    &$4\times 10^{-4}$
&$1\times 10^{-3}$    &1		\\
 5      &0.20  - 0.40    &$1\times 10^{-4}$    &$1\times 10^{-4}$
&$1\times 10^{-4}$    &1		\\
 6      &0.40  - 1.00    &$3\times 10^{-5}$    &$3\times 10^{-5}$
&$3\times 10^{-5}$    &1		\\
 7      &1.00  - 3.00    &$3\times 10^{-6}$    &$3\times 10^{-6}$
&$3\times 10^{-6}$    &$1\times 10^{-1}$	\\
 8      &3.00  - 10.0    &$1\times 10^{-6}$    &$1\times 10^{-6}$
&$1\times 10^{-6}$    &$1\times 10^{-2}$	\\
 9      &10.0  - Inf.    &$1\times 10^{-6}$    &$1\times 10^{-6}$
&$1\times 10^{-6}$    &$1\times 10^{-3}$	\\
\hline
\end{tabular}

\caption{Desired detector inefficiencies obtained from
background rejection studies
of \mbox{$K_L\rightarrow 2\pi^0$} decays.}
\label{tab:inef_far}
\end{center}
\end{table}

In addition, the inefficiencies in Table~\ref{tab:fuse_far} 
for separating fused photons 
were applied, based on our study with the existing KTeV CsI calorimeter, 
described in Section 5.1.

\begin{table}[here,top,bottom]
\begin{center}
\begin{tabular}{|l|r|r|r|}
\hline
ID      &Distance     &CsI-In/Mid  &CsI-Out	\\
        &Low-High (cm)  &Inefficiency & Inefficiency        \\
\hline\hline
 0      &0.0 -  2.5   &1.0     &1.0       \\
 1      &2.5 -  5.0   &0.1     &1.0		\\
 2      &5.0 -  7.5   &0       &0.1   	\\
 3      &7.5 -  10.   &0       &0.1    	\\
 4-9    &10. -  Inf.  &0       &0 		\\
\hline
\end{tabular}
\caption{Desired inefficiency for separating fused clusters in the CsI as a 
function of the separation of the two photons.  These numbers resulted from a 
background study
of $K_L \rightarrow 2\pi^0$ decays.}
\label{tab:fuse_far}
\end{center}
\end{table}

\parindent=0.25in
\subsubsection{Results for KAMI-Far option}

We have studied the KAMI-Far option
using the inefficiencies in Tables~\ref{tab:inef_far} and \ref{tab:fuse_far}.
After imposing these inefficiencies, Table~\ref{tab:sig_noise} 
was generated 
for the categories which contribute more than 0.01 events to the background 
level, calculated for a signal sensitivity of $3\times10^{-11}$ 
(i.e. Standard Model signal level).
Since all of the background numbers are normalized in this way, 
one can consider the table entries to be the expected Noise to Signal level.

\begin{table}[here,top,bottom]
\begin{center}
\begin{tabular}{|c|c|c|l|c|l|c|}
\hline

  ID1 &ID2 & \#background &Type1(Bin)& Default  &Type2(Bin)& Default \\
\hline\hline
20 &68  &0.089$\pm$0.002 &PV ($<$0.02)  &0.2   &COi(3-10)  &$1\times 10^{-6}$ \\
20 &27  &0.079$\pm$0.000 &PV ($<$0.02)  &0.2   &PV (1-3)   &$3\times 10^{-6}$ \\
19 &20  &0.067$\pm$0.005 &BA (10$<$)    &0.001 &PV ($<$0.02) &0.2 \\
19 &30  &0.061$\pm$0.007 &BA (10$<$)    &0.001 &CIf($<$2.5)&1.0 \\
20 &67  &0.040$\pm$0.002 &PV ($<$0.02)  &0.2   &COi(1-3)   &$3\times 10^{-6}$ \\
17 &68  &0.025$\pm$0.001 &BA (1-3)      &0.1   &COi(3-10)  &$1\times 10^{-6}$ \\
20 &28  &0.020$\pm$0.000 &PV ($<$0.02)  &0.2   &PV (3-10)  &$1\times 10^{-6}$ \\
16 &26  &0.019$\pm$0.001 &BA (0.4-1)    &1.0   &PV (0.4-1) &$3\times 10^{-5}$ \\
19 &31  &0.017$\pm$0.002 &BA (10$<$)    &0.001 &CIf(2.5-5) &0.1 \\
21 &68  &0.014$\pm$0.000 &PV (.02-.04)  &0.03  &COi(3-10)  &$1\times 10^{-6}$ \\
17 &27  &0.013$\pm$0.000 &BA (1-3)      &0.1   &PV (1-3)   &$3\times 10^{-6}$ \\
21 &27  &0.013$\pm$0.000 &PV (.02-.04)  &0.03  &PV (1-3)   &$3\times 10^{-6}$ \\
16 &27  &0.012$\pm$0.000 &BA (0.4-1)    &1.0   &PV (.02-.04)&$0.03$ \\
19 &21  &0.011$\pm$0.001 &BA (10$<$)    &0.001 &PV (.02-.04)&$0.03$ \\
\hline\hline 
\multicolumn{7}{|c|}{Total of 0.57 background events / SM signal ($3\times
10^{-11}$)}
\\
\hline
\end{tabular}
\caption{Categories of events from $2\pi^0$ decays with 2 missing photons
which contribute more than 0.01 events to the background level for
KAMI-Far.  The inefficiencies in 
Tables~\ref{tab:inef_far} and \ref{tab:fuse_far} were used to generate the
backgrounds.}
\label{tab:sig_noise}
\end{center}
\end{table}

\vspace*{0.125in}
\parindent=0.25in

In order to further reduce the background level, tighter kinematical
cuts were studied.
Signal events have a maximum $P_t$ of 231 MeV/c, while $2\pi^0$ backgrounds
have a maximum $P_t$ of 209 MeV/c.
By taking advantage of our good $P_t$ resolution, one can consider a $P_t$ cut
at 215 MeV/c. 
This cut was made before the additional photon veto inefficiency 
table was imposed.
Signal acceptance was reduced from 7.1\% to 1.2\% by this $P_t$ cut.
The level of background which results is shown in Table~\ref{tab:sig_noise1}.

\begin{table}[here,top,bottom]
\begin{center}
\begin{tabular}{|c|c|c|l|c|l|c|}
\hline

ID1&ID2 & \#background &Type1(Bin) &Default  &Type2(Bin)  &Default\\
\hline\hline
20 &68 &0.103$\pm$0.023  &PV ($<$0.02) &0.2   &COi(3-10)   &$1\times 10^{-6}$\\
17 &68 &0.052$\pm$0.012  &BA (1-3)     &0.1   &COi(3-10)   &$1\times 10^{-6}$\\
20 &67 &0.031$\pm$0.022  &PV ($<$0.02) &0.2   &COi(1-3)    &$3\times 10^{-6}$\\
21 &68 &0.017$\pm$0.004  &PV (.02-.04) & $3\times10^{-2}$  
&COi(3-10)   &$1\times 10^{-6}$\\
20 &27 &0.013$\pm$0.001  &PV ($<$0.02) &0.2   &PV (1-3)    &$3\times 10^{-6}$\\
\hline\hline
\multicolumn{7}{|c|}{Total of 0.30 background events / SM signal ($3\times
10^{-11}$)}
\\

\hline
\end{tabular}
\caption{Categories of events from $2\pi^0$ decays with 2 missing photons
which contribute more than 0.01 events to the background level 
for KAMI-Far.  A $P_t$ cut at 215~MeV/c was imposed.}
\label{tab:sig_noise1}
\end{center}
\end{table}

\vspace*{0.125in}
\parindent=0.25in

A large fraction of the remaining background was found to be the odd combination
of two photons i.e. each photon comes from a different $\pi^0$.
As a result, the two photons detected by the CsI tend to have a large energy
imbalance.  By requiring E(low)/E(high) $>$ 0.3 
the background level can be further reduced, as shown in 
Table~\ref{tab:sig_noise2}, 
while the signal acceptance is reduced to 0.88\%.

\begin{table}[here,top,bottom]
\begin{center}
\begin{tabular}{|c|c|c|l|c|l|c|}
\hline

ID1&ID2 & \#background &Type1(Bin) &Default  &Type2(Bin)  
&Default\\
\hline\hline
20 &68 &0.097$\pm$0.026  &PV ($<$0.02)  &0.2   &COi(3-10)   & $10^{-6}$ \\
17 &68 &0.045$\pm$0.012  &BA (1-3)    &0.1   &COi(3-10)   & $10^{-6}$ \\ 
20 &67 &0.021$\pm$0.021  &PV ($<$0.02)  &0.2   &COi(1-3)    &
$3\times10^{-6}$ \\
17 &67 &0.010$\pm$0.010  &BA (1-3)    &0.1   &COi(1-3)    & $3\times10^{-6}$ \\ 
26 &60 &0.021$\pm$0.015  &PV (0.4-1)  &0.2   &COi($<$0.02)  &1 \\
\hline\hline
\multicolumn{7}{|c|}{Total of 0.25 background events / SM signal ($3\times
10^{-11}$)}
\\

\hline
\end{tabular}
\caption{Categories of events from $2\pi^0$ decays with 2 missing photons
which contribute more than 0.01 events to the background level 
for KAMI-Far.  
A $P_t$ cut at 215~MeV has been imposed
along with an energy imbalance cut of E(low)/E(high) $>$ 0.3.}
\label{tab:sig_noise2}
\end{center}
\end{table}

\subsubsection{Results for KAMI-Near option}

The possibility of moving the target station downstream to a position
of z~=~120~m, 
the KAMI-Near option, was also studied.
A larger beam size of 
\mbox{1~mrad x 1~mrad = 1~$\mu$str} was used. 

Even though the beam solid angle is larger than for KAMI-Far,
the beam size at the BA is smaller.
As a result, the background level related to the inefficiency
in BA-In can be further reduced.

\vspace*{0.125in}
\parindent=0.25in

Table~\ref{tab:sig_noise3} itemizes the sources of backgrounds for the
standard $P_t$ cut at 150 MeV/c.
As shown in this table, there is no entry for BA-In.
All the background sources are related to the inefficiency of the
vacuum veto (PV) at low energy($<$ 40 MeV). 

\begin{table}[here,top,bottom]
\begin{center}
\begin{tabular}{|c|c|c|l|c|l|c|}
\hline

ID1&ID2 & \#background &Type1(Bin) &Default  &Type2(Bin)  
&Default\\
\hline\hline
20 &68   &0.086$\pm$0.003  &PV ($<$0.02)  &0.2   &COi(3-10) & $10^{-6}$ \\ 
20 &27   &0.074$\pm$0.000  &PV ($<$0.02)  &0.2   &PV (1-3)  & $3\times10^{-6}$ \\ 
20 &67   &0.035$\pm$0.003  &PV ($<$0.02)  &0.2   &COi(1-3)  & $3\times10^{-6}$ \\ 
20 &28   &0.019$\pm$0.000  &PV ($<$0.02)  &0.2   &PV (3-10) & $10^{-6}$ \\ 
21 &68   &0.014$\pm$0.000  &PV (.02-.04)  &0.03  &COi(3-10) & $10^{-6}$ \\ 
21 &27   &0.013$\pm$0.000  &PV (.02-.04)  &0.03  &PV (1-3)  & $3\times10^{-6}$ \\ 
\hline\hline
\multicolumn{7}{|c|}{Total of 0.32 background events / SM signal ($3\times
10^{-11}$)}\\
\hline
\end{tabular}
\caption{Categories of events from $2\pi^0$ decays with 2 missing photons
which contribute more than 0.01 events to the background level 
for KAMI-Near.}
\label{tab:sig_noise3}
\end{center}
\end{table}

With tighter $P_t$ cuts, the background level is reduced even further
to 0.13 events/SM signal as shown in Table~\ref{tab:sig_noise4}, 
while the acceptance is reduced from 
7.4\% to 1.0\%.

\begin{table}[here,top,bottom]
\begin{center}
\begin{tabular}{|c|c|c|l|c|l|c|}
\hline

ID1&ID2  & \#background &Type1(Bin) &Default  &Type2(Bin)  
&Default\\
\hline\hline
20 &68   &0.043$\pm$0.003  &PV ($<$0.02)  &0.2  &COi(3-10) & $10^{-6}$ \\ 
20 &67   &0.027$\pm$0.004  &PV ($<$0.02)  &0.2  &COi(1-3)  & $3\times10^{-6}$ \\ 
20 &27   &0.014$\pm$0.000  &PV ($<$0.02)  &0.2  &PV (1-3)  & $3\times10^{-6}$ \\ 
\hline\hline
\multicolumn{7}{|c|}{Total of 0.13 background events / SM signal ($3\times
10^{-11}$)}
\\
\hline
\end{tabular}
\caption{Categories of events from $2\pi^0$ decays with 2 missing photons
which contribute more than 0.01 events to the background level 
for KAMI-Near.  An additional $P_t$
cut has been imposed at 215 MeV/c.}
\label{tab:sig_noise4}
\end{center}
\end{table}

An additional energy balance cut of E(low)/E(high)$>$ 0.3 reduces
the background/Signal level to 0.077, while the signal acceptance goes down 
to 0.71\%, as shown in Table~\ref{tab:sig_noise5}.

\begin{table}[here,top,bottom]
\begin{center}
\begin{tabular}{|c|c|c|l|c|l|c|}
\hline
ID1&ID2 & \#background &Type1(Bin) &Default  &Type2(Bin)  
&Default\\
\hline\hline
20 &68  &0.026$\pm$0.015  &PV ($<$0.02) &0.2   &COi(3-10)   & $10^{-6}$ \\ 
26 &60  &0.013$\pm$0.013  &PV (0.4-1)   &0.2   &COi($<$0.02) &1 \\
21 &68  &0.010$\pm$0.004  &PV (.02-.04) &0.03  &COi(3-10)   & $10^{-6}$ \\ 
\hline\hline

\multicolumn{7}{|c|}{Total of 0.077 background events / SM signal
($3\times 10^{-11}$)}
\\
\hline
\end{tabular}
\caption{Categories of events from $2\pi^0$ decays with 2 missing photons
which contribute more than 0.01 events to the background level 
for KAMI-Near.  An additional $P_t$
cut has been imposed at 215~MeV/c along with an energy imbalance cut
of E(low)/E(high) $>$ 0.3.}
\label{tab:sig_noise5}
\end{center}
\end{table}

\subsubsection{Summary}

Table~\ref{tab:sum} summarizes the results for both the KAMI-Far
and KAMI-Near options for various kinematical cuts.
In this table, the expected
accuracy on the measurement of $\eta$, based on the number
of signal and background events, is listed as well for each case.
Since the \kpnn branching ratio is proportional to $\eta^2$,
the statistical error on the number of signal
events $(S)$ corresponds to an accuracy of
1/2$\sqrt{S}$ on $\eta$, assuming there is no background.
If we have to subtract the number of background events $(N)$,
the accuracy on $\eta$ is degraded and given by
$\sqrt{(1 + 2N/S)}/2\sqrt{S}$.
Poisson statistics are used for both signal and background.

From Table~\ref{tab:sum}, it is clear that KAMI-Near with a loose $P_t$ cut
(at 150 MeV/c) results in the most precise measurement of $\eta$.
After one year of running, 6\% accuracy can be
achieved.

\begin{table}
\begin{center}
\begin{tabular}{|l|r|r|r|}
\hline
			&Pt$ >$ 150 MeV    &Pt$ >$ 215 MeV    &Pt$ >$ 215 MeV \\
                         &           &       &E(low)/E(high)$>$0.3 \\
\hline\hline
 & & & \\  
{\bf [KAMI-Far]}    & & & \\
\hspace*{0.3in} Signal acceptance    &7.1\%    &1.2\% &0.88\% \\
\hspace*{0.3in} Signal/year in SM     &30 events &5.1 events& 3.7events\\
\hspace*{0.3in} Noise/Signal ratio   &0.57 &0.30  &0.25 \\
 & & & \\
\hspace*{0.3in} Accuracy on $\eta$ (/year)&13\% &28\% &32\% \\
 & & & \\  
\hline\hline
 & & & \\  
{\bf [KAMI-Near]}   & & & \\		 
\hspace*{0.3in} Signal acceptance   &7.4\%  &1.0\% &0.71\% \\
\hspace*{0.3in} Signal/year in SM  &124 events  &17 events &12 events\\
\hspace*{0.3in} Noise/Signal ratio&0.32  &0.13 &0.077 \\
 & & & \\
\hspace*{0.3in} Accuracy on $\eta$ (/year)&6\% &14\% &16\% \\
 & & & \\
\hline
\end{tabular}
\caption{Summary of results from the Monte Carlo simulation
for KAMI-Far and KAMI-Near.}
\label{tab:sum}
\end{center}
\end{table}

\vspace*{0.125in}
\parindent=0.25in
  
The most critical detector parameters and their design goals, based
on the above study, can be summarized as follows:

\begin{enumerate}
\item	Low energy ($<$20 MeV) photon detection in PV 	
	\hfill 	Ineff. $<$0.2 \\
\item	High energy (1-3 GeV) photon detection in PV   	
	\hfill	Ineff. $<3\times10^{-6}$ \\
\item	High energy (3-10 GeV) photon detection in CsI 	
	\hfill	Ineff. $<1\times10^{-6}$ \\
\item	Very high energy ($>$10 GeV) photon detection in BA 
	\hfill Ineff. $<1\times10^{-3}$. \\
\end{enumerate}

For the KAMI-Near option, the BA inefficiency can be relaxed 
to $1 \times 10^{-2}$ level.
However, we should also consider the fact that the neutron flux 
in the BA is higher in the case of KAMI-Near.

In Section~\ref{det_plan}, our future plan to study the feasibility of these
detector design goals will be discussed in some detail.

\subsection{Other possible background sources}

There are many possible background sources other than
$K_L \rightarrow 2\pi^0$.  None of these other backgrounds appear to be
as critical, particularly from the perspective of detector performance.
Detailed studies into these other background sources are still underway.
Careful and thorough studies have already been done
by the KEK group for their own proposal~\cite{inagaki}.
We have greatly benefited from their contribution.

\subsubsection{Background from $K_L\rightarrow 3\pi^0$}

The decay \kpipipi has a much larger branching fraction (21.6\%) than
$K_L\rightarrow 2\pi^0$. However, the final state has six photons which
make its rejection much easier.
Once care is taken to reduce $2\pi^0$ backgrounds,
it is easy to show that the rejection of $3\pi^0$ can be
achieved to the required level, thanks to the additional photons.

The most problematic background from \kpipipi decays is the case where the
decay
takes place upstream of the Mask Anti.
If four photons go undetected and two photons from two different $\pi^0$s
pass through the Mask Anti and reach the CsI calorimeter,
the vertex position reconstructed from the two mis-paired photons could 
be shifted downstream into the fiducial decay region.
To reject such events, the upstream beam pipe
is completely surrounded by the double stage Mask Anti
and the vacuum veto system.
This ensures that at least one of the 4 extra photons can be detected in 
this region.

\subsubsection{Background from $\Lambda \rightarrow n\pi^0$}

\begin{figure}
\centerline{\psfig{figure=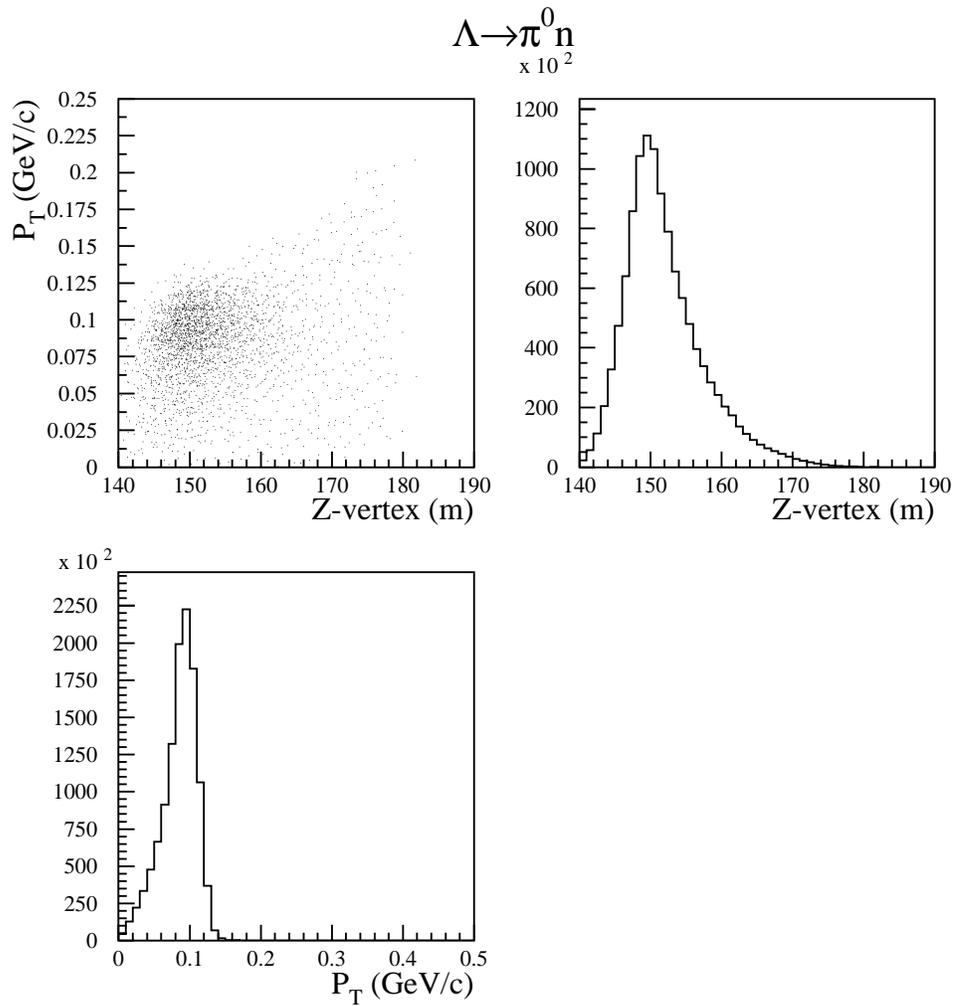,width=14cm}}
\caption{The z vertex and $P_t$ distributions for 
$\Lambda \rightarrow n\pi^0$ events from a Monte Carlo simulation.}
\label{zvpt_lambda}
\end{figure}

This decay has a large branching ratio (36\%) but the $P_t$ endpoint
is at 104 MeV/c.  
The $P_t$ distribution and  
vertex $z$ distribution are shown in Fig.~\ref{zvpt_lambda}.
By restricting the neutral beam divergence through the
use of a small beam, this background can be effectively rejected
by a $P_t$ cut around 150 MeV/c.  
 
Since lambdas have a shorter lifetime than kaons,
after 100~m or so of decay length their flux becomes
completely negligible.  Thus, in the KAMI-Far option, lambdas
are not an issue.

More careful study is required to understand the lambda background 
for the KAMI-Near scenario. 
If one of the photons from $\Lambda\rightarrow$n$\pi^0$ hits the
CsI and the other photon is missed,
an extra cluster in the CsI from accidental activity could
combine with the detected photon to mimic a $\pi^0$ with high $P_t$ 
which reconstructs inside the allowed fiducial volume.
  
A related background source is the cascade decay into $\Lambda \pi^0$,
followed by the lambda decay mentioned above.
This is troublesome even without accidentals because the final
$\pi^0$ could carry a $P_t$ as large as 230 MeV. 
This background should be removed by detecting the $\pi^0$ from the initial
cascade decay in the upstream region surrounded by the double stage Mask
Anti. 

\subsubsection{Background from $K_{e3}$ and $K_{\mu3}$ decays}

Another source of background comes from copious kaon decays to
two charged particles such as $K_L \rightarrow \pi^{\pm} e^{\mp} \nu$
and $\pi^{\pm} \mu^{\mp} \nu$.  To reject these decays,
at least one charged particle must be vetoed before it strikes the CsI.
This can be achieved at the required level by adding a charged veto
in front of the calorimeter.
The inefficiency of scintillation counters has been extensively
studied by the KEK group~\cite{inagaki} and they have demonstrated that
1 cm thick scintillators, located in front of the calorimeter 
can reduce this type of background to a negligible level.

\subsubsection{Background from nA $\rightarrow \pi^0$A}
 
Beam neutrons can interact with any material in their path
and produce $\pi^0$s copiously.
Thus, it is of critical importance to minimize material thickness in the
beam region.
In KTeV, neutron interactions in the vacuum window and downstream 
detector elements, such as the drift chambers, have proven to be the
most serious background for the \kpnn search using the $2\gamma$
decay mode of the $\pi^0$. 

To reduce this background to a manageable level in KAMI, the entire region from
the beam collimator to just upstream of the CsI calorimeter should be 
evacuated to the level of $3\times10^{-7}$~torr.  This estimate is based
on the number of neutron interactions in a vacuum of this level which would
result in 2 photons in the CsI calorimeter. 
We have already achieved $1\times10^{-6}$ torr in the KTeV vacuum decay region.
Since the vacuum decay region in KAMI is much shorter than that of KTeV,
with modest upgrades of the existing vacuum pumping system, we expect 
to achieve $3\times10^{-7}$ torr.

\section{Other Decay Modes}

There are many rare kaon decays other than \kpnn which are of substantial 
interest
for a variety of reasons.  The large kaon flux provided by the Main Injector
combined with KAMI's large acceptance for many of these channels make KAMI
the best place to study these decays.  None of the other proposed experiments
for measuring \kpnn have this capability.

In addition to $K_L\rightarrow \pi^0 \nu \overline{\nu}$, there are other 
rare kaon decays which are sensitive to direct CP violation.  
According to the Standard Model, a substantial fraction of the decays
$K_L\rightarrow \pi^0 e^+e^-$ and $\pi^0 \mu^+ \mu^-$ should be direct
CP violating and are expected to have branching ratios within reach of KAMI.

KAMI will also have the capacity to perform sensitive searches for 
other rare and forbidden decays.  These include processes
forbidden by the Standard Model, such as the lepton flavor violating
decay \kpme.  Other processes, such as
\kee, are highly suppressed in the
Standard Model and provide windows where new physics might be detected.

It will also be possible to extend the sensitivity of the 
$\epsilon^\prime/\epsilon$ measurement at the Main Injector, should it be
necessary.  A statistical accuracy of $3\times10^{-5}$ is feasible at
the Main Injector in 1~year of running based
on the order of magnitude increase in decay rates combined with the
7-fold increase in the regeneration amplitude obtained at the lower
kaon momentum.

\subsection{\kpee and \kpmm}

The physics motivation for the decay \mbox{$K_L\rightarrow\pi^0 e^+ e^-$}
has a long history in the literature.
The decay has a CP conserving component, an indirect
CP violating component and a direct CP violating component.  The
direct CP violating component is of primary interest and
could be the largest of the three~\cite{don}.  The branching ratio for this
decay is predicted to be on the order of several times $10^{-12}$ and the
current best limits on the decay are at the $10^{-9}$ level~\cite{harris1}.
Once detected, untangling the various contributions to the decay,
particularly in the presence of the attendant background from the
radiative Dalitz decay of the kaon,
\mbox{$K_L\rightarrow e^+ e^- \gamma \gamma$},
is a significant experimental challenge.  There
could be a significant electron asymmetry present, of the form \\
\begin{equation}
A = \frac{N(E_+ > E_-) - N(E_+ < E_-)}{N(E_+ > E_-) + N(E_+ < E_-)},
\end{equation}
which would signal the interference of the CP violating and CP conserving
amplitudes.  This asymmetry will, of course, be diluted by the radiative
Dalitz background. 

Estimates for the sensitivity expected in KAMI
for $K_L\rightarrow \pi^0 e^+ e^-$ are listed in Table~\ref{tab:kpee_sens}.  
Sensitivities of $1.4 \times 10^{-12}$ and $7.8 \times 10^{-14}$ are
expected for KAMI-Far and KAMI-Near, respectively.
Note that the \mbox{$0.07 \times 0.07$~m$^2$} beam holes in
the fiber planes reduce the acceptance by about 30\% for this mode.

The related mode $K_L \rightarrow \pi^0 \mu^+ \mu^-$ is of interest
for similar reasons.  However, the CP conserving amplitude may be
significantly larger for this mode than for $\pi^0 e^+ e^-$ because
there is no helicity suppression. 
The branching ratio
for this decay is also predicted to be on the order of $10^{-12}$ and the
current best limits on the decay are at the $10^{-9}$ level~\cite{harris2}.
 
Estimates for the sensitivity expected in KAMI
for $K_L\rightarrow \pi^0 \mu^+ \mu^-$ are listed in Table~\ref{tab:kpee_sens}.
Sensitivities of $1.1 \times 10^{-12}$ and $6.8 \times 10^{-14}$ are
expected for KAMI-Far and KAMI-Near, respectively.

\subsection{\kppee}

\begin{figure}
  \centerline{ \psfig{figure=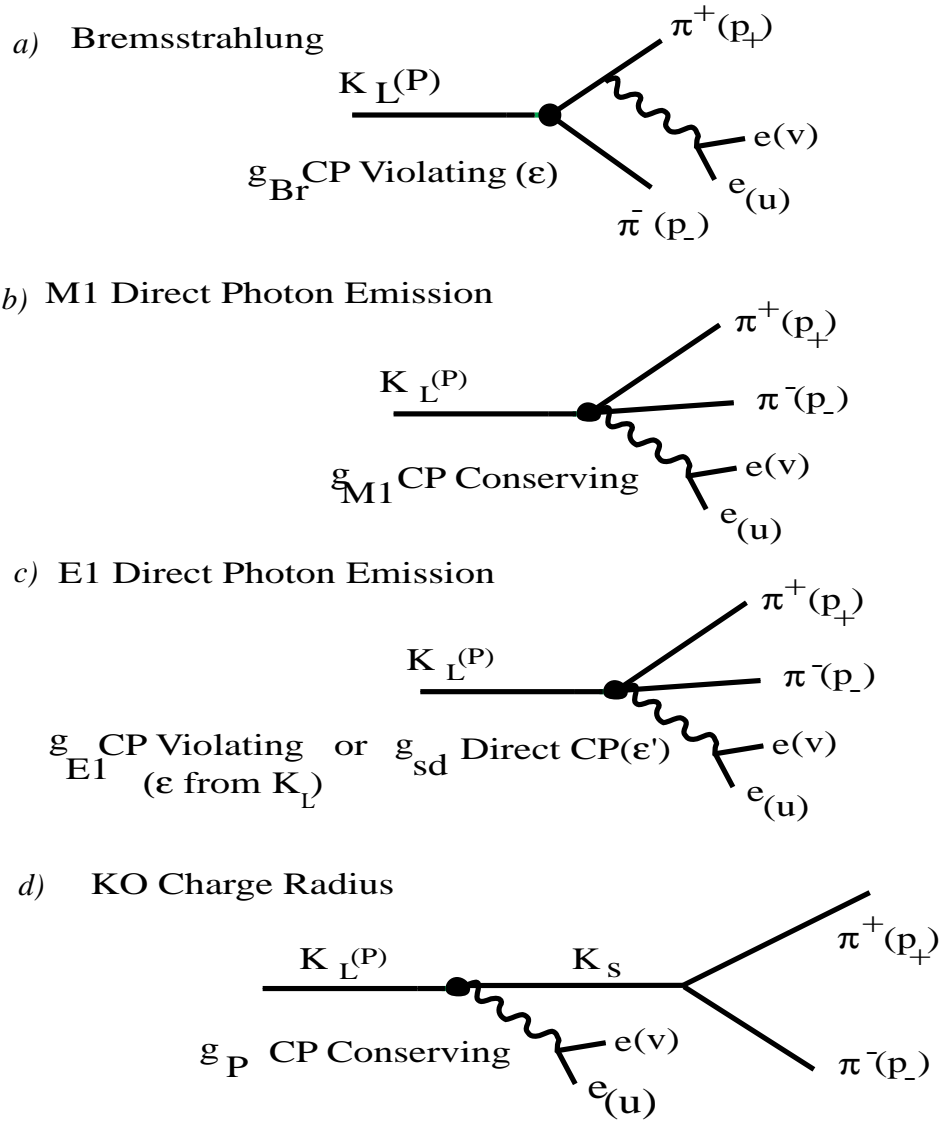,width=12.5cm} }
  \caption{Contributing diagrams to the decay
$K^{0}_{L}\rightarrow\pi^{+}\pi^{-}e^{+}e^{-}$. \label{Fig:2}}
\end{figure}

In the 1996-97 run of KTeV, the previously undetected decay
$K^{0}_{L}\rightarrow\pi^{+}\pi^{-}e^{+}e^{-}$ has been 
observed~\cite{odell}.
As mentioned in Section~\ref{sec-pipiee}, the strong interest in this 
mode is because of the prospect for observing
CP violation as predicted in Ref.~\cite{seghal}.  This decay can proceed
via the four processes shown in Fig.~\ref{Fig:2}.  The interference of the
indirect CP violation Bremsstrahlung process (Fig.~\ref{Fig:2}a) with the
CP conserving M1 emission of a virtual photon (Fig.~\ref{Fig:2}b)
is expected to generate an asymmetry in the angle $\phi$ between the 
normals to the decay planes of the $e^{+}e^{-}$ and the $\pi^{+}\pi^{-}$ in 
the $K^{0}_{L}$ center of mass.  In addition, direct CP violation effects, 
albeit small, can occur in this mode via the interference of the weak process 
of Fig.~\ref{Fig:2}c with the other three amplitudes.  

\begin{figure}
  \centerline{ \psfig{figure=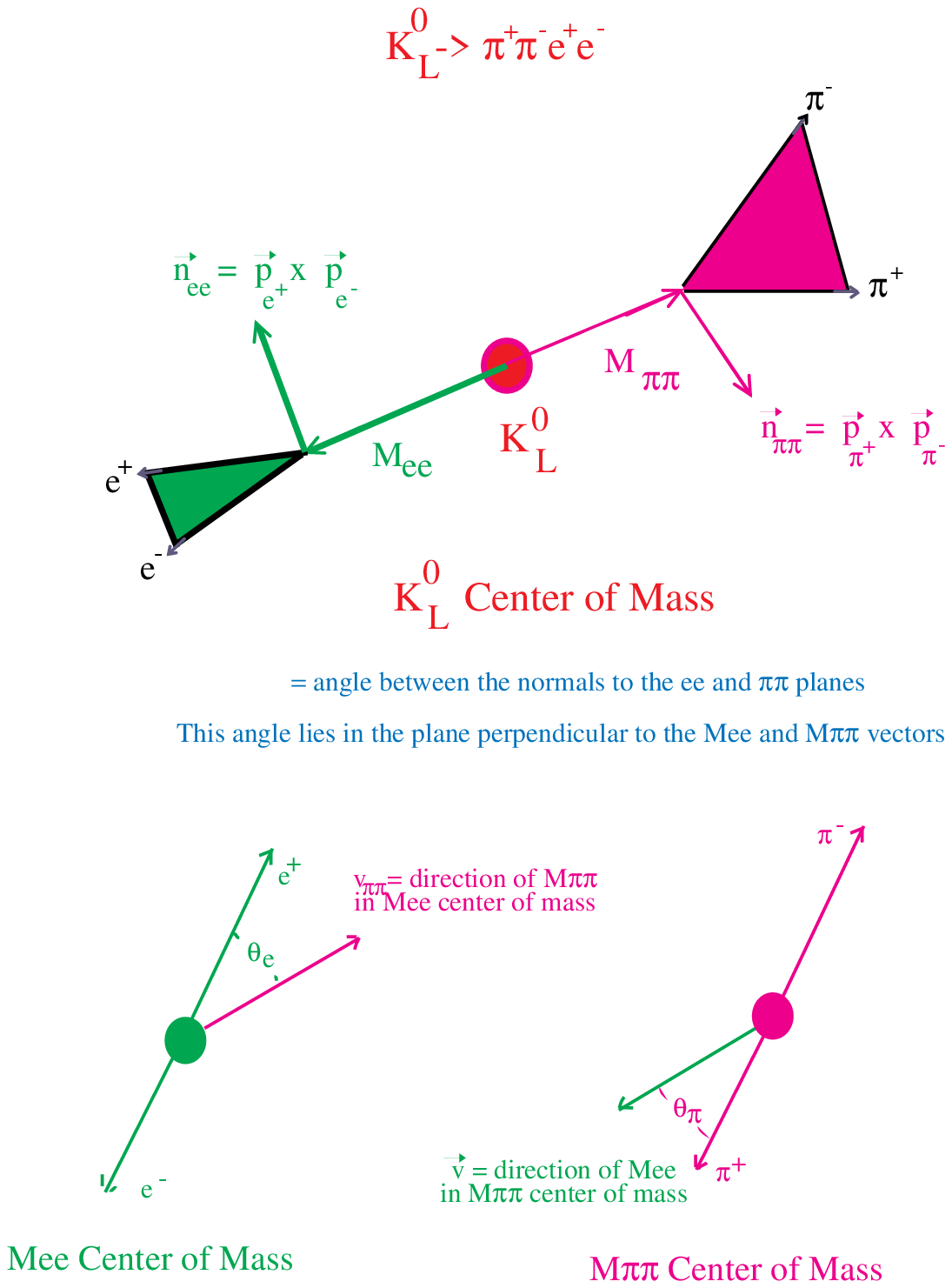,width=12.5cm} }
  \caption{Angles in which indirect and direct CP violating asymmetries may be
seen in \kppee decays. \label{Fig:3}}
\end{figure}

We show in Fig.~\ref{Fig:3} the angles in which the indirect and
direct CP violation asymmetries are expected to be observed. The angular 
distribution as a function of $\phi$ and $\theta$, where $\theta$ is the 
angle of the positron with respect to the direction of the
M$_{\pi\pi}$ cms in the M$_{ee}$ cms, is given by

\begin{eqnarray*}
 \frac{d\Gamma}{dcos\theta d\phi} = K_{1}+K_{2}cos2\theta +K_{3}sin^{2}\theta
cos2\phi + K_{4}sin2\theta cos2\phi +K_{5}sin\theta cos\phi \\
+K_{6}cos\theta
+K_{7}sin\theta sin\phi +K_{8}sin2\theta sin\phi 
+K_{9}sin^{2}\theta sin2\phi.
\end{eqnarray*}

The $K_{4}$, $K_{7}$ and $K_{9}$ terms are the ones in which CP violation is
expected to appear.  The $K_{7}$ term is where direct CP violation
would occur.  Ignoring small terms and integrating over $\theta$, the
$\phi$ angular distribution is obtained:

\begin{eqnarray*}
 \frac{d\Gamma}{d\phi} = \Gamma_{1}cos^{2}\phi + \Gamma_{2}sin^{2}\phi +
\Gamma_{3}sin\phi cos\phi.
\end{eqnarray*}

An asymmetry in the $sin\phi cos\phi$ distribution will signal the presence
of indirect CP violation.  This would  
be the fourth observation of indirect CP violation in 35 years and the 
first manifestation of CP violation in a dynamic variable.
An asymmetry of
13.1\% is expected between $sin\phi cos\phi\geq$0 and
$sin\phi cos\phi\leq$0 in the events accepted by the KTeV spectrometer.
This asymmetry will be measured in the 800~GeV/c KTeV data with a
statistical error of $\approx$1\% using the total data from the 1997 and
1999 runs ($\approx$ one half the expected statistical error of the 1997 run).

%The $sin\phi cos\phi$ angular distribution of the 1250 events in the 
%$K^{0}_{L}$ mass region is shown in Fig.~\ref{Fig:4} together with the 
%theoretical expectation calculated per the prescription of  Ref.~\cite{seghal}.
%As can be seen, there is a clear asymmetry in this angular distribution.
%The theoretical expectation matches the data quite well.  
%An asymmetry of 
%13.1\% is expected between $sin\phi cos\phi\geq$0 and 
%$sin\phi cos\phi\leq$0 in the events accepted by the KTeV spectrometer.  
%This asymmetry will be measured in the 800 GeV/c KTEV data with a 
%statistical error of $\approx$1\% using the total data from the 1997 and 
%1999 runs ($\approx$ one half the expected statistical error of the 1997 run). 

%\begin{figure}
%  \centerline{ \psfig{figure=KTEV97_phi_asym.eps,width=12.5cm} }
%  \caption{ The $sin\phi cos\phi$ angular distribution; points with
%error bars are KTEV data; points with no error bars are Monte Carlo
%simulations according to the theoretical expectations of Ref.~\cite{seghal}.
%\label{Fig:4}}
%\end{figure}

Estimates for the number of \kppee decays expected in KAMI
are listed in Table~\ref{tab:kpee_sens}.
64~k and 1.1~M events are expected for KAMI-Far and KAMI-Near, respectively.

\begin{figure}
  \centerline{ \psfig{figure=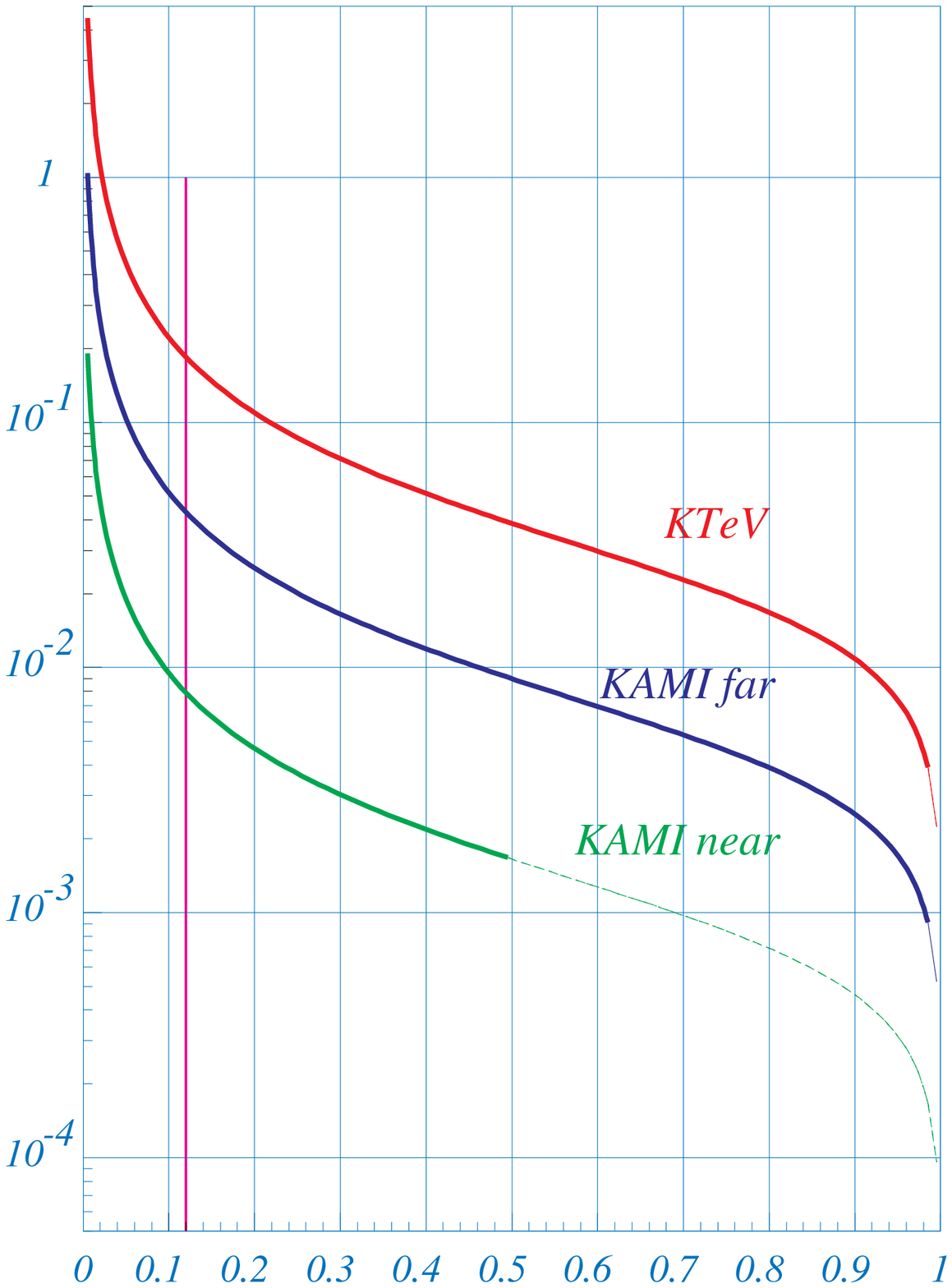,width=10cm} }
  \caption{Fractional error of the 
$sin \phi cos \phi$ asymmetry as a
function of the asymmetry level. \label{fig:5}}
\end{figure}

With the greatly increased numbers of
$K^{0}_{L}\rightarrow\pi^{+}\pi^{-}e^{+}e^{-}$ events available in either
KAMI configuration relative to that which can be accumulated in the
800~GeV/c KTeV operation in the next few years, the asymmetry in
$sin\phi cos\phi$ can be measured much better.  This is demonstrated in
Fig.~\ref{fig:5} which shows the expected error in the asymmetry 
measurement for
KTeV, KAMI-Far and KAMI-Near as a function of the level
of the asymmetry. 

Finally, although  direct CP violation effects which are due to the 
Standard Model
CKM phase in $K^{0}_{L}\rightarrow\pi^{+}\pi^{-}e^{+}e^{-}$ decays are 
expected to be small, the increased statistics available in a KAMI experiment
will allow an examination with increased  sensitivity of the more complex,
joint $\theta, \phi$ distribution to search for evidences of direct CP
violation.

\subsection{\kpme and \kme}

The decays \kpme and \kme, as well as other processes which violate
lepton number conservation, can occur in the Standard Model if the
neutrino masses are not zero or degenerate.  However, existing limits
on neutrino masses and mixing angles imply exceedingly small branching
ratios which are not observable.  Observation of either of these modes
would therefore provide clear evidence for physics outside of the Standard
Model.

An active analysis of \kpme is underway in KTeV and the 1997 data set
should result in a sensitivity of $3 \times 10^{-11}$. 
Estimates for the sensitivity expected in KAMI
are listed in Table~\ref{tab:kpee_sens}.
Sensitivities of $1.2 \times 10^{-12}$ and $7.1 \times 10^{-14}$ are
expected for KAMI-Far and KAMI-Near, respectively.

KTeV did not trigger on \kme because of data acquisition bandwidth limitations.
KAMI hopes to incorporate such a trigger if it can be shown that the
large background to this mode can be suppressed.

\subsection{\kmm and \kee}

\kmm is a flavor-changing neutral-current processes which
serves as an interesting probe of second-order weak processes in the Standard 
Model.  The decay is sensitive to V$_{td}$, the same CKM element
responsible for much of the short-distance physics of 
\mbox{$K^+ \rightarrow \pi^+ \nu \overline{\nu}$}.  The theoretical
interpretation of \kmm is unfortunately complicated by uncertainties 
in the $K_L \rightarrow \gamma^* \gamma^*$ contribution. 
However, KAMI will be able to collect large samples of \kmm,
$K_L \rightarrow \mu^+\mu^- \gamma$ and $K_L \rightarrow e^+e^-\mu^+\mu^-$ 
decays which will allow a detailed
analysis of the $K_L \gamma^* \gamma^*$ form factor.

Estimates for the sensitivities and event numbers expected in KAMI for \kmm
are listed in Table~\ref{tab:kpee_sens}.
On the order of 25~k and 427~k decays are expected
for KAMI-Far and KAMI-Near, respectively.

The physics of the decay \kee is identical to that of \kmm.  However,
helicity suppression reduces the rate of Standard Model decays relative
to \kmm by a factor of order $O(m_e^2/m_\mu^2)$.  This leaves open the
possibility of observing a non-Standard Model component to the decay 
which proceeds via a pseudoscalar interaction which is unconstrained
by helicity suppression.  The best published limit for this decay
is $4.1 \times 10^{-11}$ (90\% CL)~\cite{arisaka}.

Estimates for the sensitivities expected in KAMI for \kee
are listed in Table~\ref{tab:kpee_sens}.
Sensitivities of $2.6 \times 10^{-13}$ and $1.5 \times 10^{-14}$ are
expected for KAMI-Far and KAMI-Near, respectively.

\subsection{Other decays}

There are many other rare decays accessible by KAMI 
which address important physics issues.
For example, $K_L$ and $\pi^0$ decays 
to 4 leptons will be abundant in KAMI.  

The decay $K_L \rightarrow e^+e^-\mu^+\mu^-$ proceeds 
primarily through the two-photon
intermediate state \mbox{$K_L \rightarrow \gamma^* \gamma^*$.}  The
$K_L \gamma^* \gamma^*$ form factor must be accurately known in order to
isolate the contribution of second-order weak processes to the decay
\kmm, as discussed earlier.  Additionally, the angular distribution
between the $e^+e^-$ and $\mu^+\mu^-$ decay planes provide a mechanism
to search for a possibly large direct CP violating amplitude in the
$K_L \rightarrow \gamma^* \gamma^*$ transition.  KAMI can collect on the 
order of $10^5$ of these decays.

The angular distribution between the decay planes of the lepton pairs
in the related decay $K_L \rightarrow e^+e^-e^+e^-$ is also sensitive
to direct CP violation in the $K_L \rightarrow \gamma^* \gamma^*$ transition.
KAMI can collect on the order of $10^6$ such decays for detailed study.

The angular distribution measurement for $\pi^0\rightarrow e^+e^-e^+e^-$
offers a parity violation test in electromagnetic decays for q$^2$ in the
hundreds of MeV$^2$ range.  In fact, all four-body decays modes offer
such dynamical tests.  KAMI can collect on the order of $10^{7}$ 
$\pi^0\rightarrow e^+e^-e^+e^-$ decays.

Finally, the decay $K_L \rightarrow \pi^0 \pi^0 e^+e^-$ is the neutral
partner of the previously mentioned $K_L \rightarrow \pi^+ \pi^- e^+e^-$
decay, but does not contain the inner brem term which contributes to
the latter decay mode.  This mode offers yet another opportunity to
observe a non-Standard Model CP violation effect.  The predicted
branching ratio for this decay is on the order of $10^{-10}$~\cite{seghal1}.  
KAMI can collect on the order of $10^3$ $K_L \rightarrow \pi^0 \pi^0 e^+e^-$
decays.

As can be seen from the above discussion, the KAMI charged mode program 
alone represents a diverse and important physics program which probes
the Standard Model and issues of CP violation at significant levels.  
This program is completely compatible with the KAMI program to measure
\kpnn and is not possible to execute at any of the other proposed
\kpnn experiments.

\section{Comparison with Other Proposals}

At present, there are two other proposals to measure the rate for
\kpnn;  one from a collaboration working at KEK~\cite{inagaki} and
another from a collaboration working at BNL~\cite{chiang}. 

We will concentrate our remarks in this section on the BNL
proposal.  
This is because that proposal attempts to reach a similar
level of sensitivity to ours (roughly 30 events per year at the expected
level), on a similar time scale.  
At present, the KEK collaboration is
proposing to reach a single event sensitivity per year at the level
expected from the Standard Model, and they will be working at the new JHP
facility when it becomes available. 
	
Even so, we have learned much from our colleagues at KEK.  Their
proposal incorporates many clever ideas, and they have performed some
incisive measurements, particularly of the capabilities of various photon
vetos at low energies.  In fact, we are beginning a collaborative
effort with them on these important issues. 
	
Before giving our comments, we want to state again that reaching
the goal of measuring the rate for \kpnn at the 10\% level
will take time and will be difficult to achieve.  
In making critical
remarks about our competition, we do not mean to imply that we are sure
our approach will in fact succeed. 

Given sufficient beam flux, the problem for all attempts at measuring the
rate for \kpnn is rejection of background.  The dominant
source of background again universally appears to come from \kpipi
decays where two photons are missed in the detector.  
The proponents of
the BNL experiment have opted to make the center piece of their experiment
the ability to kinematically reject background.  Indeed, such a handle, if
viable, would be invaluable in that one could tolerate greater
inefficiency in the photon vetos. 
	
Kinematic rejection means that one measures the $K_L$ momentum
via time-of-flight and reconstructs the detected $\pi^0$ invariant mass
and direction.  Thus, one can determine the center-of-mass 
momentum of the $\pi^0$ and discriminate against the \kpipi decay
which has a unique value for this quantity. 

To make use of the measured time-of-flight of the $K_L$ (actually
of the $\pi^0$ decay photons), one is pushed, first of all, into having a
very well bunched source of decays, and, second, into working at very low
momentum.  Thus, the mean kaon momentum for the BNL experiment is only 0.7
GeV, 20 times less than in the present proposal.  BNL accelerator
physicists have not yet achieved the required bunching of the beams and
this is critical to the viability of this technique.  We will assume that
this hurdle is crossed and then list our remaining concerns.  Where direct
comparisons are made, they are to our KAMI-Far geometry which has similar
sensitivity to that of the BNL experiment.

\vspace*{0.25in}
\parindent=0.in
 
{\bf 1.}  Working in such a low energy, very large solid angle
[500 $\mu$str]
beam, something that has not been attempted before, could be problematic. 
A major problem for the KEK experiment attempting to measure \kpee,
which uses a beam with some similar characteristics to that proposed
by BNL, has been a sea of low-energy neutrons.  The BNL experiment has
collaborators from this KEK experiment so they should be well aware of
such problems. And BNL is in the process of measuring the neutron flux.
From our experience, all important effects from a neutron halo are much
reduced at higher energy where beams are naturally of smaller solid angle
and the experiment is situated further from the production target.  Better
geometry really helps, especially in a neutral beam. 

% [need to get Bob H's remarks here about Sassao's
% exp. What is their solid angle and mean momentum?]
 
\vspace*{0.15in}
\parindent=0.in

{\bf 2.}  Higher energy photons are easier to
veto than lower energy ones, as
has been seen in the earlier discussions in this proposal.  Thus, the
constraint of working at lower energies to be able to use time-of-flight
means that, in general, one needs to veto lower energy photons.  This
eats somewhat  into the advantage of needing less veto power.  (However,
with kinematic reconstruction one does suppress some low-energy photons
with a missing mass cut.  And at higher energy the faster 
$\pi^0$'s in the lab, when they decay nearly along their line-of-flight, can 
produce even lower
energy photons than for the lower energy case.)

\vspace*{0.15in}
\parindent=0.in

{\bf 3.}  As another consequence of the lower energy beam, the total
neutron
flux for the BNL experiment is much higher than for the FNAL one.  At BNL,
it is estimated to be roughly 5~GHz while at FNAL, it is 200~MHz.  
The BNL proponents
argue that most of the neutrons are of too low an energy to be of any
concern, and that they are much more spread out in time than the expected
signals from decay photons.  Nevertheless, given that both detectors have
a Back-Anti or beam catcher that must live in such environments, this is a
concern. 
 
\vspace*{0.15in}
\parindent=0.in

{\bf 4.}  Another problem for the BNL experiment (not mentioned in their
proposal)  could be anti-neutrons in the beam.  We have said that a high
neutron flux, particularly a halo, could be serious.  
The BNL group argues
that the most important problem with neutrons is the production of
$\pi^0$ from material such as residual gas in the vacuum.  They then
rightly argue that only neutrons above 800 MeV/c can produce $\pi^0$'s 
(most of the 5 GHz is below that value).  However, anti-neutrons can and
do produce $\pi^0$'s even at rest;  and if there is a component of
anti-neutrons in the halo, this could be even more serious in faking
photons in their detector.  Of course at higher energies, neutrons and
anti-neutrons behave for all intents and purposes identically.  The
increase in the cross-section for very low energy anti-neutrons means that
the BNL group should also worry about (and measure) the anti-neutron
content of their beam:  estimates would put it at a few percent of the
kaon rate. 

\vspace*{0.15in}
\parindent=0.in

{\bf 5.}  The imposed kinematic cut to reject the \kpipi events also
rejects 65\% of signal events.  Also, the requirement of measuring the
photon energies and angles means that an active converter must be employed
(see 7 below)  which costs another factor of 2 in acceptance.  These are
the two major factors contributing to the FNAL experiment having 7\%
acceptance (see Table~\ref{tab:sum}) while at BNL it is 1.6\%.  
As a result, the BNL kaon decay
rates, for similar sensitivity, are about a factor of 9 greater than for
FNAL (25 MHz vs. 2.8 MHz) as seen in Table~\ref{tab:kpnn_sens}. 

\vspace*{0.15in}
\parindent=0.in

{\bf 6.}  Another concern is associated with the geometry of the BNL 
``beam-catcher."  Again for timing reasons, their layout shows this device
situated 15~m behind the calorimeter;  and the beam pipe in this region
must be surrounded by anti-counters to catch any photons from kaon decays in
the decay region.  But the rates in these counters will be high:  about 70
MHz just from kaon decays occurring in this 15 m region alone.  Thus, the
full rate of kaon decays that the BNL experiment needs to reject is 95
MHz, more than 30 times the similar rate for our proposed experiment. 
These rates are getting to the point where one worries about serious
veto dead-time and, more importantly, inefficiencies. 

\vspace*{0.15in}
\parindent=0.in

{\bf 7.}  Additional problems arise because of the rather elaborate photon
converter the BNL group is forced to use in order to reconstruct the
$\pi^0$
decay photons.  We have already mentioned the loss in acceptance that this
entails.  But more serious is the spreading of the resulting
electromagnetic showers.  In order to obtain the desired resolution, the
BNL group must add in calorimeter channels corresponding to a region of
diameter of about 120 cm~\cite{bryman}.
The corresponding figure for the KTeV calorimeter is just 15 cm.  
Given the much higher rate conditions at BNL than at FNAL as well as the
correspondingly greater weight that a stray minimum-ionizing particle has
at lower energy, this is an area of considerable concern.  (In the first
phase of E799, we also had a converter situated in front of a precision
calorimeter, but in the end the complications associated with shower
spreading were never outweighed for the physics we were doing
by its purported advantages.) 
Additionally, because extremely good timing is required from the
calorimeter and given that their beam is a thin rectangle taking up a good
part of the width of the calorimeter, a bar geometry is being considered
with only a \mbox{y-view~\cite{bryman}.}  
Thus the effective region that is added
together to account for a single photon becomes a significant fraction of
the area of the calorimeter itself. 

\vspace*{0.15in}
\parindent=0.in

{\bf 8.}  To obtain the required vacuum, the BNL group has opted to
configure
their photon veto system outside of the vacuum.  From our experience with
counters operating in vacuum, we can obtain a sufficient vacuum without
the problems associated with a roughly 5\% $X_0$ vacuum wall between the
kaon decay and the detector.  
Having the counters in vacuum improves their
efficiency, especially in the low energy region.  This is an important
consideration in our design. 

\vspace*{0.15in}
\parindent=0.in

{\bf 9.}  Finally, the BNL experiment, if
built as configured in their proposal, would attack a  limited area of
neutral kaon physics.  It is of course good to focus on a particular
problem, particularly one with the importance of the \kpnn decay
mode.  However, it is also good to have a broad program and it appears
that there is little else the BNL experiment can address.  With the wall
of the vacuum tank and no magnetic field or tracking, they will not be
able to study any charged mode that requires background rejection.  And
the granularity of the calorimeter, as discussed above, may not allow for
any incisive study of multi-body neutral decays. 

\vspace*{0.30in}
\parindent=0.25in

In contrast, our approach is to produce a detector which 
is fully hermetic and has state-of-the-art tracking capabilities, allowing
a broad program of investigation. We have opted for high acceptance and
therefore relatively low rates and clean signatures:  a high energy, very
well defined and shielded beam with low neutron contamination, no
conversions, and vetos in vacuum.  We have outlined an R\&D program which
will reveal whether we can achieve the necessary rejection power which,
admittedly, will be very challenging. 
	
Our comments about the BNL proposal are made in the spirit of
giving the reader our honest reservations about their approach.  We have
had the advantage of having their proposal available in the preparation of
our own one.  We, of course, look forward to their comments about the
present document.

\section{Detector Development Plan}
\label{det_plan}

As KAMI is a natural extension of the ongoing KTeV experiment,
we greatly benefit from KTeV's current activities.
KTeV completed the first phase of its highly successful data taking
in September 1997, with plans to continue a second phase of the experiment
in 1999, with minor modifications.

The existing KTeV detector represents a significant investment of time,
money, and manpower.  It can evolve into a powerful
detector for KAMI in an efficient and cost effective manner.
This Section first describes the current R \& D activities within the 
KTeV group which are directly related to our plans for making the
transition from KTeV to KAMI.  Then a plan for
bringing the Main Injector beam to the
existing KTeV target and using the existing KTeV detector facility
for KAMI R\&D studies is described. 

Our ultimate goal for these proposed R\&D studies is to determine
if it is indeed possible to design a detector which meets the
specifications required for sufficient background rejection, as
described in Section~\ref{background_study}.

\subsection{Studies of data from KTeV}

\subsubsection{CsI calorimeter}

KTeV's greatest asset is the CsI calorimeter.
As already mentioned in Section~\ref{csi}, it has been demonstrated to
perform to a very high level. 
Energy resolution, position resolution and fusion rejection efficiency
have been studied with very encouraging results.
These results were incorporated into the Monte Carlo simulation programs,
which produced the background study reported in Section~\ref{background_study}. 

We will continue to study the detection inefficiency for photons of various
energies as well as the time resolution of the CsI.  Both issues can
be addressed using existing KTeV data.  These studies should conclude
shortly.

\subsubsection{Vacuum photon veto}

The vacuum veto detectors in KTeV, referred to as the Ring Counters,
were designed primarily to reduce the 3$\pi^0$ background contribution to
the 2$\pi^0$ signal.  They have performed this task well.  While
the hermeticity and level of performance required by the photon veto
detectors for KAMI are well in excess
of that required by KTeV, the KTeV Ring Counters still provide us with
a valuable tool for understanding photon veto counter performance.
A detailed analysis of the KTeV Ring Counter performance
will provide valuable information on how to design
veto detectors for KAMI.  These studies are ongoing.

\subsubsection{Back Anti}

The Back Anti (BA) resides in the neutral beam and detects photons which
pass through the beam hole of the CsI calorimeter.
Since it is exposed to an intense flux of neutrons and kaons,
degradation of its veto efficiency has been observed in KTeV.
For example, for typical E799 intensities (for rare decay
studies), the neutron flux is 44~MHz at the BA. As the existing BA has one
nuclear interaction
length, with the low energy threshold required for the \kpnn study in the
Dalitz decay mode, the signal sensitivity is reduced by 50\% due to
neutron interactions.
We are developing the best algorithm possible
to reduce backgrounds while maintaining
good signal sensitivity.

\subsection{Detector R\&D at KTeV 99}

The anticipated KTeV run in 1999 will give us an ideal opportunity to
study detector prototypes for KAMI in a realistic environment.
As described in our Letter of Intent for KTeV~99~\cite{loi99},
we propose to perform detector R \& D in two areas; photon vetos and
fiber tracking.

\subsubsection{Vacuum photon veto}

The KAMI vacuum veto is based on the existing KTeV photon veto design.
However, to achieve much better efficiency for low energy photons
as low as several MeV, finer sampling using 1~mm lead sheets (instead
of 2.8~mm in KTeV), thicker scintillator (5~mm instead of 2.5~mm in KTeV),
and more dense WLS fiber readout are currently being considered.
At the same time, we are considering the use of either extruded
or injection molded scintillator in order to reduce costs significantly.

We plan to develop a small prototype detector (about 50 cm by 50 cm
active region) based on our current design described in Section~\ref{phot_veto}.
The key issues to be studied are:

\begin{enumerate}
\item Light collection efficiency i.e. the number of photo electrons per
incident energy as a function of thickness of the scintillator
and the density of WLS fibers;

\item Detection efficiency for low energy photons (0 - 20 MeV),
as a function of lead sheet thickness and tilt angle;
 
\item Detection efficiency for high energy photons ($>$ 1 GeV),
as a function of total depth;

\item Pulse shape and time resolution; and
 
\item Mechanical design, including support structure for the lead sheets and
 the interface between the WLS fiber routing and the vacuum pipe.
\end{enumerate}

\subsubsection{Back Anti}

The Back Anti is critical for detecting photons down the CsI beam
holes from $K_L \rightarrow 2\pi^0$ and $3\pi^0$ decays.
It covers the largest acceptance of all the photon vetos and it
must operate in an environment with a significant neutron flux.
 
Under KAMI operating conditions,
the neutron flux is expected to exceed 100~MHz.
Therefore it becomes critical to design a BA which is neutron
transparent, yet has high rejection power for photons.
One way to achieve such rejection is through
fine-grained depth segmentation, with active sampling every 3-4 radiation
lengths.  The energy threshold on each individual section can be tuned
to maximize photon rejection and to distinguish photons from neutrons.
Fast timing resolution is also useful
to distinguish out-of-time neutron interactions, once the beam is debunched
as expected for KAMI.  
For this reason, we are considering a Cherenkov medium
such as lucite or quartz as the active material, as described in
Section~\ref{ba}. 

During the KTeV 1999 run, a neutron flux
of close to 100~MHz is expected.
To study \kpnn using the Dalitz mode, we must use the BA to veto photons
in this environment. 
Therefore, 
we plan to develop a KAMI-compatible Back Anti so that
we can fully test its functionality in a realistic environment
during the KTeV 1999 run.

\subsubsection{Fiber tracking}

The scintillating fiber tracker is another major piece of new hardware 
necessary for KAMI.
We have been following the developments made by the D0 collaboration closely,
and we have benefited from their experience~\cite{ruchti}.
The required number of channels is about the same as for the D0 fiber tracker,
though the mechanical design is quite different.
The KAMI fiber tracking planes will be operated inside of the vacuum
decay volume and will have holes cut out of their center to allow
beam to pass through.
We are considering 500 $\mu$m diameter fibers, assuming that the 
photon yield is sufficient. 

We plan to develop our own prototype device consisting of thin fibers
and a mechanical structure which allows operation in the vacuum.
It is our hope that we can borrow resources from
the D0 collaboration, in particular, the VLPC readout and 
associated cryogenics.

Members of KTeV have recently started working on the D0 fiber tracker in order
to gain critical experience with this new technology which will
eventually be applied to KAMI.

\subsection{R \& D with Main Injector beam}

Physics data taking for the 1999 KTeV run will be completed by the end of FY99.
The next natural step is to bring the Main Injector beam to the
KTeV target station.  In addition to learning a great deal about the
transport of 120 GeV beam, we could begin to understand the performance of
individual detector elements with the lower energy kaons.

\subsubsection{120 GeV beam study with the KTeV target station}

The highest priority will be the measurement of kaon and neutron production
rates with various targeting angles using the Main Injector beam.
The KTeV spectrometer magnet can be run at a lower field setting in order
to measure the kaon flux.
The neutron flux can be measured using the new BA which will be developed and 
used for the KTeV 1999 run.
The beam profile can be studied by using reconstructed kaon decays.

In principle, we can study all the beam parameters necessary for the 
KAMI-Far option without any serious upgrade of the KTeV spectrometer.

\subsubsection{Detector study}

Prototypes of the vacuum veto detector as well as the fiber tracker
will continue to be studied after the KTeV 99 run.
Fully reconstructed kaon decays can be used to understand their performance
in detail.
As mentioned several times, understanding the photon detection efficiency
at low energy ($<$20 MeV) and high energy ($>$1GeV) is most critical
for background-free detection of \kpnn.
We plan to complete all of the studies by the end of the year 
2000 with Main Injector beam.

\newpage
\section{Cost Estimate and Schedule}

\subsection{Cost estimate}
\label{cost}

We have just started a rough cost estimate based on our experience
with KTeV.
This section contains estimates only for the major items.
We expect to complete a more through study by the time of the submission
of a proposal.

\vspace*{0.15in}
\parindent=0.25in

The total cost of the vacuum veto detector is estimated to be \$4.4M.
The cost is broken down in Table \ref{tab:scint_cost}.

The total cost of the fiber tracking system is dominated by the cost of the
VLPC detectors and by the associated electronics and cryogenics.
A cost breakdown appears in Table \ref{tab:fiber_cost}.  This estimate
is for reading out both ends of the fibers.

\begin{table}[here,top,bottom]
\begin{center}
\begin{tabular}{|l|r|r|r|}
\hline
Item                    &Unit cost       &Amount     &Total     \\
\hline\hline
Plastic Scinti.(5 mm t)  & \$6k/ton         &100 ton         & \$0.60M \\
Lead Sheet (1 mm t)      & \$3k/ton         &200 ton         & \$0.60M \\
WLS Fiber (1 mm $\phi$)  & \$0.8/m          &1500 km         & \$1.20M \\
Photo tube (2",Linear)  & \$400/tube       &1000            & \$0.40M \\
Photo tube (2", FM)     & \$1000/tube      &100             & \$0.10M \\
Vacuum tank etc.        &                  &                & \$1.50M \\
\hline\hline
Total Cost              &                  &                & \$4.4M  \\
\hline
\end{tabular}
\caption{Cost breakdown for the vacuum veto system.}
\label{tab:scint_cost}
\end{center}
\end{table}

\begin{table}[here,top,bottom]
\begin{center}
\begin{tabular}{|l|r|r|r|r|}
\hline

Item                    &Unit cost       &No. of Unit     &Total        \\
\hline\hline
VLPC                    & \$25/ch          &98658           & \$2.5M    \\
Preamp etc.             & \$25/ch          &98658           & \$2.5M    \\
Fiber plane etc.        & \$100K/module        &5           & \$0.5M    \\
Optical cables          & \$100K/module        &5           & \$0.5M    \\
\hline\hline
Total cost              &                  &                & \$6.0M    \\

\hline
\end{tabular}
\caption{Cost Breakdown of the KAMI fiber tracking system.}
\label{tab:fiber_cost}
\end{center}
\end{table}

The overall cost estimate for the entire detector is listed in 
Table~\ref{tab:total_cost}.
It totals \$15.6~M. 
This includes the cost of detector upgrades for both the KAMI-Far
and KAMI-Near options.
However, the cost for the new target station for the KAMI-Near option
is not included.

\parindent=0.25in

\begin{table}[here,top,bottom]
\begin{center}
\begin{tabular}{|l|r|r|r|r|}
\hline

Item                    &Unit cost       &Units     &Total        \\
\hline\hline
Beam Collimator        &         &       &  \$0.5M \\
Mask Anti              & \$200K  &  2    &  \$0.4M \\
Photon Veto            & \$3-700K&  7    &  \$4.4M \\ 
Fiber Tracker          & \$1M    &  5    &  \$6.0M \\
Magnet regaping        &         &       &  \$0.3M \\
Charged hodoscope      &         &       &  \$0.1M \\
Vacuum window          &         &       &  \$0.1M \\
CsI Anti               & \$200K  &  2    &  \$0.4M \\
CsI upgrade            &         &       &  \$0.2M \\
Back Anti              &         &       &  \$0.2M \\
Muon range counter     & \$50K   & 10    &  \$0.5M \\
DAQ/Electronics        &         &       &  \$2.0M \\
Others                 &         &       &  \$0.5M \\
\hline\hline 
Total cost             &         &       & \$15.6M  \\

\hline
\end{tabular}
\caption{The overall cost estimate for the KAMI detector.}
\label{tab:total_cost}
\end{center}
\end{table}

\subsection{Cost estimate for detector R \& D}

We have outlined the detector R \&D plan in Section~\ref{det_plan}. 
This can be achieved at a modest cost spread over the next three years. 
Table~\ref{tab:RD_cost} shows the itemized yearly budget to carry out 
this R\&D program.
Also shown here is the cost of the Back Anti for the KTeV 99 run,
since this development is directly related to KAMI.
 
\begin{table}[here,top,bottom]
\begin{center}
\begin{tabular}{|c|l|r|c|}
\hline
 
Year &       Items                           & Cost  & Subtotal\\
\hline\hline
     &   &    & \\
1998 & {\bf Vacuum Photon Veto Prototype}      &   & \$60K    \\
 & \hspace*{0.5in} Extruded (or injection molded) scintillator & \$25K &\\
 & \hspace*{0.5in} WLS fiber 				       & \$10K &\\
 & \hspace*{0.5in} Lead sheet                                  & \$5K &\\
 & \hspace*{0.5in} Mechanical structure                        & \$10K &\\
 & \hspace*{0.5in} PMT and readout electronics                 & \$10K &\\
 &      &       & \\
    & {\bf Back Anti, EM Section (used in KTeV 99)}  & & \$50K  \\
 & \hspace*{0.5in} Quartz (or lucite) plates                   & \$30K &\\
 & \hspace*{0.5in} Light guide                                 & \$10K &\\
 & \hspace*{0.5in} Mechanical modificaion to KTeV Back Anti    & \$10K &\\
 &      &       & \\
    & {\bf Back Anti,Hadronic Section (used in KTeV 99)}   & & \$40K  \\
 & \hspace*{0.5in} Scintillator plate                          & \$10K &\\
 & \hspace*{0.5in} Light guide                                 & \$5K &\\
 & \hspace*{0.5in} Iron plate                                  & \$5K &\\
 & \hspace*{0.5in} Support frame                               & \$10K &\\
 & \hspace*{0.5in} PMT and readout electronics                 & \$10K &\\
 &      &       & \\
\hline
     &   &    & \\
1999 & {\bf Fiber Tracker Prototype}                            & & \$60K  \\
 & \hspace*{0.5in} Thin fiber ribbon                           & \$10K &\\
 & \hspace*{0.5in} Supporting frame                            & \$10K &\\
 & \hspace*{0.5in} Vacuum feed through                         & \$5K &\\
 & \hspace*{0.5in} Clear light guide                           & \$5K &\\
 & \hspace*{0.5in} VLPC, readout electronics and cryogenics    & \$30K &\\
     &   &    & \\ 
 & {\bf Vacuum Photon Veto Prototype (Second version) } 
     & & \$50K \\
     &   &    & \\
 & {\bf Readout electronics R\&D }   & & \$30K  \\
     &   &    & \\
\hline
     &   &    & \\
2000 & {\bf Vacuum Veto (Pre-production version) }    & & \$50K \\
     &   &    & \\
     & {\bf Fiber Tracker (Pre-production version) } & & \$50K \\
     &   &    & \\ 
\hline\hline
 & {\bf Total Cost}                        &       &  \$390K \\
\hline
\end{tabular}
\caption{Cost estimate for KAMI detector R \& D.}
\label{tab:RD_cost}
\end{center}
\end{table}

\subsection{Schedule}

In order to finalize the detector design and operating conditions,
we consider the following two phases of R \& D as the most critical:

\begin{enumerate}
\item Detailed study of the three new major detectors systems; 
vacuum photon veto, Back Anti and fiber tracker;  

\item Understand and optimize the kaon beam line, including target 
position, targeting angle and collimator design for both \kpnn 
and charged decay modes.
\end{enumerate}

We expect to complete all necessary studies by the end of year 2000
so that we can start construction of major hardware in 2001.
Table \ref{tab:milestones} lists the milestones necessary to meet this goal.
Here we assume that we start from a dedicated \kpnn run with 
the KAMI-Far configuration at the earliest possible date, 
since it requires the least expensive upgrades to the existing KTeV detector.
Our final goal is to construct the KAMI-Near configuration with
full tracking capability by the year 2005.
We believe that this plan is compatible with Fermilab's plans over
this time period.

\begin{table}[here,top,bottom]
\begin{center}
\begin{tabular}{|c|l|}
\hline
Year            & Milestone \\
\hline\hline
                & \\ 
1997 & {\bf Submission of EOI.} \\
                & \\
1998 & Construction of the prototype of Vacuum veto and BA.   \\
     &{\bf Submission of the scientific proposal.}\\
       
                & \\
1999 & KTeV 99 run.\\
     & First tests of photon veto, BA, fiber tracker prototypes.\\
     & {\bf Submission of the Technical Design Report.}\\
                & \\
2000 & First delivery of the Main Injector beam onto the KTeV target.\\
     & Neutral beam study.\\
     & Tests of detector prototypes (veto, BA, tracker) continue.\\
     & Final design of the detector completed.\\
                & \\
2001 & Photon veto construction. \\
                & \\
2002 & Partial construction of fiber tracker.\\
                & \\
2003 & {\bf First physics data taking of \kpnn with KAMI-Far.}\\
     & Full construction of fiber tracker.\\
                & \\
2004 & Second year of KAMI-Far operation, including charged modes.\\
     & KAMI-Near target station construction.\\
                & \\
2005 & {\bf First physics data taking with KAMI-Near.}\\
                & \\
2006 & KAMI-Near data taking continues... \\
                & \\
\hline
\end{tabular}
\caption{KAMI milestones.}
\label{tab:milestones}
\end{center}
\end{table}

\newpage
\section{Conclusion}

The neutral kaon program at Fermilab has a long and distinguished
history.  We believe the future is equally bright.
With the advent of the Main Injector, 
we anticipate a great opportunity for incisive exploration 
of CP violation and other 
important physics in the kaon sector. 

In this Expression of Interest, we have demonstrated that this bright 
future is made possible by combining the existing KTeV infrastructure with 
new state-of-the-art detector technology which has been largely developed 
at Fermilab for other experiments.

\vspace*{0.15in}
\parindent=0.25in
 
The physics importance of \kpnn is compelling and it appears feasible 
to detect a large number of events which will allow a measurement of
the $\eta$ parameter 
with an accuracy which is compatible with the $sin(2\beta)$ 
measurements proposed at B factories.
  
Many other important decay modes are also accessible at sensitivity levels
between $10^{-13}$ and $10^{-14}$.  These measurements are not possible at
the dedicated \kpnn experiments proposed at other Labs.

\vspace*{0.15in}
\parindent=0.25in
   
The KAMI project represents an efficient and cost effective plan
for extending Fermilab's neutral kaon program into the future.
We plan to continue our current studies and will engage 
in an aggressive R\&D program
in order to make the best use possible of the formidable neutral
kaon factory which will be provided by the Main Injector. 

\vspace*{0.5in}
\section{Acknowledgements}

As we prepare this Expression of Interest to map out the future of
the neutral kaon program at Fermilab, we would like to acknowledge
the work of all of our KTeV Collaborators who have made the present
possible.  Without their dedicated efforts to make KTeV a success,
it would not be possible to progress towards the future.  We would also
like to acknowledge the efforts of the numerous engineers and technicians 
from all of the collaborating institutions who helped to design and 
build the KTeV detector, much of which will be used for KAMI.  We
would like to thank Fermilab for its continuing support of our program.
In particular, we would like to thank the Computing Division, Beams Division
and Particle Physics Division for their invaluable support before, during
and after the first KTeV run.  Many thanks also to Elizabeth Pod from
the University of Chicago for her help with several of the figures in
this document.

\end{document}